\newif\ifARXIV
\ARXIVfalse

\newif\ifAppendix
\Appendixfalse

\documentclass[sigconf,screen]{acmart}
\usepackage{graphicx}

\newif\ifSPACEHACK
\SPACEHACKfalse

\newif\ifDEBUG
\DEBUGfalse

\newif\ifANONYMOUS
\ANONYMOUSfalse

\newif\ifHISTOGRAMEQUALIZATION
\HISTOGRAMEQUALIZATIONfalse

\newif\ifJOURNALEXTENSION
\JOURNALEXTENSIONfalse

\usepackage{amsmath}
\usepackage{graphicx}
\graphicspath{{./images/}}

\usepackage{xcolor}
\usepackage{booktabs} % For formal tables \toprule
\usepackage{xspace} % Put an \xspace within numeric macros
\usepackage{enumitem}
\usepackage{array}
\usepackage{caption}
\usepackage{subcaption}
\usepackage{diagbox}
\usepackage{multirow}
\usepackage{lscape}
\usepackage{hyperref}
\usepackage{wrapfig}
\usepackage{circledsteps}
\usepackage{algorithm}
\usepackage{algpseudocode}
\usepackage{listings}
\usepackage{adjustbox}
\usepackage{microtype}
\usepackage{xurl}

\usepackage[noadjust]{cite}

\definecolor{codecomment}{RGB}{0,128,0}
\definecolor{codestring}{RGB}{128,0,0}

\lstdefinestyle{algorithmstyle}{
  language=Python,
  basicstyle=\small\ttfamily,
  keywordstyle=\bfseries,
  commentstyle=\color{codecomment},
  stringstyle=\color{codestring},
  numbers=left,
  numberstyle=\tiny\color{gray},
  frame=single,
  breaklines=true,
  captionpos=b,
}

\newcommand{\circledstep}[1]{\Circled{#1}}
\makeatletter
\def\cl@chapter{}
\makeatother
\usepackage{cleveref}

\crefformat{section}{\S#2#1#3}
\crefname{figure}{Figure}{Figures}
\crefname{appendix}{Appendix}{Appendices}
\crefname{table}{Table}{Tables}
\crefname{algorithm}{Algorithm}{Algorithms}
\crefname{listing}{Listing}{Listings}
\crefname{theorem}{Theorem}{Theorems}
\crefname{thm}{Theorem}{Theorems}
\crefname{lemma}{Lemma}{Lemmata}
\crefname{equation}{Eqt.}{Eqts.}
\crefformat{Grammar}{Grammar #1}

\usepackage{enumitem}
\usepackage{booktabs}
\usepackage{svg}
\usepackage{listings}

\newcommand{\ie}{\textit{i.e.,} }
\newcommand{\eg}{\textit{e.g.,} }
\newcommand{\etal}{\textit{et al.}\xspace}
\newcommand{\etals}{\textit{et al.'s}\xspace}

\usepackage{soul}

\newcommand{\TODO}[1]{\hl{#1}}
\ifDEBUG
    \newcommand{\JD}[1]{\textcolor{purple}{[JD: #1]}}
    \newcommand{\AG}[1]{\textcolor{olive}{[AG: #1]}}
    \newcommand{\WJ}[1]{\textcolor{cyan}{[WJ: #1]}}
    \newcommand{\GKT}[1]{\textcolor{brown}{[GKT: #1]}}
    \newcommand{\MH}[1]{\textcolor{pink}{[MH: #1]}}
    \newcommand{\NMS}[1]{\textcolor{orange}{[NMS: #1]}}
    \newcommand{\WPM}[1]{\textcolor{red}{[Trey: #1]}}
    \newcommand{\TRS}[1]{\textcolor{olive}{[Taylor: #1]}}
    \newcommand{\PJ}[1]{\textcolor{blue}{[PJ: #1]}}
    \newcommand{\AT}[1]{\textcolor{magenta}{[AT: #1]}}
    \newcommand{\YHL}[1]{\textcolor{green}{[YHL: #1]}}
    \newcommand{\JW}[1]{\textcolor{magenta}{[Joe: #1]}}
    \newcommand{\NJE}[1]{\textcolor{green}{[Nick E: #1]}}
    \newcommand{\AS}[1]{\textcolor{magenta}{[Anusha: #1]}}
    \newcommand{\EK}[1]{\textcolor{magenta}{[Erik: #1]}}

\else
    \newcommand{\JD}[1]{}
    \newcommand{\AG}[1]{}
    \newcommand{\WJ}[1]{}
    \newcommand{\GKT}[1]{}
    \newcommand{\NS}[1]{}
    \newcommand{\NV}[1]{}
    \newcommand{\NMS}[1]{}
    \newcommand{\WPM}[1]{}
    \newcommand{\TRS}[1]{}
    \newcommand{\MH}[1]{}
    \newcommand{\YHL}[1]{}
    \newcommand{\PJ}[1]{}
    \newcommand{\AT}[1]{}
    \newcommand{\JW}[1]{}
    \newcommand{\NJE}[1]{}
    \newcommand{\AS}[1]{}
    \newcommand{\EK}[1]{}
\fi

\newcommand{\code}[1]{{\small\texttt{#1}}\xspace}

\usepackage{tcolorbox}

\newcommand{\myparagraph}[1]{\noindent\hspace{0.15cm}\textbf{#1:}}
\newcommand{\inlinequote}[1]{``\emph{#1}''}

\newcounter{finding}

\newcommand{\newfinding}{
  \refstepcounter{finding} % Increments the counter and allows referencing
  \textbf{Finding \thefinding.} % Displays the counter
}

\usepackage[subtle]{savetrees}

\newcommand\PyTorchClosedIssues{20,782\xspace}
\newcommand\PyTorchModuleOnnxIssues{959\xspace}
\newcommand\PyTorchPostTLFilterIssues{327\xspace}

\newcommand\TFClosedIssues{792\xspace}
\newcommand\TFClosedIssuesFilter{649\xspace}
\newcommand\TFPostTLFilterIssues{242\xspace}

\newcommand\HFTorrentTotalRepos{63,182\xspace}

\newcommand\TotalSamples{200\xspace}
\newcommand\KappaCoeffCauses{0.90\xspace}
\newcommand\KappaCoeffSymptoms{0.95\xspace}

\newcommand\TotalSamplesAdded{111\xspace}
\newcommand\YearAdded{5\xspace}
\newcommand\Years{(2018-2022)\xspace}
\newcommand\AvgSamplesAdded{22\xspace}

\newcommand\TorchRecall{$0.81$\xspace}
\newcommand\TorchPrecision{$0.94$\xspace}
\newcommand\TFRecall{$0.82$\xspace}
\newcommand\TFPrecision{$0.93$\xspace}
\newcommand\HandLabelledIssues{$50$\xspace}
\newcommand\RecallPrecisionThreshold{$\geq0.8$\xspace}

\newcommand\TotalDriftSamples{20\xspace}
\newcommand\DriftTesting{1.0\xspace}
\newcommand\DriftTestingYearlyCauses{0.94\xspace}
\newcommand\DriftTestingYearlySymptoms{1.0\xspace}

\newcommand\GenLowerBound{15\xspace}
\newcommand\GenUpperBound{100\xspace}
\newcommand\GenIncrements{5\xspace}
\newcommand\TFOps{56\xspace}
\newcommand\PTOps{71\xspace}
\newcommand\PTOpsRemoved{1\xspace}
\newcommand\TFOpsRemoved{3\xspace}

\newcommand\OpsetAdditions{149\xspace}
\newcommand\OpsetUpdates{217\xspace}

\newcommand\HFArchitectures{58\xspace}
\newcommand\HFDistinctArch{112\xspace}

\newcommand\CrashPercent{56\%\xspace}
\newcommand\WrongModelPercent{33\%\xspace}
\newcommand\BadPerfPercent{2\%\xspace}
\newcommand\HangPercent{0\%\xspace}
\newcommand\BuildFailPercent{3\%\xspace}
\newcommand\UnreportedFailPercent{8\%\xspace}

\newcommand\ShenCrashPercent{63\%\xspace}
\newcommand\ShenWrongModelPercent{28\%\xspace}
\newcommand\ShenBadPerfPercent{2\%\xspace}
\newcommand\ShenHangPercent{1\%\xspace}
\newcommand\ShenBuildFailPercent{1\%\xspace}
\newcommand\ShenUnreportedFailPercent{6\%\xspace}

\newcommand\ShenIncompExt{16\xspace}
\newcommand\ShenIncompInt{5\xspace}
\newcommand\ShenIncompRes{4\xspace}
\newcommand\ShenTypeNode{13\xspace}
\newcommand\ShenTypeConv{20\xspace}
\newcommand\ShenTypeTens{42\xspace}
\newcommand\ShenAlgo{59\xspace}
\newcommand\ShenShape{67\xspace}
\newcommand\ShenAPI{35\xspace}
\newcommand\ShenOthers{98\xspace}
\newcommand\ShenTotalCauses{359\xspace}

\newcommand\ShenIncompExtPercent{4\%\xspace}
\newcommand\ShenIncompIntPercent{1\%\xspace}
\newcommand\ShenIncompResPercent{1\%\xspace}
\newcommand\ShenTypeNodePercent{4\%\xspace}
\newcommand\ShenTypeConvPercent{6\%\xspace}
\newcommand\ShenTypeTensPercent{12\%\xspace}
\newcommand\ShenAlgoPercent{16\%\xspace}
\newcommand\ShenShapePercent{19\%\xspace}
\newcommand\ShenAPIPercent{10\%\xspace}
\newcommand\ShenOthersPercent{27\%\xspace}
\newcommand\ShenTotalCausesPercent{100\%\xspace}

\newcommand\IncompExt{28\%\xspace}
\newcommand\IncompInt{1\%\xspace}
\newcommand\IncompRes{0\%\xspace}
\newcommand\TypeNode{23\%\xspace}
\newcommand\TypeConv{3\%\xspace}
\newcommand\TypeTensor{2\%\xspace}
\newcommand\Algo{12\%\xspace}
\newcommand\Shape{11\%\xspace}
\newcommand\API{6\%\xspace}

\newcommand\BinnedOther{32\xspace}
\newcommand\BinnedOtherTF{17\xspace}
\newcommand\BinnedOtherPT{15\xspace}
\newcommand\BinnedOtherPercentage{16\%\xspace}

\newcommand\BinnedJointOtherCrashTF{8\xspace}
\newcommand\BinnedJointOtherCrashPT{8\xspace}
\newcommand\BinnedJointOtherWrongModelTF{0\xspace}
\newcommand\BinnedJointOtherWrongModelPT{3\xspace}
\newcommand\BinnedJointOtherBadPerfTF{0\xspace}
\newcommand\BinnedJointOtherBadPerfPT{0\xspace}
\newcommand\BinnedJointOtherBuildFailTF{3\xspace}

\newcommand\BinnedJointOtherUnreportedTF{6\xspace}
\newcommand\BinnedJointOtherUnreportedPT{2\xspace}
\newcommand\BinnedJointOtherOtherTF{9\xspace}
\newcommand\BinnedJointOtherOtherPT{4\xspace}
\newcommand\TotalSuccessForDiff{4,819\xspace}

\newcommand\TotalAttempted{2,202\xspace}

\newcommand\TotalHFModels{1,761\xspace}

\newcommand\TotalSynModel{3,544\xspace}
\newcommand\TotalRealModels{1,605\xspace}

\newcommand\SuccessConvNN{2,742\xspace}
\newcommand\TrueSuccessHFPTM{1,361\xspace}
\newcommand\TrueSuccessHFPTMPercent{75\%\xspace}
\newcommand\TrueSuccessHFTFM{1,212\xspace}
\newcommand\TrueSuccessHFTFMPercent{68\%\xspace}

\newcommand\CrashesHFTFHugging{456\xspace}
\newcommand\CrashesHFTFHuggingPercent{26\%\xspace}
\newcommand\CrashesHFPTHugging{342\xspace}
\newcommand\CrashesHFPTHuggingPercent{20\%\xspace}
\newcommand\CrashesHFPT{20\xspace}
\newcommand\CrashesHFPTPercent{2\%\xspace}
\newcommand\CrashesHFTF{65\xspace}
\newcommand\CrashesHFTFPercent{4\%\xspace}

\newcommand\CrashesORTHFPT{27\xspace}
\newcommand\CrashesORTHFPTPercent{2\%\xspace}
\newcommand\CrashesORTHFTF{19\xspace}
\newcommand\CrashesORTHFTFPercent{1\%\xspace}

\newcommand\SuccLoadedPTSyn{389\xspace}
\newcommand\FailureHFPTSemantic{11\xspace}
\newcommand\FailureHFPTSemanticPercent{1\%\xspace}
\newcommand\FailureHFTFSemantic{9\xspace}
\newcommand\FailureHFTFSemanticPercent{1\%\xspace}
\newcommand\FailureHFSemanticTotal{20\xspace}

\newcommand\FailureNNPTSemantic{94\xspace}

\newcommand\TotalTFConGen{1,820\xspace}
\newcommand\TotalPTConGen{1,724\xspace}

\newcommand\ConvCrashTFConGen{800\xspace}
\newcommand\ConvCrashTFConGenPercent{44\%\xspace}
\newcommand\ConvCrashPTConGen{2\xspace}
\newcommand\ConvCrashPTConGenPercent{1\%\xspace}
\newcommand\TotalConvCrash{802\xspace}

\newcommand\ORTCrashTFConGen{757\xspace}
\newcommand\ORTCrashTFConGenPercent{42\%\xspace}
\newcommand\ORTCrashPTConGen{741\xspace}
\newcommand\ORTCrashPTConGenPercent{42\%\xspace}

\newcommand\SuccLoadedPTConGen{1,781\xspace}

\newcommand\SemanticTFConGen{220\xspace}
\newcommand\SemanticTFConGenPercent{12\%\xspace}
\newcommand\SemanticPTConGen{100\xspace}
\newcommand\SemanticPTConGenPercent{6\%\xspace}

\newcommand\SuccessfulTFConGen{43\xspace}
\newcommand\SuccessfulTFConGenPercent{2\%\xspace}
\newcommand\SuccessfulPTConGen{881\xspace}
\newcommand\SuccessfulPTConGenPercent{51\%\xspace}

\newcommand\FailureSemanticTotalSyn{320\xspace}

\newcommand\IncorrectTotalPercentage{6}

\newcommand\IncorrectTotalSynPercentage{9}
\newcommand\IncorrectTotalRealPercentage{1}
\newcommand\ConverterFailures{85\xspace}
\newcommand\ConverterAttempted{3,522\xspace}
\newcommand\ConverterFailuresPercentage{2\xspace}
\newcommand\ConverterSuccessPercentage{90\xspace}
\newcommand\UnsucessfulUniqueCrashes{4\xspace}
\newcommand\SuccessLoadNN{1,244\xspace}

\newcommand\IncorrectCrashTotalPercent{11}

\newcommand\ORTUnique{11\xspace}

\newcommand{\NumInteropUsers}{92\xspace}
\newcommand{\NumONNXUsers}{38\xspace}
\newcommand{\NumSurveyRes}{108\xspace}
\newcommand{\ResponseRate}{4\%\xspace}
\newcommand{\SurveyTotalResNum}{117\xspace}

\newcommand{\NumofORGUsers}{1,985\xspace}
\newcommand{\NumofPROUsers}{228\xspace}
\newcommand{\SurveyCompensation}{\$10\xspace}

\setcopyright{rightsretained}
\acmDOI{10.1145/3650212.3680374}
\acmYear{2024}
\copyrightyear{2024}
\acmISBN{979-8-4007-0612-7/24/09}
\acmConference[ISSTA '24]{Proceedings of the 33rd ACM SIGSOFT International Symposium on Software Testing and Analysis}{September 16--20, 2024}{Vienna, Austria}
\acmBooktitle{Proceedings of the 33rd ACM SIGSOFT International Symposium on Software Testing and Analysis (ISSTA '24), September 16--20, 2024, Vienna, Austria}
\acmSubmissionID{issta24main-p1204-p}
\received{2024-04-12}
\received[accepted]{2024-07-03}

\begin{document}

\newcommand{\MyTitle}[1]{}
\newcommand{\MyShortTitle}[1]{}

\renewcommand{\MyTitle}{Interoperability Failures in Deep Learning Model Converters: A Case Study in the ONNX Ecosystem}
\renewcommand{\MyShortTitle}{\MyTitle}

\ifARXIV
\renewcommand{\MyTitle}{Analysis of Failures and Risks in Deep Learning Model Converters: \\
% Software: 
A Case Study in the ONNX Ecosystem
}
\renewcommand{\MyShortTitle}{ONNX Survey and Failure Analysis}
\else
    \renewcommand{\MyTitle}{Analysis of Failures in ONNX Deep Learning Model Converters}
    \renewcommand{\MyTitle}{Interoperability Failures in Deep Learning Model Converters: A Case Study in the ONNX Ecosystem}
    \renewcommand{\MyTitle}{Interoperability Failures in ONNX Model Converters}
    \renewcommand{\MyTitle}{Interoperability Failures in ONNX Deep Learning Model Converters}
    \renewcommand{\MyTitle}{Interoperability Failures in ONNX Deep Learning Converters}
    \renewcommand{\MyTitle}{A Failure Analysis of ONNX Deep Learning Model Converters}
    \renewcommand{\MyTitle}{A User Survey and Failure Analysis of ONNX Model Converters}
    \renewcommand{\MyTitle}{Interoperability in Deep Learning: \\ A User Survey and Failure Analysis of ONNX Model Converters}
    %\renewcommand{\MyShortTitle}{Interoperability Failures in Deep Learning Model Converters}
    %\renewcommand{\MyShortTitle}{Interoperability Failures in Deep Learning Model Converters}
    
    %\renewcommand{\MyTitle}{Interoperability in Deep Learning: \\ }
    %\subtitle{A User Survey and Failure Analysis of ONNX Model Converters}
    
    \renewcommand{\MyTitle}{Interoperability in Deep Learning: \\ A User Survey and Failure Analysis of ONNX Model Converters}
    \renewcommand{\MyShortTitle}{Interoperability in Deep Learning}
\fi

\ifARXIV
    %\authorrunning{Jajal \etal}% Part of LEFT running header
    %\shortauthors{Jajal \etal}
\fi

% \title[\MyShortTitle]{\MyTitle}
\title{Interoperability in Deep Learning: A User Survey and Failure Analysis of ONNX Model Converters}

% \ifARXIV
% \author{Purvish Jajal}
% \email{pjajal@purdue.edu}
% \affiliation{%
%   \institution{Purdue University}
%   \country{USA}
% }

% \author{Wenxin Jiang}
% \email{jiang784@purdue.edu}
% \affiliation{%
%   \institution{Purdue University}
%   \country{USA}
% }

% \author{Arav Tewari}
% \email{tewari6@purdue.edu}
% \affiliation{%
%   \institution{Purdue University}
%   \country{USA}
% }

% \author{Erik Kocinare}
% \email{ekocinar@purdue.edu}
% \affiliation{%
%   \institution{Purdue University}
%   \country{USA}
% }

% \author{Joseph Woo}
% \email{woo56@purdue.edu}
% \affiliation{%
%   \institution{Purdue University}
%   \country{USA}
% }

% \author{Anusha Sarraf}
% \email{asarraf@purdue.edu}
% \affiliation{%
%   \institution{Purdue University}
%   \country{USA}
% }

% \author{Yung-Hsiang Lu}
% \email{yunglu@purdue.edu}
% \affiliation{%
%   \institution{Purdue University}
%   \country{USA}
% }
  
% \author{George K. Thiruvathukal}
% \email{gkt@cs.luc.edu}
% \affiliation{%
%   \institution{Loyola University Chicago}
%   \country{USA}
% }

% \author{James C. Davis}
% \email{davisjam@purdue.edu}
% \affiliation{%
%   \institution{Purdue University}
%   \country{USA}
% }

% \else

\author{Purvish Jajal}
\orcid{0000-0002-1199-6363}
\affiliation{%
  \institution{Purdue University}
  \city{West Lafayette}
  \country{USA}
}
\email{pjajal@purdue.edu}

\author{Wenxin Jiang}
\orcid{0000-0003-2608-8576}
\affiliation{%
  \institution{Purdue University}
  \city{West Lafayette}
  \country{USA}
}
\email{jiang784@purdue.edu}

\author{Arav Tewari}
\orcid{0000-0002-1512-858X}
\affiliation{%
  \institution{Purdue University}
  \city{West Lafayette}
  \country{USA}
}
\email{tewari6@purdue.edu}

\author{Erik Kocinare}
\orcid{0009-0007-9151-5008}
\affiliation{%
  \institution{Purdue University}
  \city{West Lafayette}
  \country{USA}
}
\email{ekocinar@purdue.edu}

\author{Joseph Woo}
\orcid{0009-0006-7686-4157}
\affiliation{%
  \institution{Purdue University}
  \city{West Lafayette}
  \country{USA}
}
\email{woo56@purdue.edu}

\author{Anusha Sarraf}
\orcid{0009-0000-1384-1990}
\affiliation{%
  \institution{Purdue University}
  \city{West Lafayette}
  \country{USA}
}
\email{asarraf@purdue.edu}

\author{Yung-Hsiang Lu}
\orcid{0000-0002-5491-7661}
\affiliation{%
  \institution{Purdue University}
  \city{West Lafayette}
  \country{USA}
}
\email{yunglu@purdue.edu}

\author{George K. Thiruvathukal}
\orcid{0000-0002-0452-5571}
\affiliation{%
  \institution{Loyola University Chicago}
  \city{Chicago}
  \country{USA}
}
\email{gkt@cs.luc.edu}

\author{James C. Davis}
\orcid{0000-0003-2495-686X}
\affiliation{%
  \institution{Purdue University}
  \city{West Lafayette}
  \country{USA}
}
\email{davisjam@purdue.edu}
% \fi

% \date{Received: date / Accepted: date}
\renewcommand{\shortauthors}{Jajal, Jiang, Tewari, Kocinare, Woo, Sarraf, Lu, Thiruvathukal, Davis}

\begin{abstract} \label{sec: abstract}
Software engineers develop, fine-tune, and deploy deep learning (DL) models using a variety of development frameworks and runtime environments.
\textit{DL model converters} move models between frameworks and to runtime environments.
Conversion errors compromise model quality and disrupt deployment.
However, the failure characteristics of DL model converters are unknown, adding risk when using DL interoperability technologies.

This paper analyzes failures in DL model converters.
We survey software engineers about DL interoperability tools, use cases, and pain points (N=\NumInteropUsers).
Then, we characterize failures in model converters associated with the main interoperability tool, ONNX (N=200 issues in PyTorch and TensorFlow).
Finally, we formulate and test two hypotheses about structural causes for the failures we studied.
We find that the node conversion stage of a model converter accounts for $\sim$75\% of the defects, and that 33\% of reported failure are related to semantically incorrect models.
The cause of semantically incorrect models is elusive, but models with behaviour inconsistencies share operator sequences.
Our results motivate future research on making DL interoperability software simpler to maintain, extend, and validate.
Research into behavioural tolerances and architectural coverage metrics could be fruitful.
\end{abstract}



\maketitle

% \vspace{-0.18cm}
\section{Introduction} \label{sec:Intro}
% \vspace{-0.05cm}

Deep Learning (DL) achieves state-of-the-art performance in many domains~\citep{DlAutonomousDrivingSurvey, DlMedicalImaging}.
Software engineers engage in many activities for deep learning, including developing, re-using, fine-tuning, and deploying DL models~\citep{Amershi2019SE4MLCaseStudy, Han2021PTM, jiang_empirical_2023_1, jiang2023challengespracticesdeeplearning}.
% The DL (DL) software ecosystem is heterogeneous with a variety of software and tools available for the development and use of DL applications. 
They use tools at each stage: 
  DL frameworks for development
    (\eg
    PyTorch~\citep{pytorch});
    %TensorFlow~\citep{tensorflow}),
  DL model registries for re-use
    (\eg
      HuggingFace~\citep{HuggingFaceWeb1}),
      %ONNX Model Zoo~\citep{XXX});
  and DL compilers for deployment platforms
    (\eg
      TVM~\citep{TVM_Chen_Moreau_Jiang_Zheng}).
      %ONNX Runtime~\citep{Li_Liu_Liu_Sun_You_Yang_Luan_Gan_Yang_Qian_2021},
      %Glow~\citep{Glow}).
%\cref{fig:onnx_ecosystem} depicts some tools that engineers use for each activity.
% Engineers may use different tools and frameworks in each activity.
Preferably, these tools would be \emph{interoperable}, so that DL models can move seamlessly from one to another.
% \emph{Model converters} offer interoperability between DL frameworks and DL compilers (\cref{fig:onnx_ecosystem}).
Model conversion errors disrupt engineering workflows or compromise the resulting models~\citep{ConvertingDLempirical}.
High-quality model converters are crucial to the deep learning ecosystem.

%The DL ecosystem relies on interoperability software called model converters.

\begin{figure}[t]
    \centering
    \includegraphics[width=0.78\columnwidth]{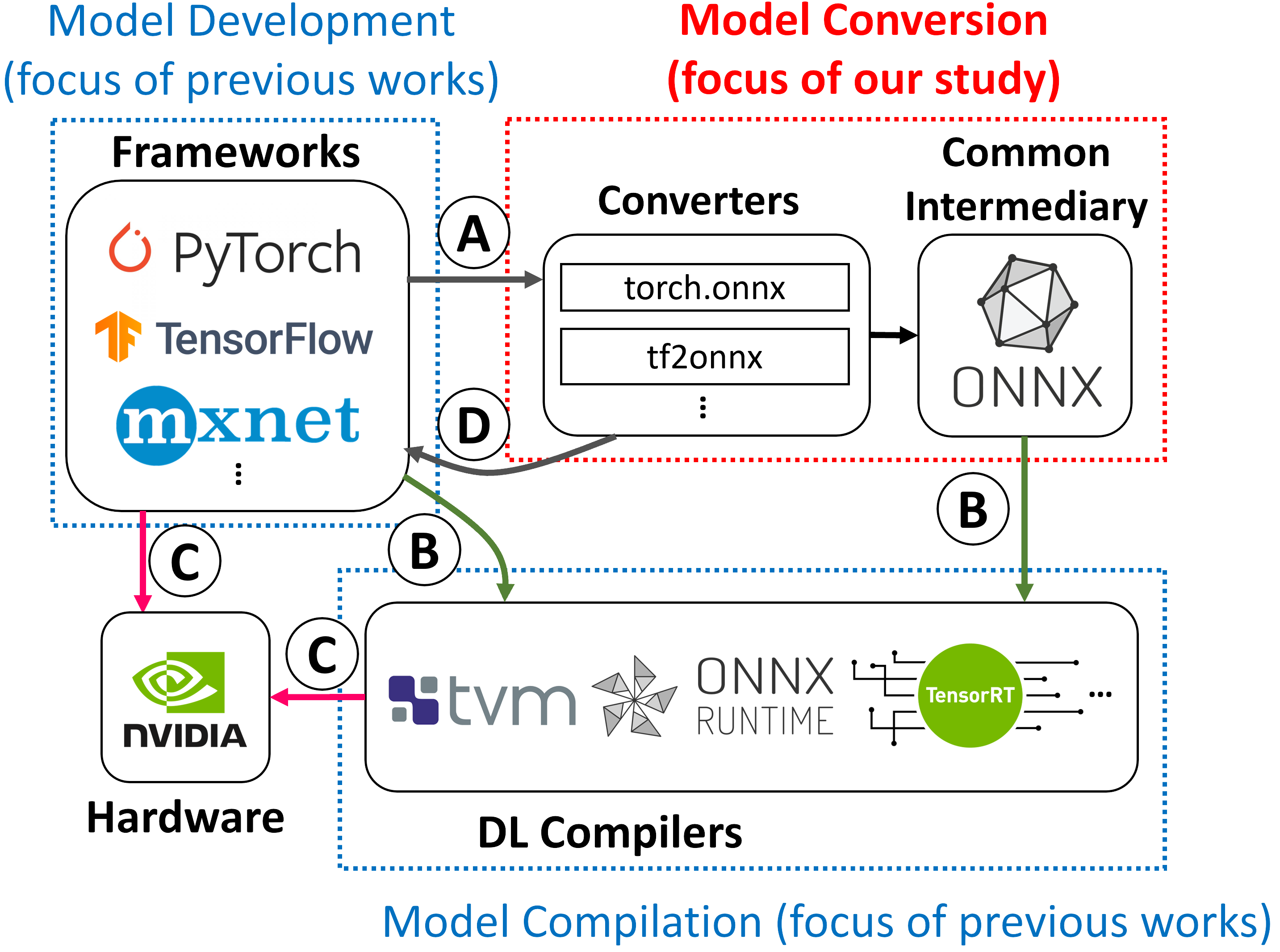}
    \caption{
        % \small
        Paths from model development to deployment on hardware.
        Model interoperability facilitates reuse across frameworks and deployment environments~\citep{davis2023reusing}.
        %focuses on model reuse and is defined as the ability of deep learning (DL) software to exchange DL models.
        %The ONNX Ecosystem is an exemplar of a DL interoperability framework.
        %The arrows indicates paths that practitioners can take through the ONNX ecosystem to deploy a model. 
        \circledstep{A} represents \textbf{\textit{model conversion}} to a common intermediary. 
        \circledstep{B} represents \textbf{\textit{compilation}}.
        \circledstep{C} represents \textbf{\textit{model deployment}}.
        \circledstep{D} represents \textbf{\textit{model conversion}} to a framework.
        %The red box indicates the focus of this study. 
        %The ecosystem consists of software interoperable with ONNX.
        %Per~\cref{sec:Background},
        % Downward path
        %     %from Frameworks to DL compilers
        %     uses \emph{pairwise mappings}.
        % %Each DL compiler supports DL models from one or more frameworks via customized loaders.
        % Rightward path
        %     %from Frameworks to DL compilers
        %     uses a \emph{common intermediary} (ONNX).
        %Frameworks export to a common format for use by a DL compiler.
        % Frameworks are compatible through the use of converters, which convert to ONNX representation.  
        % Compilers themselves contain converters but these converters either convert directly from framework representations of models or from ONNX. 
        %Converters in DL compilers have been studied by Liu \etal~\citep{NNSmith}. 
        % \JD{FOR YHL: This figure is yuge. Scale it down and use wrapfig to weave it with the text. Or make it wide instead of tall.}
    }
    \label{fig:onnx_ecosystem}
% \end{wrapfigure}
% \vspace{-0.5cm}
\end{figure}

Researchers have characterized failures in most of the DL ecosystem, but not in model converters.
As depicted in~\cref{fig:onnx_ecosystem}, previous works have considered
  DL development frameworks~\citep{Chen2022DLFrameworkBug_1, PerformanceAccuracyBugsDLFrameworks, islam2019comprehensive, Pham2020AnalysisofVarianceinDLSWSystems, tan_exploratory_2022}
  % \JD{I think there is relevant work from UIowa (Hridesh Rajan) and Purdue (Lin Tan). Cite a little here.}\WJ{UIowa: \citep{islam2019comprehensive, Islam2020RepairingDNN:FixpatternsChallenges}; Lin: \citep{Pham2020AnalysisofVarianceinDLSWSystems} \citep{Guo2019DLDevelopmentandDeploymentAcrossDifferentFrameworksandPlatforms} includes development and deployment. Also consider this one ``An Exploratory Study of Deep Learning Supply Chain''}
  %DL registries~\citep{Jiang2022SCORED,jiang_empirical_2023_1, montes_discrepancies_2022},
  and
  DL deployment compilers and runtimes~\citep{DLcompilerbugs, Guo2019DLDevelopmentandDeploymentAcrossDifferentFrameworksandPlatforms}. 
In contrast, prior research on DL model converters is limited to measuring conversions of 5 DL models~\citep{ConvertingDLempirical}.
We lack systematic knowledge of failure symptoms, causes, and patterns in DL model converters.

In this work, we analyze failures of DL model converters.
We first survey software engineers 
% (N=\NumSurveyRes),
(N=\NumInteropUsers),
focusing on their experiences with DL interoperability tools, their use cases, and failure modes encountered.
Then we analyze failures in DL model converters to the ONNX IR (Open Neural Network eXchange's Intermediate Representation), the most prominent interoperability target for DL models.
%\emph{First}, we study \emph{how} ONNX model converters fail using a failure analysis.
% \emph{First}, we conduct a retrospective failure analysis of the ONNX model converters for PyTorch and TensorFlow.
We sample 200 closed GitHub issues (100 per converter)
and determine failure symptoms, causes, and locations. %, and the relationship between these factors.
  % and 
  % trends over time, especially the correlation of failures with changes in the ONNX specification.
Finally, we examine two possible root causes of model converter failures, testing hypotheses about specification updates and model types. %on opsets, model types, and operator/layer sequences.
% converting and evaluating real and synthetic models.
% \emph{Second}, we measure the present frequency of failure by converting and evaluating real and synthetic models. 

% \WJ{TODO: 1-2 sentences about survey results.}

Our survey results show that ONNX is the most popular interoperability tool. It is primarily utilized for model deployment and framework conversion, with crashes and performance degradation being the most reported problems.
Our failure analysis found that:
  common symptoms are crashes and incorrect model behaviors;
  common causes are incompatibility and type problems; 
  and these issues tend to occur in a converter's graph translation and graph optimization components.
%Crashes can be caused by both incompatibility and type problems, while wrong models are attributable to type problems, algorithmic errors, and shape problems.
Most results were consistent between both systems examined as well as with prior studies.
% Converter failures were correlated, weakly, with changes in the ONNX specification.
Finally, in our root-cause examination of \emph{why} converters fail, we describe some model characteristics that are correlated with converter failures. %no evidence of evaluate two hypotheses falsifying one and finding equivocal results on the other. 
% 11\% of tested models failed via crash or incorrect behavior, primarily affecting the synthetic models.
% Finally, in our measurement of present failure frequency, we found that 11\% of tested models failed via crash or incorrect behavior, primarily affecting the synthetic models. % 2\% of real-world models experience failure, while synthetic models were rare yet incorrect model behavior was relatively common.

% \vspace{0.05cm}
\noindent
\hspace{0.015cm}
\textbf{Our contributions are:}
% \vspace{-0.12cm}
\begin{itemize}[leftmargin=0.5cm]
    % \item A model of the ONNX ecosystem,  \cref{sec:ONNX:ONNX_Ecosystem}.
    %\item A depiction ONNX usage patterns, \cref{sec:ONNX:ONNX_workflows}.
    %\item A model of ONNX model converters, identifying and synthesizing the parallels with DL compilers, \cref{sec:dl_model_converters:converters} and \cref{sec:dl_model_converters:compilers}.
    %\item We report a failure analysis of DL interoperability software, specifically the PyTorch and TensorFlow ONNX converters.
    \item We survey \NumInteropUsers engineers and report common interoperability tools, use-cases, and pain point (\cref{sec:theme0}). 
    \item We analyze failures in two DL model converters: PyTorch and TensorFlow into ONNX. We taxonomize and measure the distribution of failure symptoms, causes, and locations (\cref{sec:theme1}).
    % \item We describe how model converters are currently tested (\cref{sec:res:RQ4}). In their end-to-end tests, they examine few real models and no synthetic models. We show that this is an oversight, finding 11 defects (5 new) using a corpus of real and synthetic models (\cref{sec:res:RQ3}). %across \code{torch.onnx}, \code{tf2onnx}, and ONNX Runtime. 
    \item We find that defective converter behavior is correlated with unusual models, but not with changes in the ONNX specification nor with the use of individual model layers or sequences (\cref{sec:theme2}).% simple paths analysis of appear to be correlated with new versions of the ONNX specification, nor with simple  report that the causes of erroneous converter behavior remain elusive \JD{Update this bullet based on the new framing} (\cref{sec:res:RQ3}). Individual operators are not predictive of failure and sequences of operators may be correlated. 
    %\item We categorize how model converters are tested.
    % \item Measured current frequency of failures in model converters.
    % \item We provide directions for future research grounded in our observations. \JD{FOR YHL: This is a pretty weak one, suggest you cut it}
\end{itemize}

%\myparagraph{Significance}
% \vspace{-0.05cm}
\noindent
\hspace{0.015cm} \textbf{Significance for software engineering}:
% The impact of this study 
DL interoperability tools, especially model converters, underpin many deployment pipelines.
We conducted the first systematic study of DL interoperability tools and model converters through a user survey and failure analysis.
Understanding \emph{how} and \emph{why} these tools fail will help software engineers make informed judgments about their robustness. 
%Characterizing the failures in light of the test suites guides the future development of these converters.  
% First, model converters are key to many software pipelines --- understanding their failure modes and failure frequency enables debugging and helps engineers make informed judgments about their dependencies.
%Second, we provide evidence that the DL ecosystem is evolving to the ``common intermediary'' pattern of interoperability (\cref{B:interoperability}), which may affect the focus of future research. 
% Second, we introduce a methodological innovation:
  % we combine traditional failure analysis (failure characteristics given that a failure has occurred) with a prospective estimate of failure frequency borrowed from the software testing literature.
%Our method informs engineers about both failure criticality \emph{and} likelihood, allowing them to judge the risk of a dependency. % depending on a component.

%\vspace{-0.05cm}
\section{Background and Related Work} \label{sec:Background}

Here we
  define DL model converters via the concept of \textit{interoperability},
  and
  discuss prior failure studies of DL ecosystem components.

\iffalse
Model sit between DL frameworks and compilers in the DL ecosystem.
%providing interoperability between DL frameworks and compilers.
We describe them with the software engineering concept of \emph{interoperability}. %, contextualized to DL.
\fi

% \vspace{-0.05cm}
\subsection{DL Model Conversion as Interoperability} \label{B:Interoperability}

%\JD{I think we have too much background. Let's abridge 2.1 as follows: (1) remove the subsubsections, (2) merge 2.1.1 into one paragraph, (3) make ``Evolution of DL ecosystem'' much more focused, and (4) move selected material from \$3.1 into that part as well --- for this, I want Figure 2 (ONNX as an example) and Table 3 (this should be a general system model if possible, so let's confirm that the other IR converters use a similar design). That will allow us to present \$3.1 much more briefly (in \$2 we talk about converters/interop in DL, and in \$3 we just give the novel bits/evidenec that we should focus on ONNX because it's popular)}
% \subsubsection{General Patterns of Interoperability} \label{B:Interoperability-General}

%Interoperability has several definitions and dimensions that vary based on system domain~\citep{Kosanke_2006}.
\iffalse
ISO/IEC/IEEE define interoperability as the ``\emph{degree to which two or more systems...can exchange [and use] information}''~\citep{ISO/IEC/IEEE}.
%Valle \etal observe that
Interoperability often focuses on the data being exchanged between systems~\citep{Valle_Garcés_Nakagawa_2019}.
It reduces component coupling, increasing engineering efficiency by separating concerns~\citep{Wiederhold_1995,GoF_Gamma_Helm_Johnson_Vlissides_1994}.
%Wiederhold advocates for intermediate representations and middleware to reduce component coupling~\citep{Wiederhold_1995}, also called the ``Adapter pattern'' by Gamma, Helm, Johnson, and Vlissides~\citep{GoF_Gamma_Helm_Johnson_Vlissides_1994}.
\fi
In the context of DL, \textbf{interoperability} focuses on model reuse and is defined as \emph{the ability of DL software to exchange DL models (\ie deep neural networks/DNNs)}~\citep{enchanceInterop}.
% We define \textbf{DL interoperability} as \emph{the ability of DL software to exchange DL models (\ie deep neural networks/DNNs)}.
Wegner describes two patterns for interoperability:
  creating \emph{pairwise mappings} between systems;
  and
  introducing a \emph{common intermediary} understood by all participants~\citep{Wegner_1996}. 
% These patterns are illustrated in~\cref{fig:conv_mech} in the context of DL. %, engineers use both of Wegner's interoperability mechanisms~\citep{Wegner_1996} to make those systems interoperable. %in DL interoperability.
These patterns are illustrated in~\cref{fig:onnx_ecosystem} in the context of DL.
% in the context of DL: 
%The \emph{pairwise mapping} is illustrated by the \emph{downward path} --- \emph{Framework} to \emph{DL Compilers}.  
%Whereas the \emph{common intermediary} is illustrated by the \emph{rightward path} --- \emph{Framework} to \emph{Common Intermediary} to \emph{DL Compilers}. %, engineers use both of Wegner's interoperability mechanisms~\citep{Wegner_1996} to make those systems interoperable. %in DL interoperability.
The two patterns trade-off customizability for scalability.

Two kinds of DL systems interoperate on DL models:
  \textit{frameworks} for DL model development,
  and
  \textit{runtimes} for DL model deployment. %, see~\cref{fig:onnx_ecosystem}.
%DL frameworks create DL models and runtimes/compilers are used to deploy them. The internal representations of models vary based on the software using them due to variable computational graph implementations and goals, as such, interoperability is required. 
%The types of interoperability most relevant to DL interoperability are syntactic and semantic interoperability as these are related to the communication and exchange of data --- in this case DL models. 
%DL 
%The DL ecosystem uses both of Wegner's interoperability patterns. %in DL interoperability.
% As depicted in~\cref{fig:conv_mech}, the DL ecosystem uses both of Wegner's~\citep{Wegner_1996} interoperability patterns. %in DL interoperability.
%The general interoperability techniques apply to DL interoperability. 
%We depict these mechanisms in~\cref{fig:conv_mech}. % depicts interface bridging and standardization in DL software. 
% Pairwise mappings (\cref{fig:conv_mech:bridging}) occur in DL compilers such as TVM~\citep{TVM_Chen_Moreau_Jiang_Zheng}, which map from framework representations of models into representations internal to the compiler.
In the downward path of~\cref{fig:onnx_ecosystem},
pairwise mappings occur in DL compilers such as TVM~\citep{TVM_Chen_Moreau_Jiang_Zheng}, which map from framework representations into internal compiler representations for each supported framework.
%This is depicted in \cref{fig:conv_mech:bridging}, the arrows indicate framework-runtime bridges with each runtime supporting models from a set of frameworks, this results in $M \times N$ mappings between $M$ frameworks and $N$ runtimes. 
Along the rightward path,
common intermediaries such as ONNX~\citep{ONNXFixed} and MMDnn~\citep{enchanceInterop} give standard representations for DL models.
% Common intermediaries (\cref{fig:conv_mech:standardization}) such as ONNX~\citep{ONNXFixed} and MDnn~\citep{enchanceInterop} provide common representations for DL models. 
In this pattern, each framework and runtime has a one- or two-way adapter to the common intermediary. 
\myparagraph{DL Model Converters}
Model converters fill a purpose similar to compiler front-ends~\citep{Li_Liu_Liu_Sun_You_Yang_Luan_Gan_Yang_Qian_2021}. 
They transform a model from a DL framework into a high-level IR representing the model's computations and control flow. 
Graph-level optimizations are applied before further conversion to a low-level IR for hardware optimization and code generation. %upon which hardware-level optimizations are applied and through which code generation occurs for hardware targets. 
\cref{tab:ONNXConverterStages} summarizes a typical design. 

{
\renewcommand{\arraystretch}{0.9}
\begin{table}
\caption{
% \small
    Stages of a DL model converter to the ONNX intermediate representation, based on \code{torch.onnx}, \code{tf2onnx}, and \code{mxnet.onnx}.
}
  % Reduce font size?
  % \small
\footnotesize
\begin{tabular}{lp{5.7cm}}
\toprule
\textbf{Component} & \textbf{Definition} \\
\toprule
Load Model           &  
Framework representation $\rightarrow$ ONNX graph.
Tracing used for dynamic graphs (\eg PyTorch). \\
%    Loading is easy for frameworks with static graphs (\eg TensorFlow).
%    For dynamic graphs (\eg PyTorch), graph is traced by model execution. \\ % and this is straightforwrad. Tracing is required for models from frameworks that use dynamic computational graphs, \eg PyTorch. This step executes the model and records all operations that occur within it. The ``traced" graph is then converted.
Node conversion & Graph nodes replaced by ONNX equivalents. \\ %This results in a graph that contains ONNX nodes. \\
Optimization & Nodes (\eg operator fusion), dataflows (\eg DCE). \\ %dead code elimination). \\
Export & Model serialized into protocol buffer (protobuf). \\ %~\citep{ProtocolBuffers}. \\  %, so the typical final step is writing the optimized graph to a protocol buffer \citep{ProtocolBuffers}.
Validate & %As an optional final step, the model is validated.
    Syntactic checks (compliance with spec) and semantic checks (behavioral changes). \\
    %Testing proceeds in two ways: 
    %  (1) Syntactic validity (compliance with ONNX specification); %, typically using the ONNX model checker;
    %  and
    %  (2) Semantic validity (equivalence with original). \\ % performing inference with the two models and checking if the models give \emph{similar} outputs, given the same inputs.
\bottomrule
\end{tabular}
\label{tab:ONNXConverterStages}
% \vspace{-0.3cm}
%\end{wraptable}
\end{table}

% \begin{wrapfigure}{r}{0.35\textwidth}
\begin{figure}
    \centering
    \includegraphics[width=0.75\columnwidth]{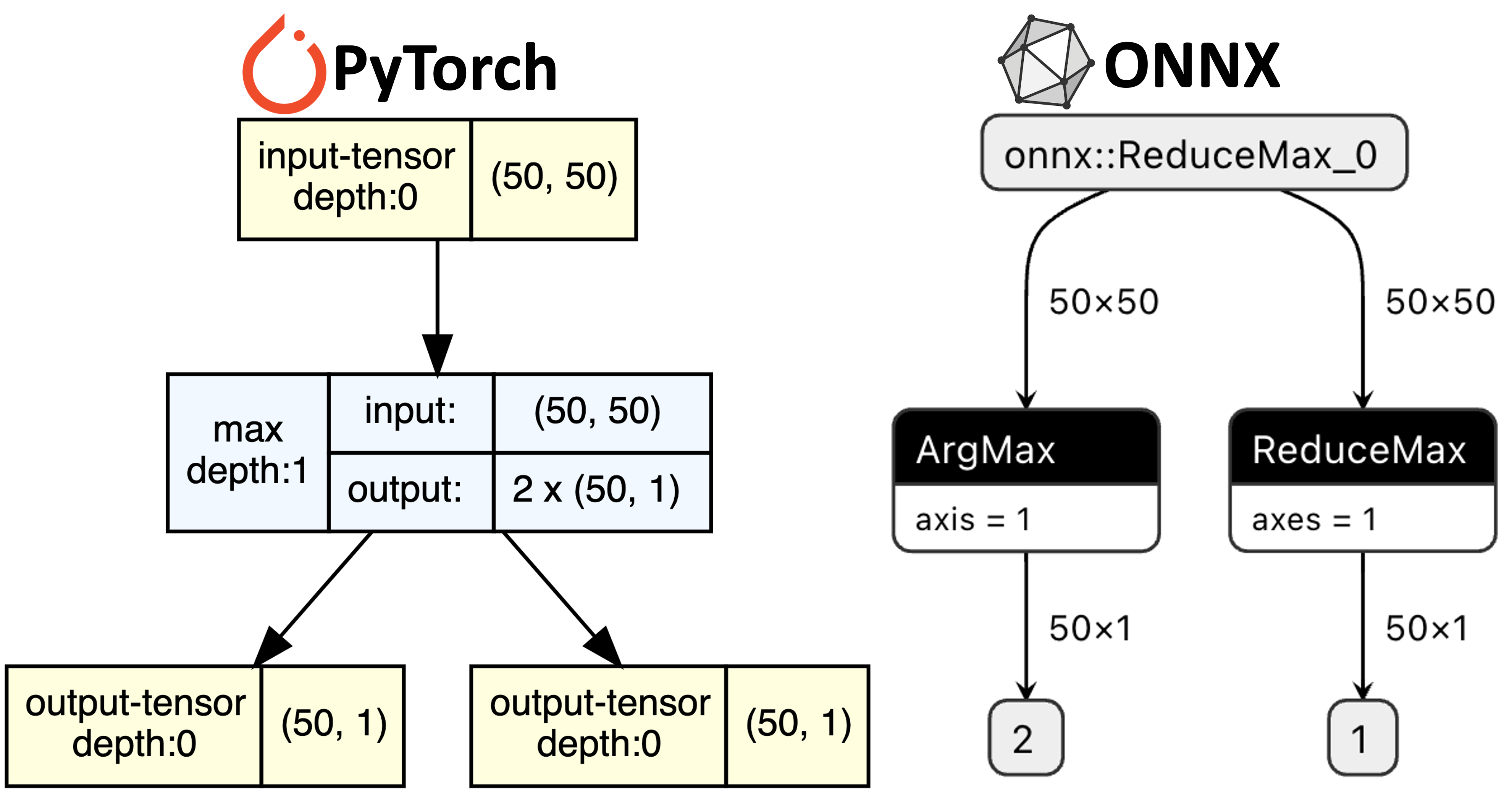}
    \caption{
        % \small
        PyTorch model converted to ONNX~Intermediate Representation. %, illustrating the conversion of operators performed by a DL model converter.
        The PyTorch model calculates the per-row maximum using \code{torch.max}.
        %The PyTorch model is an \code{nn.Module} with the subsequent \code{torch.max(input, dim=1, keep\_dim=True)} in its \code{forward} call.
        In ONNX, this uses the operators \code{ArgMax} plus \code{ReduceMax}. 
        % \code{ReduceMax} output is 50x1 (green box) but could be erroneously transposed (red box).
        %As an example conversion error, the ReduceMax output might be transposed (red box).
        %The red box indicates an incorrect conversion.
        %The green box indicates a correct.
    }
    \label{fig:graph_comp}
    % \vspace{-0.45cm}
% \end{wrapfigure}
\end{figure}
}

%Much like how the front end of a compiler converts a model (in framework representation) to a high-level IR that is used for code generation, a model converter converts a model into a high-level IR (ONNX IR) that is used for interoperability.
%Though this is the case, converters differ from compilers in that the high-level IRs used by compilers are for the purpose of compilation and as such remain internal to the compiler, whereas for ONNX the IR is serialized into a protocol buffer to allow for portability.  

\cref{fig:graph_comp} illustrates a model converted from PyTorch to ONNX.
Conversion is challenging, as noted by AirBus~\citep{onnx_airbus,gauffriau2023formal} and others~\citep{MLEAP2023}, because it maps between graphs expressed with different operators and different semantics.
Model conversion can produce models that are incompatible with runtimes or have different behaviours.
%models incompatible with runtimes, or even change model behaviors.
%introduce errors that cause incorrect behavior.
% \footnote{See \url{https://github.com/pytorch/pytorch/issues/74732}.}
%(1) Wrong Output, see \href{}{\emph{Inference result is different between Pytorch and ONNX model}}. 
%In this example, after the conversion of the model to ONNX the classification of the ``smile" was substantially different between the ONNX and the original model.
For a \textit{compatibility} issue, in PyTorch \#78721 a converted ONNX model had a type mismatch~\citep{pytorchissue78721}. % in the inputs to the \code{Concat} operators~\citep{pytorchissue78721}.
For a \emph{behavioral} issue, in PyTorch \#74732 a converted ONNX model's prediction 
%of ``Smiling or not?''
changed~\citep{pytorchissue74732}.
% \footnote{See \url{https://github.com/pytorch/pytorch/issues/78721}.}
%(2) Incorrect Model, see \href{}{\emph{Scripted reshape incorrect if shape is dynamically calculated}}. In this example, ONNX model fails to load to the ONNX Runtime due to a type mismatch in the inputs to the \code{Concat} operators. 
% : a PyTorch model is converted to its equivalent converted ONNX model. 

After conversion to an IR, models can be rendered back to DL frameworks or deployed to hardware.
Framework-to-framework conversion can bypass issues in model reengineering~\citep{jiang2023challengespracticesdeeplearning}.
%\JD{Cite the CV Reengineering arXiv and the MaskFormer arXiv}
Deployment from an IR allows DL runtimes to optimize against the IR rather than the many DL frameworks.

% The \code{torch.max} operation is equivalent to the \code{ArgMax} and \code{ReduceMax} operators in the ONNX model. 

\iffalse
%\begin{wrapfigure}{l}{0.5\textwidth}
\begin{figure}[h]
    \centering
    \includegraphics[width=0.88\textwidth]{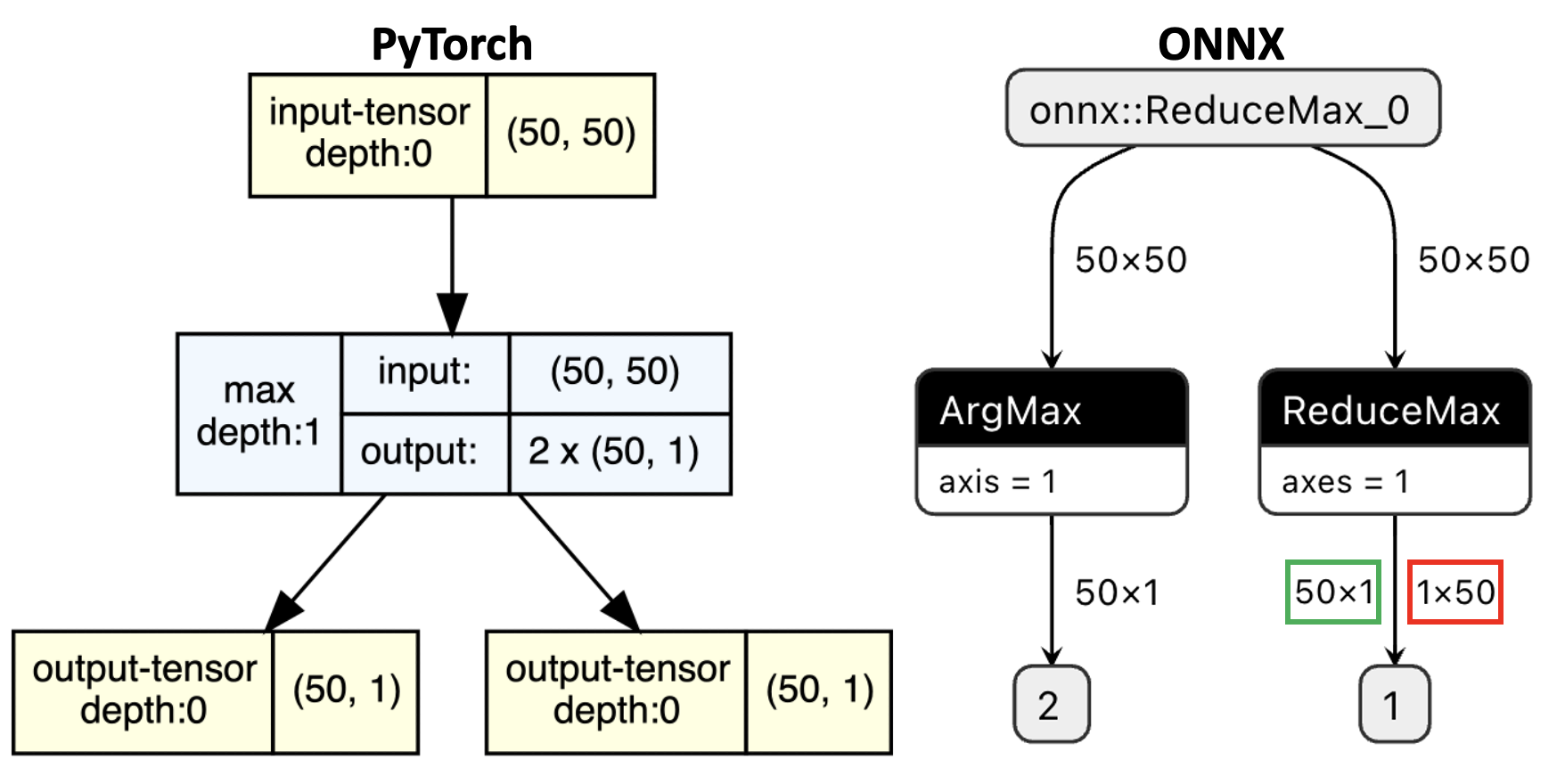}
    %\vspace{-0.1cm}
    \caption{
        Comparison of original PyTorch model to converted ONNX model, illustrating the conversion of operators performed by a DL model converter.
        The PyTorch model is an \code{nn.Module} with the subsequent \code{torch.max(input, dim=1, keep\_dim=True)} in its \code{forward} call.
        The PyTorch \code{torch.max} operation becomes \code{ArgMax} and \code{ReduceMax} in ONNX. 
        The red box indicates an incorrect conversion.
        The green box indicates a correct.
        % \TODO{Update figure to show where something incorrect might happen}
        % \JD{Also make it smaller}
    }
    \label{fig:graph_comp}
    %\vspace{-0.2cm}
%\end{wrapfigure}
\end{figure}

\else

\fi

\subsection{Failure Studies of DL Components} \label{sec:Background-FailureStudiesinDL}
%\JD{This subsection heading is uncoordinated with \$2.1. 2.1 went ``interop in general and then specific to DL''. Can we do the same in 2.2 --- ``failrues in interop in general and then specific to DL''?}

Software engineering failures inform development and maintenance~\citep{petroski1994design,anandayuvaraj_reflecting_2023,amusuo_reflections_2022}.
%Software failures are crucial for the advancement of software evolution and maintenance~\citep{petroski_design_1994, anandayuvaraj_reflecting_2023m tan_bug_2014, amusuo_reflections_2022}, as evidenced by comprehensive analyzes in various domains, including open source bug reports~\citep{islam2019comprehensive, TFprogramBugs}, Stack Overflow discussions~\citep{Zhang2019CommonChallengesinDevelopingDLApplications}, deployment failures~\citep{failuresDLJobs}, and interview studies~\citep{jiang2023challengespracticesdeeplearning}.
% Studies of software engineering failures influence software evolution, defect detection, and program repair~\citep{tan_bug_2014,amusuo_reflections_2022}.
% Many researchers have examined failures in DL software~\citep{Injadat2021MLtowardsIntelligentSystems, Wang2019SurveyofDLMLFrameworkandLibrary, tan_exploratory_2022}, drawing from open-source bug reports~\citep{islam2019comprehensive, TFprogramBugs}, Stack Overflow~\citep{Zhang2019CommonChallengesinDevelopingDLApplications}, deployment failures~\citep{failuresDLJobs}, and interview studies~\citep{jiang2023challengespracticesdeeplearning}.
%
Previous failure analyses of DL interoperability software focused on the ``Development'' and ``Runtime'' components of~\cref{fig:onnx_ecosystem}.
%Interoperability defects do occur in these contexts, but since they use Wegner's pairwise mapping pattern, little is learned about the common intermediary pattern.
%Specifically, the work focused on fundamental components of DL software such as DL frameworks and compilers. 
Chen \etal studied 800 defects from 4 DL frameworks to obtain testing guidelines~\citep{Chen2022DLFrameworkBug_1}.
Shen \etal 
studied DL compiler defects~\citep{DLcompilerbugs},
and
%studied 603 defects in three DL compilers and provided guidelines for DL compiler bug detection and debugging~\citep{DLcompilerbugs}.
% DL compilers have been one of the most fundamental and important DL software infrastructures which are still under-studied~\citep{DLcompilerbugs}.
% We conduct a failure analysis of popular DL converters.
% Moreover, DL compilers have been one of the most fundamental and important DL software infrastructures which are tightly relevant to the model deployment and optimization~\citep{DLcompilerbugs}.
% --- see~\cref{fig:conv_mech}.
others have tested DL compilers~\citep{NNSmith, Xiao_Liu_Yuan_Pang_Wang_2022}.
Some of this work has incidentally examined interoperability failures~\citep{NNSmith,morovati2024bug};
  for example,
  Shen \etals study of DL compilers included ``front ends'' offering interoperability through pairwise (DL framework specific) and common intermediary (ONNX) model loaders~\citep{DLcompilerbugs};
  %and Liu \etals work found defects in PyTorch's ONNX converter.
% and incidentally used ONNX in the process.
%Their study of DL compilers also incidentally considered the common intermediary pattern, since some DL compilers can ingest both framework representations (\eg from PyTorch) and intermediate representations (\eg ONNX). 
%Although their study was on DL compilers, approximately half of the conversion defects they discovered were related to the PyTorch ONNX exporter, motivating our focused study of this interoperability pattern.
%In summary, this line of prior work has examined failures in DL interoperability software that follows the pairwise mapping pattern but has not examined the common intermediary pattern. % but not the common intermediary pattern.
%No prior work has focused on the analogous failures in DL interoperability software that follows Wegner's common intermediary pattern.
%However, the challenges of intermediate representation were not the focus of those works.

The two prior failure studies of DL model converters focused on testing and fault localization.
% Research on DL model converters
%the common intermediary pattern in DL
% is limited. % (``Converters'' in ~\cref{fig:onnx_ecosystem}).
%Even though there has been much study of failures in DL frameworks, applications, and compilers, DL interoperability software is yet understudied. 
% Interoperability software is a fundamental component of DL. It acts as the interface between software and allows them to work together. 
% which can significantly benefit the deployment and maintenance of DL software. 
Openja \etal converted 5 popular DL models~\citep{ConvertingDLempirical}.
They considered only the current state of model converters, not past failures.
%It is unclear whether their results on 5 models will generalize.
Louloudakis \etal studied behavioral issues resulting from framework-to-framework conversion~\citep{louloudakis2023deltann}.
They found failures in 10 out of 36 conversions. %study the conversion of 3 models $\times$ 12 framework pairs and find failures in 10 out of 36 conversions.
They created a fault localization and repair pipeline to localize and fix discrepancies~\citep{louloudakis2023fault}.

% The engineering community has reported many issues about interoperability software (\ie ONNX)~\citep{ONNXBlog} and its converters~\citep{ConvertingDLempirical}. 
% These problems can significantly affect the maintainability and reliability of DL software. 

In light of this literature, the main contribution of our work is the first systematic analysis of failures in DL model converters.
As a conceptual contribution, we frame DL model converters as a class of interoperability software.
We consider \textit{how} and \textit{why} they fail.
\section{Research Questions \& Study Design} \label{sec:RQsAndStudyDesign}

 \iffalse
This section states our RQs (\cref{sec:RQs}) and study design (\cref{sec:StudyDesign}).
\cref{fig:methods} gives a summary.
\fi

\begin{figure}[b]
%\begin{wrapfigure}{r}{0.7\columnwidth}
    \centering
    \includegraphics[width=0.8\columnwidth]{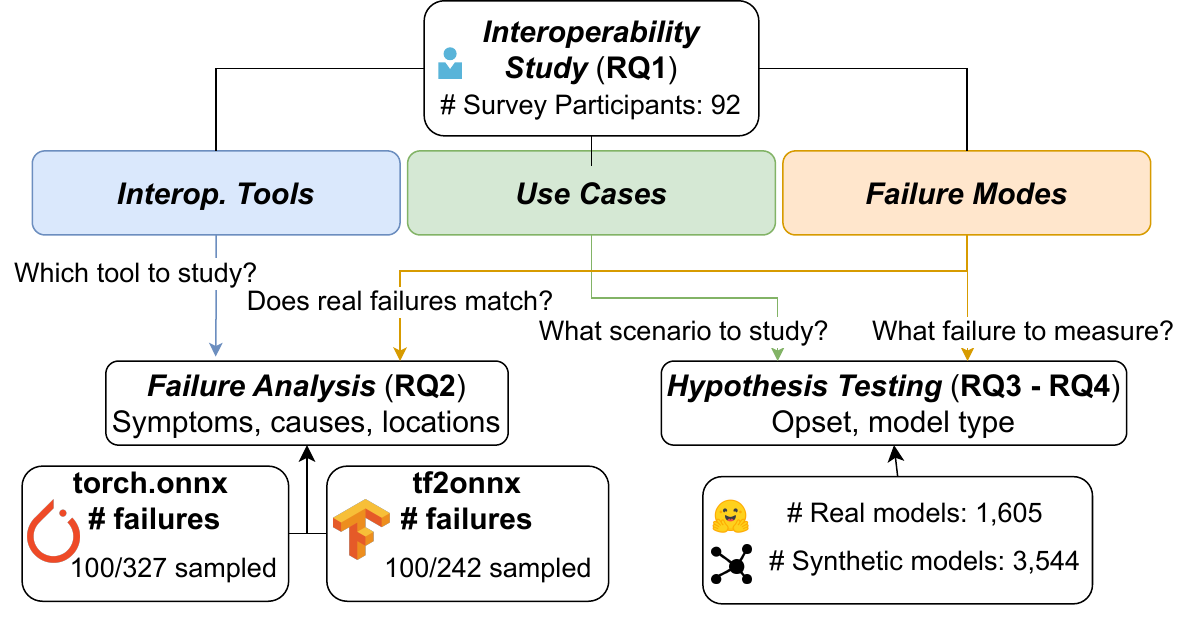}
    \caption{ 
    % \small
    Goal, research questions, methods, and data sources.
    % \JD{Update the figure so ``Methods'' becomes clearer, right now it's not clear}
    % \WJ{Add survey?}
    % \TODO{Update this figure}
    % \JD{This figure needs to be updated to indicate just RQ3-4 (We merged RQ4 and RQ5)}
    % \JD{I think the $N$ in this figure may be incorrect (RQ1)}
    }
    \label{fig:methods}
    % \vspace{-0.45cm}
\end{figure}

\cref{fig:methods} shows our RQs (\cref{sec:RQs}) and study design (\cref{sec:StudyDesign}), detailed next.
%Summarized in This section states our RQs (\cref{sec:RQs}) and study design (\cref{sec:StudyDesign}).
% gives a summary.

%\end{wrapfigure}

% \vspace{-0.15cm}
\subsection{Research Questions} \label{sec:RQs}

We summarize the related work in \cref{sec:Background} as follows:
The DL ecosystem is growing more complex (\cref{B:Interoperability}), motivating 
% participants to 
a shift 
% from the pairwise mapping pattern 
to common intermediaries such as ONNX. % pattern of interoperability.
% ONNX is the most prominent common intermediary (\cref{sec:B-ONNX}).
Analyzing failures in this emerging pattern will inform ecosystem participants of risks and opportunities for improvement.
Prior work examined failures in DL frameworks and runtimes, but has paid little attention to interoperability %interoperability or 
% focus on the pairwise mapping pattern 
%did not examine it comprehensively
(\cref{sec:Background-FailureStudiesinDL}).
% There is only one prior study of the common intermediary pattern; it was exploratory, not systematic.
% In this study, we utilize a mixed-methods approach to understanding model converters. 
% First, we conduct a failure analysis of the PyTorch-ONNX and TensorFlow-ONNX model converters. 
% We then empirically measure failures in these model converters. 
% Such framing is conducive to providing a full picture of failures in systems. 
%We focus on ONNX (\cref{sec:B-ONNX}), and analyze failures in this DL interoperability software (\cref{sec:rov:rq}). 
%to 
% specifically in
%the ONNX ecosystem (\cref{sec:B-ONNX}).

Our study fills this knowledge gap by analyzing failures in DL interoperability software.
We specifically focus on ONNX, the leading DL interoperability framework.
Our research proceeds in three themes:
  (1)~why and how engineers use interoperability tools;
  (2)~a failure analysis of the most popular interoperability tool;
  and
  (3)~evaluating hypotheses about the root causes of failures.

\iffalse
% \emph{how} converters fail and \emph{why} converters fail.
To understand \emph{how} converters fail, we answer the standard questions of software failure analysis: failure symptoms, causes, and locations.
% However, we observe that such analysis is conditional on failure, \ie we learn $Prob(\mathrm{failure\ characteristic}\mid\mathrm{failure})$.
However, this method is hampered by the unsystematic nature of issue reports.
To complement this, we systematically evaluate failures using a diverse set of input models and examine the characteristics of the test suites that allow failures to escape.
% To understand current system quality, we also need the likelihood of failure, \ie $Prob(\mathrm{failure}\mid\mathrm{use})$. 
% Our \emph{prospective} question measures failure likelihood on common and systematically generated inputs.
We ask:
\fi

\noindent
\textbf{Theme 1: Interoperability User Survey (\cref{sec:theme0})}
\begin{itemize}[topsep=0pt]
    \item[\textbf{RQ1}]{How and why do engineers use interoperability tools?}     
\end{itemize}
% \hspace{0.025cm}
% \textbf{Theme 2: The \emph{How} of failure} 

\noindent\textbf{Theme 2: Failure Analysis (\cref{sec:theme1})}

\begin{itemize}[topsep=0pt]
    \item[\textbf{RQ2}]{What are the failure characteristics in DL interoperability software --- symptoms, causes, and locations?}
    % \JD{I revised RQ2 a little more, PTAL}
    % \item[\textbf{RQ2}]{How are failure characteristics related to one another?}
    % Commenting out post-ICSE'24 review.
    % \item[\textbf{RQ3}]{To what extent do changes in the ONNX specification correlate with model converter failures?}
\end{itemize}
% \hspace{0.025cm}
% \textbf{Theme 3: The \emph{Why} of failure}

% \JD{Should this be `causes' rather than `cases'?}
% \JD{Plural hypothEs, right?}

\noindent
\textbf{Theme 3: Hypotheses on ONNX Failure Causes (\cref{sec:theme2})}

% \JD{Typeset these as RQs not H here, for consistency and following community norms. Otherwise it will be hard to follow.}
\begin{itemize}[topsep=0pt]
    %\item[\textbf{RQ3}] Do \textit{operator sets} (opsets) affect failure cases?
    \item[\textbf{RQ3}] Does ONNX evolution affect converter failure rates?
    \item[\textbf{RQ4}] Do model types affect converter failure rates?
    % \item[\textbf{RQ5}] Do \GKT{consider adding: compositions of} \textit{operator/layer sequences} affect failure cases?
    % \item[\textbf{RQ3}] To what extent do changes in the ONNX specification correlate with model converter failures?
    % \item[\textbf{RQ4}] What model attributes result in failures?
    % \item[\textbf{RQ5}] Does converter testing cover failing models?
\end{itemize}

% \vspace{-0.05cm}
\subsection{Study Design} \label{sec:StudyDesign}

\cref{fig:methods} relates research questions to methods.
%\JD{GKT asked for more details here, but I thought that was the wrong direction. This is a signposting paragraph. So I trimmed it instead to make it more skeletal.}
%We apply techniques from software failure analysis and software testing to answer our research questions.
%\WJ{Why full perspective needs mixed-method? Maybe cite 
%\citep{Aranda2009SecretLifeofBugs} which indicate the pitfall of only studying github issues.
%All methods are applied to the same two projects (\cref{sec:Methods-RepositorySelection}) \WJ{because ...}.
% For retrospective failures
For RQ1 we use a user survey (\cref{sec:theme0}).
The survey results inform the remainder of the work.
For RQ2 (\cref{sec:theme1}) and RQ3-4 (\cref{sec:theme2}) we apply methods from mining software repositories and software testing.
%we apply methods from software failure analysis: selecting and analyzing relevant failures from GitHub issues.
We analyze GitHub issues (RQ2),
  correlate issue frequency to ONNX versions (RQ3),
  and
  evaluate ONNX on a range of model types (RQ4).

\section{Theme 1: Interoperability Study}
\label{sec:theme0}
To understand the use of DL interoperability, we surveyed DL practitioners on DL development and deployment practices. 
We selected the survey methodology to gather insights across diverse practices and experiences~\citep{easterbrook2008selecting}. 
Our method is given in~\cref{sec:mot:meth}, results in~\cref{sec:mot:results}.

% \vspace{-0.05cm}
\subsection{Methodology}
\label{sec:mot:meth}

\subsubsection{Instrument Design}

We followed Kitchenham \& Pfleeger's guidelines to develop our survey instrument~\citep{kitchenham2008personalSurvey}.
\cref{tab:ExampleSurveyQuestions} illustrates the result.
We asked general questions about DL interoperability, and specific questions about ONNX (interview data~\citep{shankar2024we} and GitHub suggest it is most popular).
We removed some demographic questions after the pilot study (e.g., ML/SE expertise, organization types) to reduce survey time in order to increase response rates, as is recommended by survey guidelines such as~\citep{ghazi2018survey}. 
The pruning removed both demographic and technical questions, and reduced the survey time from 15-20 minutes (with which we had a 1\% response rate from 130 initial invitations) to 5-8 minutes (yielding a 4\% response rate from subsequent invitations). We made sure that we did not prune too far, so that the collected demographic data (reported in Table 3) is consistent with prior empirical software engineering works such as~\citep{jiang_empirical_2023_1, michael2019regexes, ernst2015measure}. 
% For higher response rate, the survey was short and mostly closed-ended.
\iffalse
The initial version asked questions aligned with our RQ:
  about subjects' usage of interoperability tools;
  model development and deployment context;
  which tools they used and for what purposes;
  and challenges and best practices. % in utilizing interoperability tools.
This last group of questions were specific to the use of ONNX, since GitHub popularity data suggests that it is currently the most popular DL interoperability tool in the rapidly evolving DL realm.
\fi
We iterated internally, piloting with 3 external participants to check instrument clarity.
%We used their feedback to improve the instrument's clarity.

We sent the survey out in batches to allow for iteration.
We examined the results halfway to determine whether the instrument aligned with a larger sample of users' behavior.
We observed an unexpectedly high incidence of one use case,
  and so in the second half we added an open-ended question to clarify this use case.
%\cref{tab:ExampleSurveyQuestions} shows the topics and example questions in our survey.

{
\renewcommand{\arraystretch}{0.6}
\renewcommand{\arraystretch}{0.4}
\begin{table}[h]
\centering
\caption{
  % \small
  Survey excerpt.
  *Shown to second half of participants.
  } 
\label{tab:ExampleSurveyQuestions}
\small
\begin{tabular}{
    p{0.19\columnwidth}p{0.71\columnwidth}
}
\toprule
    \textbf{Topic} &
    \textbf{Example questions}
    \\
\midrule
Demographics & (1) How long have you worked on ML/DL projects? (2) What deployment environments are targeted?\\
\\
Interoperability tool usage & (1) What do you consider when choosing between deploying from a DL framework vs. via a tool like ONNX? (2) Do you use ONNX as part of your model development and deployment process?\\
\\
Use cases & (1) For what purpose do you use interoperability tools? (2*) If you use ONNX for framework-to-framework conversion, please describe your use case further? \\
\\
Failure modes (ONNX-specific) & (1) Do you commonly encounter problems while working with ONNX models? (2) If you encounter such problems, how do you address them?\\
\bottomrule
\end{tabular}
\end{table}
}

\subsubsection{Population and Sampling}
\label{sec:meth:sampling}
With approval from our Institutional Review Board (IRB),
  survey respondents were recruited from Hugging Face users. 
The Hugging Face ecosystem is the primary location for deep neural network-based model development and re-use~\citep{jiang_empirical_2023_1, jones2024we}, so its users are a suitable population for questions about DL interoperability tools.
%Part of our survey study is associated with other pre-trained model practices.
To increase the likelihood of responses from experienced software engineers,
  we collected email addresses from users with PRO accounts (\ie paid),
  and from accounts in organizations marked as \textit{company}, \textit{community}, and \textit{non-profit} (excluding types such as \textit{education}).
Participants received \SurveyCompensation gift cards.
%We specifically filtered for 
%To filter for practitioners, we collected users from \code{company}, \code{community}, and \code{non-profit} organizations.
%As indicated in prior work, most of these users are DL engineers who are actively developing and deploying DL models~\citep{jiang_empirical_2023_1}.

We targeted a confidence level of 90\% with a 10\% margin of error.
With an estimated Hugging Face user population of 1.2 million~\citep{ForbesHuggingFace2024, Gargi2024HuggingFace},
%Hugging Face's estimated user base was $\sim$1.2 million at the time of survey deployment \JD{CITE}.
% demonstrating a rising trend over the years. 
 a sample size of 69 respondents was needed.
% We collected additional emails from the organization members\footnote{\url{https://huggingface.co/organizations}}.
% PRO accounts mean that the users actively pay for the features provided by Hugging Face.
We sent out surveys in batches until reaching our desired sample size. 
In total, we distributed our survey to \NumofPROUsers PRO users and \NumofORGUsers organization members.
% \JD{For *this* paper, do not mention any respondents (or data) who did not fill the interop part of the survey. The ICSE'25 submission is about the naming page. The ISSTA'24 submission is about the interop page. There is no overlap between the two parts, so report only the interop-relevant part.}
We received a total of \NumInteropUsers valid responses (\ResponseRate response rate).
All questions were optional to improve the response rate, so we do not have responses from all subjects on all questions.

%\SurveyTotalResNum users responded to our survey (\NumSurveyRes actively interacted with Hugging Face). Our survey study resulted in a response rate of \ResponseRate.
%Out of  \SurveyTotalResNum users, there were \NumInteropUsers (77\%) users who filled out the page about interoperability tools with any information.
\cref{tab:demographic} shows respondent demographic information.

{
\renewcommand{\arraystretch}{0.4}
\begin{table}[h]
\centering
\caption{
  % \small
  Participant demographics (N=\NumInteropUsers but 3 skipped this).
  } 
\label{tab:demographic}
\small
\begin{tabular}{
    p{0.20\columnwidth}p{0.71\columnwidth}
}
\toprule
    \textbf{Kind} &
    \textbf{Distribution of Responses}
    \\
\midrule
ML experience  & $<$1 yr. (8) ; 1-2 yr. (17) ; 3-5 yr. (32) ; $>$5 yr. (32)\\
\\
SE experience  & $<$1 yr. (4) ; 1-2 yr. (12) ; 3-5 yr. (20) ; $>$5 yr. (53)\\
\midrule
Org. size  &  Small, $<50$ employees (48) ; Medium, $<250$ employees~(17) ; Large, $>250$ employees (24) \\
\\
Deployment environment  & Web application (59) ; Cloud and data center (52) ; Desktop application (19) ; Mobile (14) ; IoT/embedded systems (14) ; Other (4)\\
% \midrule
% \#used PTMs  & 0 (0), 1-5 (38), 6-10 (18), 11-20 (19), $>$ 20 (28)\\
% \\
% \#created PTMs  & 0 (34), 1-5 (37), 6-10 (15), 11-20 (5), $>$ 20 (12)\\ 
\bottomrule
\end{tabular}
\end{table}
}

\subsubsection{Analysis}
% \JD{@Wenxin I simplified this, PTAL and remove this comment if it looks OK.}
Most questions were closed-ended (multiple-choice, checkbox).
Qualitative analysis was needed for 3 open-ended questions, about
  use cases,
  model deployments,
  and
  interoperability problems.
%All questions in our survey are optional, allowing us to analyze the data based on the number of responses for each question.
Two authors reviewed the data and agreed that the responses to these questions were short and did not involve much subjectivity.
Therefore, all data were analyzed by one author. 
%, as they consist of short answers that are adequately objective.
% \JD{A single-researcher analysis is a threat to validity, so you need to justify why you think the data is sufficiently objective (\eg ``short answers'' or ``minimal subjective content'').}

% \vspace{-0.05cm}
\subsection{RQ1: Engineers and Interoperability Tools}\label{sec:mot:results}

%\JD{Let's figure out how to get LATEX to do a counter on Finding numbers so we reduce one source of error in copy/paste. (Also on my wish list is to give the finding number a label so we can cref it, to close the loop on the associated errors)}
%\GKT{If nobody solves this, let me know. It is a simple matter of creating a counter and incrementing it on each finding.
% \GKT{@JD and all: Please use the "\\newfinding" macro for each finding. I think I got them all.}
\begin{tcolorbox} [width=\linewidth, colback=yellow!30!white, top=1pt, bottom=1pt, left=2pt, right=2pt]
%\textbf{Finding 1}:
\newfinding ONNX is the most popular interoperability tool, used by approximately 42\% of respondents. Meanwhile, 41\% of respondents do not use interoperability tools. 
% \GKT{We need to be clear what others are represented and also show that they are not strong contenders.}
% ONNX plays an important role in the development and deployment processes of practitioners.
% \JD{I think the finding is ``ONNX is the most popular interoperability tool, used by approximately XX\% of respondents''. We do not have data saying it plays an *important* role or a way to define what important woudl be.}
%\textbf{Finding 2}:

\newfinding Model deployment and framework-to-framework conversion are the primary use cases for interoperability tools, both used by over half of the interoperability-using respondents.
% The most common use cases for interoperability tools are model deployment (54\%) and framework-to-framework conversion (41\%).
% \JD{This is phrased a little funny -- these are the only plausible use cases, right? So basically we are saying it is evenly split, rather than ``these are the two most popular''? Or do I misunderstand?}

\newfinding Many respondents (59\%) encounter ONNX problems.
Crashes and performance differences are the most common failure modes.
Each was encountered by $\sim$35\% of respondents.
% \JD{Do we have data about the other 40\% of ONNX problems?}

\end{tcolorbox}

% \subsubsection{How do practitioners develop models?}
% Q60

\subsubsection{Interoperability Tool Usage} \label{sec:mot:results:interopTools}
The majority of respondents utilized the PyTorch or TensorFlow frameworks for model development.
% (see \PJ{add figure}).
Most of the respondents utilized PyTorch (89\%).
% (89/\NumFrameworkData).
TensorFlow is the second most used framework (37\%).
% (33/\NumFrameworkData).
JAX, MLX, and other frameworks made up 20\%
% 18/\NumFrameworkData 
responses. 
The data show that the respondents often use multiple frameworks or have moved from one framework to another (\eg moving from TensorFlow to PyTorch),
perhaps motivating their use of interoperability tools. 
% This indicates a flexible "toolbox" approach, suggesting a shifting landscape with transitions between frameworks, such as TensorFlow to PyTorch.
% \GKT{Good, but the sum is greater than 100\% so this means that many respondents (as I would expect) are using both. We need a synthetic statement to indicate that this is part of the "multiple tools in the toolbox" thinking and as a motivator for ongoing study. After all, there will likely be a successor to both (including some of the ones presently not well-represented, \eg JAX). It could also suggest that some have worked with one but are moving to others, \eg moving from TensorFlow to PyTorch. In any event, this is a strong motivator for interoperability as well, since you want to stay abreast of likely flux in this field.}

% Q61, Q62, Q63
Approximately 42\% (39/\NumInteropUsers) of respondents reported using ONNX.
Only 15 respondents used other interoperability tools such as MMdnn and NNEF.
The rest (41\%) do not use interoperability tools. 
In our survey, although we did not gather extensive data on participants' specific roles, insights from their deployment considerations suggest they are primarily model developers.
They either lack model deployment responsibilities or deploy the models directly. %themselves without intermediary processes.
% \JD{Also note that XX of them do not use interoperability tools?}

\subsubsection{Use Cases} \label{sec:mot:results:interopUseCases}
Interoperability tools are used equally to interoperate between frameworks and to deploy models. 
% \JD{Report bigger numbers before smaller numbers}
69\% (37/54) of respondents report using model conversion for deployment of DL models.
Whereas 52\% (28/54) of respondents report using framework-to-framework conversion (\eg PyTorch to TensorFlow).

For the respondents who use interoperability tools for framework-to-framework conversion, most of them use the tools to \inlinequote{integrate the models in non-Python code}.
For example, one respondent wants to \inlinequote{interact with the base CUDA system}. % and to build pipelines for converting the models}.
Another respondent \inlinequote{exports [pre-trained models] to ONNX for import into Axon, an Elixir/Nx based DL framework that often [lacks] native pre-trained models}.

% \JD{Do we have quotes or examples by use case? Table 2 is focused on the deployment use case, but that focus seems odd because we have equal amount doing framework-to-framework.}

\cref{table:Survey-DeployConsideration} presents the deployment considerations of practitioners when using interoperability tools.
Many practitioners report deploying directly from frameworks when they want easy deployment or expect to update models often.
%They find framework deployment easy (no conversion required).
%A respondent indicated that \inlinequote{deploying directly is simpler}.
The broader organization's practices may play a role: \inlinequote{if I have control over the entire training to deployment pipeline, it is easier to use PyTorch exports}.

{
% \small
\renewcommand{\arraystretch}{0.5}
\begin{table}[h]
\centering
\caption{
% \small
  Induced themes for model deployment consideration when using interoperability tools, such as ONNX.
  Based on code-able responses from 32 participants.
  }
\label{table:Survey-DeployConsideration}
\small
\begin{tabular}{p{0.6\columnwidth}c}
\toprule
\textbf{Theme}  & \textbf{\# Participants (\%)} \\
\midrule

Simplicity & 8 / 32 (25\%) \\
Deployment requirements & 7 / 32 (22\%) \\
Inference speed & 5 / 32  (16\%) \\
Portability & 5 / 32 (16\%) \\
Other (\eg maintenance, stability)& 2 / 32 (6\%) \\
\midrule
Deploy directly from the DL frameworks & 13 / 32 (35\%) \\

\bottomrule
\end{tabular}
\end{table}
}

Interoperability tools are preferred when respondents care about performance gains, portability,  language compatibility, or if they require exotic configurations.
Examples were:
  \inlinequote{[PyTorch in browser is] too big},
  and
%One respondent reports PyTorch in browser is \inlinequote{
  \inlinequote{[ONNX offers] portability over performance}.
%Another preferred ONNX for \inlinequote{portability over performance}.
% \GKT{While not ML, this is part of a broader trend in the HPC world, \url{https://p3hpc.org/}. I have been looking at this as part of my work on OneAPI and data-parallel C++. P3 = portability + performance + productivity. I think ONNX is also about productivity.}

\subsubsection{Failure Modes} \label{sec:mot:results:interopFailureModes}
We ask practitioners about their experiences using ONNX.
59\% (32/54) of practitioners report encountering problems when using ONNX, while 41\% (22/54) report that they do not commonly encounter problems with ONNX models. 
Crashes (19/32, 59\%) and performance differences (19/32, 59\%) are the most prominent.
Often models do not convert, or there are performance differences between the original and converted models.
There are also some other problems (6/32, 19\%) mentioned by the respondents.
Examples included
  \inlinequote{ONNX doesn't support Fourier layers}
  and
  \inlinequote{driver problems during...deployment on user machines}.

\cref{table:Survey-AddressIssues} presents practitioners' strategies for resolving issues with the ONNX converter.
Many (44\%) turn to community resources, including GitHub issues and Stack Overflow posts, for help. 
One-third (31\%) verify the conversion process through testing, while 19\% explore solutions by changing ONNX versions.
%\PJ{@Wenxin Did they say ONNX version or Opset?} \WJ{@Purvish Have we addressed this issue? Do you need more quotes here to motivate the other two themes?}.
Some had experienced no good solution, resolving ONNX issues \inlinequote{case-by-case}.
 %\inlinequote{good solution to deploy models to user machines...[ONNX issues are solved] case-by-case}.
% These are problems are addressed through community engagement, model debugging, redesign, or conversion configuration.

{
% \small
\renewcommand{\arraystretch}{0.5}
\begin{table}[h]
\centering
\caption{
  Induced themes for strategies to address ONNX issues.
  Based on code-able responses from 16 participants.
  }
\label{table:Survey-AddressIssues}
\small
\begin{tabular}{p{0.6\columnwidth}c}
\toprule
\textbf{Theme}  & \textbf{\# Participants (\%)} \\
\midrule

Seeking help from the community & 7 / 16 (44\%) \\
Test with executions & 5 / 16  (31\%) \\
Version changes  & 3 / 16 (19\%) \\
Other (\ie documentation, config)& 3 / 16 (19\%) \\
\bottomrule
\end{tabular}
\end{table}
}

% \vspace{-0.1cm}
\section{Theme 2: Failure Analysis (of ONNX)}
\label{sec:theme1}

The second theme of this work is a failure analysis of deep learning interoperability software, specifically, deep learning model converters.
We apply the method of Failure Analysis~\citep{petroski1994design,leveson1995safeware,amusuo_reflections_2022,anandayuvaraj_reflecting_2023}, which characterizes and reports the distributions of past failures to revise engineering methodologies and prioritize research targets.

Many model converters have been proposed, but our survey data show which converters are of practical interest (\cref{sec:mot:results:interopTools}):
  those for the ONNX framework ($\sim$half of respondents use it); 
  and
  specifically those associated with the PyTorch and TensorFlow model development frameworks (the two most common).

\myparagraph{About ONNX} \label{sec:FailureAnalysis-AboutONNX}
The ONNX (Open Neural Network eXchange) specification 
% ``extensible computational graph'' 
provides a common representation that DL frameworks and runtimes use to represent DL models~\citep{ONNX_IR_2022}.
%The ONNX specification is known as the \emph{ONNX Intermediate Representation (IR)}~\citep{ONNX_IR_2022}.
% A variety of software utilizes the ONNX IR, and these range in nature: various frameworks, compilers, and model visualization tools provide export to, use of, and visualization of models in ONNX form (ONNX models). We provide an overview of the ecosystem of ONNX in \cref{sec:ONNX:ONNX_Ecosystem}. ONNX can be used in two ways, primarily, \cref{sec:ONNX:ONNX_workflows} presents these.
% \subsubsection{ONNX IR}\label{sec:ONNX:ONNX_IR}
%The ONNX intermediate representation (IR) consists of three components:
The ONNX specification has three main components:
  (1) A definition of a computational graph;
  (2) definitions of standard data types;
  and
  (3) definitions of built-in operators such as \code{ArgMax}~\citep{ONNX_IR_2022}.
The ONNX specification changes regularly to keep up with DL framework evolution~\citep{shridhar2020interoperating}. 
These changes consist of \emph{adding} new types and operators, and \emph{updating} the behavior of existing operators.
These changes are versioned within \emph{operator sets}.
%converters my susceptible to becoming incompatible with the ONNX specification and thus must respond to changes to keep up to date.
%ONNX operators are 
As of 2023 there have been 18 operator sets, totaling \OpsetAdditions additions and \OpsetUpdates updates to the available operators.  
\subsection{Methodology} \label{sec:Methods-Theme1}
% \subsection{Theme 1: \emph{How converters fail}} \label{sec:Methods-Theme1}

\subsubsection{Data Selection} \label{sec:Methods-DataSelection}

\begin{figure}
    \centering
    \includegraphics[width=0.93\columnwidth]{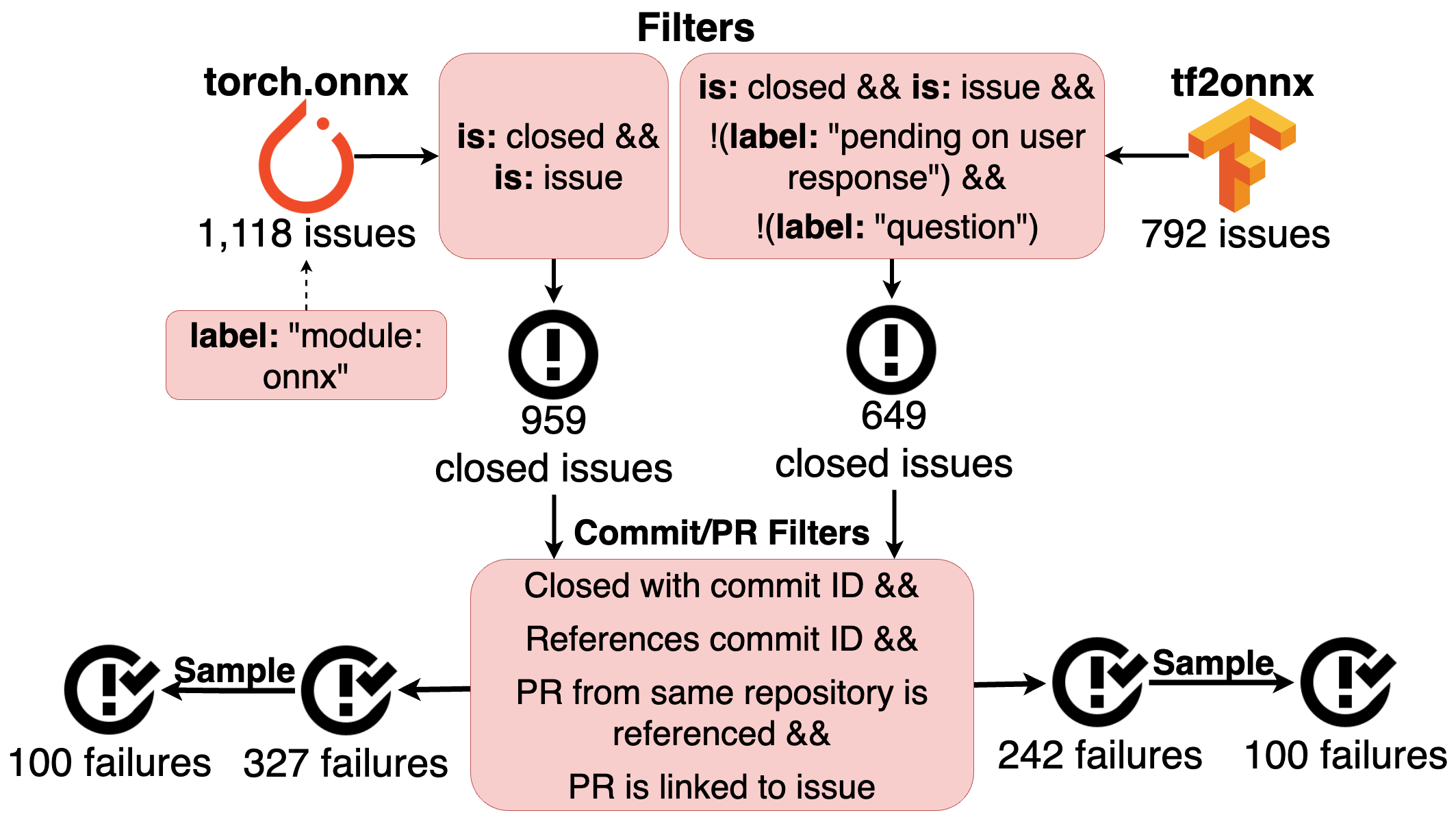}
    \caption{
    % \small
    Filtering of issues for each repository studied. \textit{Filters} are using GitHub search predicates. 
    \textit{Commit/PR filters} are applied to issue timeline events.
    Data were collected on Jan. 6, 2023.
    100 issues per repository were analyzed.
    }
    \label{fig:filtering}
    %\vspace{-0.25cm}
\end{figure}

\cref{fig:filtering} depicts our data selection method.
We describe the five main stages and rationales next.

\myparagraph{(1) Repositories}
For the reasons noted above, we studied the DL model converters from PyTorch and TensorFlow into the ONNX IR (\code{torch.onnx} and \code{tf2onnx}, respectively).
\iffalse
Our specific choice of the PyTorch and TensorFlow model converters and the ONNX IR was based on \textit{popularity} and \textit{data availability.} 
PyTorch and TensorFlow are the most popular DL frameworks, both in general~\citep{Most_used_frameworks_among_developers_globally_2022} and in our survey (\cref{sec:mot:results:interopTools}).
The ONNX IR is the most popular conversion target per our survey.
\fi
We note that among ONNX model converters, those for PyTorch and TensorFlow have the most failure data available on GitHub (\cref{tab:repos}).

%We considered converters for the top two most popular ML/DL frameworks --- TensorFlow (DL), and PyTorch (DL).
%\cref{tab:repos} summarizes popularity information of the ONNX converters for these three DL frameworks.
%In terms of available failure data, the ONNX model converters for TensorFlow and PyTorch have comparable issue counts on GitHub ($\sim$800-1000 issues).
%\GKT{Are any others represented besides TensorFlow and PyTorch? I think few reviewers will take issue with our choice of these two but wanting to be 100\% sure. Separately, are there any other converters besides those for TensorFlow or PyTorch. A brief mention here could be useful. \eg SciKit Learn, etc.? On this point, can we mention the list of converters available in ONNX?}
% \GKT{A reader may say, yes, but if you look at PyTorch vs. TensorFlow, TensorFlow is one order magnitude lower than PyTorch. SciKit Learn is one order lower than TensorFlow. I think it is ok to limit our study to these two.}
% \JD{GKT's comment will be resolved once we stop including ``stars'' as data}
%We therefore studied failures in the ONNX converters, the PyTorch-ONNX converter, and the Tensorflow-ONNX converter. 
%Our converter choice is motivated by their popularity (Table \ref{tab:repos}) and

% \begin{wraptable}{r}{0.70\columnwidth}
\renewcommand{\arraystretch}{0.7}
\begin{table}
\caption{
% \small
    Popularity and data availability of ONNX converters. 
    %We studied torch.onnx and tf2onnx.
    %Framework$\rightarrow$ONNX converters are studied and are shown at the top. 
    %ONNX-Framework converters are shown at the bottom.
    Framework$\rightarrow$ONNX converters (top) have notably more activity than ONNX$\rightarrow$Framework converters (bottom).
    % Details of ONNX converters for top DL frameworks.
    % The studied repositories are shown at the top. 
    *: \code{torch.onnx} is a component in the PyTorch repository, so stars are skewed.
    PyTorch issues are filtered for those in the converter.
    Data from April 11, 2024.
    % Data from July 7, 2023. 
    Parenthesized numbers are issues at time of data collection (July 7, 2023).
}
  % Reduce font size?
  \footnotesize
\begin{tabular}{ccrrr}
\toprule
\textbf{Project} & \textbf{Input}   
& \textbf{Stars}  
& \textbf{Forks}  
& \textbf{Closed Issues} \\ \hline
\textbf{torch.onnx}
& PyTorch, Caffe2       
& *77,429 
& 20,938 
& *1,286 (959)          \\
\textbf{tf2onnx}
& TensorFlow, Keras    
& 2,200 
& 425    
& 848 (792)          \\
Paddle2ONNX & PaddlePaddle
 & 637
 & 145
 & 192 \\
sklearn-onnx   & Sci-kit Learn 
 & 506  
 & 94     
 & 328           \\
\midrule
onnx2torch     
& ONNX
& 370    
& 35     
& 32   \\
onnx-tf        
& ONNX     
& 291  
& 25
& 129           \\
onnx-coreml
& ONNX
& 385
& 79
& archived \\
\bottomrule
\end{tabular}
\label{tab:repos}
\end{table}
% \end{wraptable}

Using the PyTorch and TensorFlow converters covers two relevant differences within DL and interoperability.
Within DL, these converters include both common representations of computational graphs: static (TensorFlow) and dynamic (PyTorch)~\citep{Semi-Static-Hattori_Sawada_Hamaji_Sakai_Shimizu_2020, Automatic-Differentiation-Baydin_Pearlmutter_Radul_Siskind}.
Within interoperability, these converters include both kinds of converter owners.
Converters are naturally owned either by the upstream producer of the data or the downstream consumer of the data.
In our case,
  \code{torch.onnx} gives an example of upstream ownership (it is owned by the PyTorch engineers),
  while
  \code{tf2onnx} gives an example of downstream ownership (it is owned by the ONNX engineers).

\ifHISTOGRAMEQUALIZATION
For sampling (\cref{sec:meth:sampling}),
  RQ1 and RQ2 are answered via a standard random sample;
  RQ3 uses time-aware sampling.
\fi

\myparagraph{(2) Filters for Relevance}
We follow prior work~\citep{autonomous_bugs, webassembly_bugs, well_typed_bugs}, and study closed issues because these issues typically contain greater information about failure causes and symptoms. 
For each repository, we collected closed GitHub issues related to ONNX conversion.
%ONNX-related issues are obtained using a set of GitHub filters (query terms).
%Then we filter these issues to find failures with sufficient information for analysi".
For \code{torch.onnx}, the PyTorch repository marks ONNX converter issues with the label \emph{module: onnx}.
We collect all closed issues with this label, yielding the search filter: \emph{is:issue label:``module: onnx'' is:closed}. 
For \code{tf2onnx}, all issues are relevant. 
We remove issues labeled ``pending on user response'' and ``question'' because these issues may not be failures.
This yields in
  \PyTorchModuleOnnxIssues issues in \code{torch.onnx} (from \PyTorchClosedIssues issues, not all ONNX-related)
  and
  \TFClosedIssuesFilter issues in \code{tf2onnx} (from \TFClosedIssues issues).

\myparagraph{(3) Filters for Data Availability}
% \JD{@Wenxin please fill the citation in here}
Following prior work~\citep{islam2019comprehensive, autonomous_bugs}, we subsequently filtered for GitHub issues that contained enough information for failure analysis. % (\eg having links to the corresponding pull requests that repair them). 
We filter issues for sufficient information (\eg the issue is resolved with a commit and pull request).
This filtering is conducted upon the timeline events for each GitHub issue and the filtering criteria are given in~\cref{fig:filtering}.
% These Commit/PR filters are given in~\cref{fig:filtering}.
This filter yields
  \PyTorchPostTLFilterIssues issues in \code{torch.onnx}
  and
  \TFPostTLFilterIssues issues in \code{tf2onnx}.
  
\myparagraph{(4) Filter Validation}
We piloted filters to ensure they captured relevant issues (recall) but not irrelevant issues (precision).
We hand-labelled \HandLabelledIssues issues per repository prior to filtering, applied the filter, and measured recall and precision.
If recall and precision \RecallPrecisionThreshold we consider our filter to acceptably balance manual work against bias.
For \code{torch.onnx}, we measured recall of \TorchRecall and precision of \TorchPrecision.
For \code{tf2onnx}, we measured recall of \TFRecall and precision of \TFPrecision.

\myparagraph{(5) Sampling}
Given the similar number of issues after filtering (\cref{fig:filtering}),
we randomly sampled and analyzed 100 issues per repository, or roughly one-third of relevant issues. 
This quantity is comparable to the proportion ($\sim$44\%) analyzed in prior work~\citep{DLcompilerbugs}.

\subsubsection{Data Analysis} \label{sec:meth:taxonomy}

We describe converter failures in 3 dimensions: \emph{location}, \emph{symptoms}, and \emph{causes}.

\textit{Location:} We map failures to the converter stages in~\cref{tab:ONNXConverterStages}: Load Model, Node Convert, Optimization, Export, and Validate. 

%As discussed in~\cref{sec:Background-FailureStudiesinDL}, the closest related work is Shen \etals study of DL runtimes~\citep{DLcompilerbugs}, which incidentally observed ONNX-related failures.
%They presented taxonomies of failure symptoms and causes.

\textit{Symptoms \& Causes:}
We adapted taxonomies from Shen \etal~\citep{DLcompilerbugs}, the closest related work, as discussed in~\cref{sec:Background-FailureStudiesinDL}.
Conceptually, we chose to reuse deep learning compiler taxonomies due to the expected similarity between compiler front-ends and converters. 
This choice also follows the recent advice of Amusuo \etal, who called for greater reuse of taxonomies in failure analysis research, where feasible~\citep{amusuo_reflections_2022}. 
To assess whether the taxonomies are applicable to converters, we trialed the taxonomies on a sample of 30 randomly chosen issues from each repository.
% As that work concerns DL compilers rather than model converters, we trialed the taxonomies on a sample of 30 randomly chosen issues from each repository.
% We conducted a pilot study to determine the appropriateness of Shen \etals taxonomies to describe ONNX converter failures. 
%We perform a  pilot study to test the efficacy of this adapted taxonomy in describing converter failures. 
%Specifically, after raw data collection (\cref{sec:Methods-DataSelection}),
%we randomly sampled 30 issues from each repository.
Two researchers classified symptoms and causes. % with the Shen taxonomies.
Computing inter-rater reliability via the Kappa coefficient~\citep{cohen1960coefficient_1},
  we found
  $\kappa$=\KappaCoeffCauses (causes)
  and
  $\kappa$=\KappaCoeffSymptoms (symptoms),
  \ie ``strong agreement''~\citep{mchugh2012interrater} for both taxonomies.
% We investigate conversion failures from two aspects: the \emph{causes} and \emph{symptoms}. 
% Due to the aforementioned similarities with DL compilers, we adapt the taxonomies presented in \citep{DLcompilerbugs}.
\ifJOURNALEXTENSION
Some wording was inapplicable to ONNX converter failures, and we adapted as needed. %ir taxonomies, adding and subtracting categories or changing words as needed.
\fi

% \cref{tab:taxonomy_symptoms} presents the taxonomy of failure symptoms.
The taxonomy we used for symptoms is: crash, wrong model, bad performance, build failure, and unknown/other.
In a \textit{crash}, conversion errors out.
A \textit{wrong model} is syntactically sound but semantically incorrect.
\textit{Bad performance} indicates unexpectedly high time or memory cost, \eg worse than on a previous version. % of \emph{model converter} is much higher than developer/user expectations (\eg the performance achieved on the previous version during regression testing or performance achieved by certain hardware).
\textit{Build failure} means the user could not install the converter.

The taxonomy of failure causes is more complex --- see~\cref{tab:taxonomy_causes_concise}.

{
\begin{table}[h]
    \caption{Taxonomy of failure causes, adapted from~\citep{DLcompilerbugs}, with concise definitions. Italics and Strikeout indicate changed or deleted wording, respectively. Bold indicates additions.}
    \footnotesize
    \begin{tabular}{lp{0.5\columnwidth}}
    \toprule
    \textbf{Cause} & \textbf{Definition} \\
    \midrule
    Incompatibility - Internal  & API compatib. issues in model converters. \\
    Incompatibility - External & Compatibility issues with third-party libraries (e.g., ONNX, TensorFlow). \\
    Type Problem - Node  & Issues related to atomic DL operators in model conversion. \\
    Type Problem - Tensor & Issues related to the types of tensors. \\
    Type Problem - Conventional & Issues with types of conventional variables in software systems. \\
    \st{Tensor} Shape Problem & \textit{Input/output tensor shape issues}. \\
    Alg. Error - Optimization Error & Issues in  model converter optimizations. \\
    Alg. Error - \emph{Tracing Issue} & Issues tracing dynamic models in computational graphs. \\
    \st{Alg. Error - Non-optim. Code Logic} & \st{Bugs outside DL compiler optimizations.} \\
    \textbf{Testing} & Test errors, incl. flaky or missing tests. \\
    \bottomrule
    \end{tabular}
    \label{tab:taxonomy_causes_concise}
\end{table}
}

\myparagraph{Interrater Agreement}\label{sec:meth:class}
In our pilot study of applying Shen \etals taxonomies, we observed strong agreement between raters.
Therefore, we relied on a single rating (one rater analyzing all issues) of the \TotalSamples sampled issues.
To assess the risk of taxonomic drift~\citep{wicks2017coding},
a second rater analyzed  one tranche of randomly selected failures.
In those \TotalDriftSamples samples, the two raters had perfect agreement/$\kappa$=\DriftTesting.
%We therefore believe the single-rater analysis was sufficiently objective.
We conclude the single-rater analysis was sound. % sufficiently objective.

\ifHISTOGRAMEQUALIZATION
  and
  one tranche of issues for RQ3 (20 samples, Kappa of \DriftTestingYearlyCauses/Causes, \DriftTestingYearlySymptoms/Symptoms).
\fi
%Using the taxonomies presented in \cref{sec:tax:causes} and \cref{sec:tax:symptoms}, we have a single rater classify 200 issues to answer \textbf{RQ1-2}. the samples according to \cref{sec:samp}. 
%Due to the substantial agreement between raters, as measured in \cref{sec:tax:creation}, we proceed with a single rater for these classifications.
%The rater classified 200 issues for \textbf{RQ1-2} and an additional 111 issues for \textbf{RQ3}. 

% Removed post-ICSE'24 review.
% \subsubsection{Correlation Analysis for RQ3} \label{sec:meth:correlation}

% Defects are expected when requirements change~\citep{sommerville2015software}.
% We therefore theorize that changes in the ONNX specification
% % cause
% % exhibit great potential to introduce
% cause
% failures in DL model converters.
% This theory predicts the following hypothesis:
%   \emph{Larger changes in the ONNX specification correlate with greater incidence of failures in DL model converters.}
% The expected resulting model converter failures would be incompatibility and type problems (defined in \cref{tab:taxonomy_causes}).

% To test this hypothesis:
%   (1) We need the size of changes to the ONNX specification.
%     To obtain this, we examine the ONNX release notes and measure the number of operator changes.
%   (2) We need the number of failures associated with that change.
%     To obtain this, we examine the issue creation times of the failures sampled for RQ1-2.
%     We attribute each issue to the nearest previous ONNX release.

% \vspace{-0.05cm}
\subsection{RQ2: The Characteristics of Failures} \label{sec:res:RQ1}

\begin{tcolorbox} [width=\linewidth, colback=yellow!30!white, top=1pt, bottom=1pt, left=2pt, right=2pt]

\newfinding
\ul{Location}: Most failures are in \emph{Node Conversion} (74\%).

\newfinding
\ul{Symptom}: The most common symptoms in DL model converters are \emph{Crash} (\CrashPercent) and \emph{Wrong Model} (\WrongModelPercent). %, similar to DL compilers~\citep{DLcompilerbugs}.

\newfinding
\ul{Causes}:
\emph{Crashes} are largely due to 
 \emph{Incompatibilities} and \emph{Type Problems}.
\emph{Wrong models} are largely due to 
 \emph{Type Problems} and \emph{Algorithmic Errors}.

\iffalse
\newfinding
Common failure causes are \emph{Incompatibility} and \emph{Type problems}, each $\sim$25\% of causes. 
\fi

% \textbf{Finding 5}: \emph{Wrong Models} are largely due to \emph{Type Problems}. 
% The \code{tf2onnx} converter also produces wrong models due to algorithmic errors in the graph optimization component.

%A further 10\% are in \emph{Graph optimization} (mostly in \code{tf2onnx}).
\end{tcolorbox}

% \noindent We characterize failures in their symptoms, causes, and locations. 

\subsubsection{\normalsize Failure Locations} \label{sec:RQ1:loc}
\code{tf2onnx} and \code{torch.onnx} have similar failure location distributions (\cref{tab:location}). % presents the distribution of failure locations.
The most common location of failures is the \emph{Node Conversion} stage: 148/200 failures occur in this stage, with a similar proportion in \code{tf2onnx} (70\%) and \code{torch.onnx} (78\%).
%For both the \code{tf2onnx} and \code{torch.onnx} converters, \emph{Node Conversion} is the most common failure location, with 78 and 70 failures respectively.
%The majority of these failures are found to be caused by \emph{Incompatibility} or \emph{Type Problems}.
The \emph{Graph Optimization} stage is the second most common location of failures, with 19 in total. 
The distribution of failures in this stage differs between converters, with $\sim$3x more in \code{tf2onnx}.
%\JD{I don't understand the next sentence. ``in line with'' is unclear. Are the TF graph optimization errors due to algorithmic? If so, write that.}
%These results are in line with \cref{sec:RQ1:causes} and indicate that the many \emph{Algorithmic Errors} in \code{tf2onnx} occur during the optimization stage.

{
\begin{table}[hbt!]
\centering
%\begin{wraptable}{r}{0.50\columnwidth}
    \caption{
    % \small
        Failure locations (cf.~\cref{tab:ONNXConverterStages}). 
        The majority of failures occur during \emph{Node Conversion} in each ONNX converter.
        %\code{tf2onnx} and \code{torch.onnx} failure are largely located within the \emph{Node Conversion} and \emph{Graph Optimization} stages.
        %\code{tf2onnx} failure in the \emph{Graph Optimization} stage more often than \code{torch.onnx}.
    }
    % Space
    \small
    \begin{tabular}{l |rr || r}
    \toprule
    \textbf{Location} &  \textbf{TF} &  \textbf{PT} &  \textbf{Total} \\
    \midrule
    Load Model            &        5 &        6 &     11 (6\%)\\
    \textbf{Node Conversion}   &       \textbf{70} &       \textbf{78} &    \textbf{148 (74\%)} \\
    (Graph) Optimization    &       14 &        5 &     19 (10\%) \\
    Export (Protobuf)              &        1 &        0 &      1 (1\%)\\
    Validation            &        0 &        3 &      3 (2\%)\\
    (Not Distinguishable) &       10 &       8 &      18 (9\%)\\
    \midrule
    \emph{Total}        &      \emph{100} &      \emph{100} &    \emph{200 (100\%)} \\
    \bottomrule
    \end{tabular}
    \label{tab:location}
\end{table}
}
%\end{wraptable}
%

\subsubsection{Failure Symptoms}
The distributions of symptoms for both \code{tf2onnx} and PyTorch are similar --- see~\cref{tab:symptoms}. 
The most common failure symptoms are \emph{Crash} and \emph{Wrong Model}, comprising $\geq$85\% of failures in each converter.
The \emph{Bad Performance} and \emph{Build Failure} make up around 5\% of failure symptoms.
%There were no cases of the \emph{Hang} symptom.

Our results show the similarity of the failure symptom distribution between DL model converters (our work) and DL compilers (Shen \etal~\citep{DLcompilerbugs}).
% interoperability systems 
\ifJOURNALEXTENSION
, who studied DL  compilers.
that use the pairwise mapping pattern of interoperability (\cref{B:Interoperability})
\fi
In the last two columns of~\cref{tab:symptoms}, we see
the \emph{Crash} and \emph{Wrong Model} symptoms make up the majority of symptoms across both DL converters and DL compiler front-ends. %for not just \code{tf2onnx} and PyTorch failures individually but also for DL compiler failures as well. 
%This finding lends empirical support to our comparison between model converters and DL compilers in \cref{sec:dl_model_converters:compilers}.
The \emph{Wrong Model} and \emph{Crash} symptoms appear characteristic of interoperability in this context, regardless of implementation.
% In the vein of the comparisons between DL compilers and converters, the \emph{Build Failure} results are \emph{not} replicated for converters. 
% This is because both converters are Python programs and as such do not need to be "built". The build failures that do occur are relegated to the installation of their dependencies. 
% \JD{The previous explanation about build failures sounds more like an accident than a meaningful difference. I think it's better to cut it if all we have to say is ``language difference''.}

% \begin{wraptable}{r}{0.6\columnwidth}
{
\renewcommand{\arraystretch}{0.7}
\begin{table}[h!]
    \centering
    \caption{
    % \small
    Distribution of symptoms.
    \emph{TF}: \code{tf2onnx}.
    \emph{PT}: \code{torch.onnx}.
    \emph{DL Comp.}: Symptoms of DL compiler failures in compiler front-ends, per Shen \etal~\citep{DLcompilerbugs}. 
    Percentages are rounded.
    % \JD{The caption here cites SHen2021 but the table inline cites DLcompilerbugs}
    % \JD{Numbers should eventually be macros}
    % \JD{Round percentages to the nearest ones place (32.5 becomes 33, etc.). Not enough data points to justify fine-grained reporting.}
    }
    % Space
    \small
    \begin{tabular}{l | rr || c || c}
    \toprule
    \textbf{Symptom}      &  \textbf{TF} &  \textbf{PT} &  \textbf{Total} & \textbf{DL Comp.} \citep{DLcompilerbugs}\\
    \midrule
    \textbf{Crash}           &       \textbf{50}  &       \textbf{62} &      \textbf{112 (\CrashPercent)} & \textbf{226 (\ShenCrashPercent)}\\
    \textbf{Wrong Model}     &      \textbf{35}  &       \textbf{30} &     \textbf{65  (\WrongModelPercent)} & \textbf{100 (\ShenWrongModelPercent)}\\
    Build Failure   &        3  &        2 &       5 (\BuildFailPercent) &  3 (\ShenBuildFailPercent)\\
    Bad Performance &       2   &        1 &      3  (\BadPerfPercent) &  6 (\ShenBadPerfPercent)\\
    Hang            &        0  &        0 &       0  (\HangPercent) &   4 (\ShenHangPercent)\\
    Unreported      &       10  &        5 &      15 (\UnreportedFailPercent) &  20 (\ShenUnreportedFailPercent)\\
    \midrule
    \emph{Total}           &      \emph{100}  &      \emph{100} &    \emph{200 (100\%)} & \emph{359 (100\%)} \\
    \bottomrule
    \end{tabular}
    \label{tab:symptoms}
% \end{wraptable}
\end{table}
}

\subsubsection{Failure Causes}\label{sec:RQ1:causes}

We report the joint distribution of failure causes by symptom in~\cref{tab:cau_sym_abridged}. % to facilitate debugging efforts.
The most common failure causes are the \emph{Incompatibility} and \emph{Type Problem}, with more than 50\% of failures exhibiting these causes for both converters. 
\emph{Algorithmic Errors} and \emph{Shape Problems} each contribute $\sim$20\% of cases. 
By symptom,
  crashes were caused by incompatibility and type problems,
  while 
  wrong models were caused by type problems and algorithmic errors.
%The \emph{Concurrency}, \emph{Incorrect Numerical Computation}, and \emph{Documentation} causes never occurred.
%\GKT{I like this analysis! Can you provide brief explanations for each of the unobvious ones. For example, what does shape mean here? Is this like Numpy's shapes on n-dimensional arrays? For numerical computations, can an example or two be provided? What do algorithmic errors mean at the IR level? Isn't it kind of a black box? Or is it the incorrect combination of operators?}
% The \emph{Shape Problem} cause distribution is relatively identical for both converters. 

For most causes, we see a similar distribution in each studied converter.
For \emph{Algorithmic Error}, we observe that \code{tf2onnx} has three times as many as \code{torch.onnx}.
The disparity in \emph{Algorithmic Errors} varies by the subclass, with the majority in \code{torch.onnx} being related to the loading of models namely tracing.
An example is PyTorch \#84092: \emph{trace not working on autocast}. 
In contrast, in \code{tf2onnx} the majority are related to optimizations,  such as
  tf2onnx \#226 (\textit{incorrect reshape}) 
  and
  tf2onnx \#1719 (\textit{incorrect folding}).
We conjecture that this difference may have a deeper cause:
  recall that the PyTorch converter is owned by the PyTorch engineers,
  while the TensorFlow converter is owned by the ONNX team (\cref{sec:FailureAnalysis-AboutONNX}).
This difference in ownership may reduce the team's understanding of the implications of optimizations, leading to worse outcomes in TensorFlow.

\ifJOURNALEXTENSION
Model converters do not utilize concurrency-oriented structures. for the former, importantly, this is because model converters do not utilize concurrency-oriented structures --- in fact, the \emph{Concurrency} cause does not appear the additional 117 failures studied for \textbf{RQ3} either (\cref{sec:res:RQ3}).
\fi

\ifJOURNALEXTENSION
 We discuss possible causes for this in \cref{sec:disc:conv_own}.
 \fi

{
\renewcommand{\arraystretch}{0.7}
\begin{table}
    % \footnotesize
    %\small
    \caption{
    % \small
    Joint distribution of primary causes and symptoms.
    %We bold notable occurrences and drop the \emph{Concurrency} and \emph{Documentation} rows because they contain all zeros. 
    The majority of \emph{Crashes} result from \emph{Incompatibilities} and \emph{Type Problems}.
    \emph{Algorithmic Errors} that result in \emph{Wrong Models} occur more often in \code{tf2onnx}.
    The top-5 causes in terms of frequency are shown, with the rest binned as \emph{Other}.
    Rare symptoms are likewise binned as \emph{Other}.
    %The \emph{Others} are \emph{Concurrency}, \emph{Docs}, \emph{Incorrect Numerical Computation}, \emph{Misconfig}, \emph{Testing}, \emph{Incorrect Exception Handling}, \emph{Typos}, and \emph{Incorrect Assignment}.
    }
\footnotesize
\small
\begin{tabular}{l|rr|rr|rr||rr}
\toprule
%\multirow{2}{*}{\backslashbox[15em]{Causes}{Symptoms}} & \multicolumn{2}{c|}{Crash} & \multicolumn{2}{c|}{Wrong Model} & \multicolumn{2}{c|}{Bad Performance} & \multicolumn{2}{c|}{Build Failure} & \multicolumn{2}{c||}{Unreported} & \multicolumn{2}{c}{Total} \\
\multirow{2}{*}{\textit{Cause} $\textbackslash$ \textit{Symptom}} & \multicolumn{2}{c|}{\textit{Crash}} & \multicolumn{2}{c|}{\textit{Wrong Model}} & \multicolumn{2}{c||}{\textit{Other}} & \multicolumn{2}{c}{\textit{Total}} \\
{} &    TF &  PT &         TF &  PT &              TF & PT &            TF & PT           \\
\midrule
Incompatibility                 &    \textbf{19} &  \textbf{28} &          4 &   3 &                         2 &  1 &    \emph{25} &   \emph{32} \\
Type Problem                    &     \textbf{8} &  \textbf{14} &         \textbf{17} &  \textbf{13} &                         0 &  2 &    \emph{25} &   \emph{29} \\
Algorithmic Error               &     4 &   3 &         \textbf{10} &   \textbf{3}         & 4 &  0 &    \emph{18} &    \emph{6} \\
Shape Problem                   &     5 &   4 &          4 &   7 &                         0 &  1 &     \emph{9} &   \emph{12} \\
API Misuse                      &     6 &   5 &          0 &   1 &                         0 &  0 &     \emph{6} &    \emph{6} \\
Other                          &    \BinnedJointOtherCrashTF &   \BinnedJointOtherCrashPT &          \BinnedJointOtherWrongModelTF &   \BinnedJointOtherWrongModelPT &              \BinnedJointOtherOtherTF & \BinnedJointOtherOtherPT & \textit{\BinnedOtherTF} & \textit{\BinnedOtherPT} \\ % \BinnedJointOtherBadPerfTF &  \BinnedJointOtherBadPerfPT &  \BinnedJointOtherBuildFailTF &  \BinnedJointOtherBuildFailTF &    \BinnedJointOtherUnreportedTF & \BinnedJointOtherUnreportedPT  &  \emph{\BinnedOtherTF} &    \emph{\BinnedOtherPT} \\
% Misconfiguration                &     1 &   0 &          0 &   0 &               0 &  0 &             2 &  2 &          1 &  0 &     4 &    2 \\
% Incorrect Exception Handling    &     1 &   1 &          0 &   1 &               0 &  0 &             0 &  0 &          0 &  0 &     1 &    2 \\
% Typo                            &     0 &   2 &          0 &   0 &               0 &  0 &             1 &  0 &          0 &  0 &     1 &    2 \\
% Testing                         &     0 &   1 &          0 &   1 &               0 &  0 &             0 &  0 &          1 &  1 &     1 &    3 \\
% Incorrect Numerical Computation &     0 &   0 &          0 &   0 &               0 &  0 &             0 &  0 &          0 &  0 &     0 &    0 \\
% Incorrect Assignment            &     0 &   2 &          0 &   0 &               0 &  0 &             0 &  0 &          0 &  0 &     0 &    2 \\
% Concurrency                     &     0 &   0 &          0 &   0 &               0 &  0 &             0 &  0 &          0 &  0 &     0 &    0 \\
% Documentation                   &     0 &   0 &          0 &   0 &               0 &  0 &             0 &  0 &          0 &  0 &     0 &    0 \\
\midrule
Total                           &    \emph{50} &  \emph{62} &         \emph{35} &  \emph{30} &                \emph{15} &  \emph{9} &   \emph{100} &  \emph{100} \\
\bottomrule
\end{tabular}
%}
\label{tab:cau_sym_abridged}
\end{table}
}

% \vspace{-0.1cm}
\section{Theme 3: Investigating Deeper Causes}
\label{sec:theme2}

% \JD{@Wenxin I added this transition paragraph}
% \WJ{Looks good}
In the final theme of this work, we investigate the possibility that ONNX converter errors have a shared \textit{structural cause}.
By this, we mean a latent cause beyond the code-level causes in the taxonomy of~\cref{tab:taxonomy_causes_concise}.
If so, ONNX users could use this factor to better assess their risk for the associated failure modes. 
Based on our survey data (\cref{sec:theme0}) and failure analysis (\cref{sec:theme1}), we formulate and test two hypotheses of structural failure causes.
\textit{The RQs and hypotheses are}:

% Does ONNX evolution affect converter failure rates?
{\textbf{RQ3: Does ONNX evolution affect converter failure rates?}}
We hypothesize 
  \textit{\textbf{H$_{{RQ}_3}$}: Changes in ONNX operator sets are correlated with increased defects}.
%The basis for this hypothesis is that defects are expected when requirements change~\citep{sommerville2015software}. 
This hypothesis is based in data from our survey and failure analysis.
In the survey, 19\% of respondents reported changing ONNX versions as a possible solution (\cref{table:Survey-AddressIssues}).
This suggests that conversion defects (not just compatibility issues) may be localized by version. 
In the failure analysis, we found that crashes were largely due to \textit{Incompatibilities} or \textit{Type Problems} (\cref{tab:cau_sym_abridged}). 
These failure modes relate to API compatibility
  % (incompatibilities)
and conversion correctness,
  % (type problems),
which can be affected by changes in the ONNX opset. %, \ie the addition, removal, or amendment of operators.

{\textbf{RQ4: Do model types affect converter failure rates?}}
We hypothesize 
  \textit{\textbf{H$_{{RQ}_4}$}: Failures are caused by model structure, \ie models with particular layers (node) or layer sequences are more prone to defects}. 
This hypothesis is primarily based on our failure analysis.
The majority of failures occurred during \textit{Node Conversion} (\cref{tab:location}), indicating that nodes and node sequences
 %(since a node sequence can map to a single node)
may be problematic.
Survey data also had a relevant anecdote: one respondent solved conversion issues by re-implementing models in pure tensor operations, implying that conversion may fail due to the use of certain operations.

% \textbf{RQ5:} \textit{Do operator/layer sequences affect failure cases?}
% \JD{We hypothesize that \textit{\textbf{H$_3$}: foo}.}
% The basis for this hypothesis is that...

% \vspace{0.15cm}
\noindent
In the remainder of this section, we describe the method and result for testing each hypothesis.
Ultimately, we find evidence to support \textit{\textbf{H$_{{RQ}_4}$}}. 
This result will help guide future validation efforts (\cref{sec:dis:conv_test}).
\subsection{RQ3: Does ONNX Evolution Lead to Failures?}
\label{sec:H1}

% \WJ{@Purvish}
\begin{tcolorbox} [width=\linewidth, colback=yellow!30!white, top=1pt, bottom=1pt, left=2pt, right=2pt]
%\textbf{Finding XX}: 
\newfinding H$_{{RQ}_3}$ is rejected: Changes in ONNX operator sets are \textbf{\textit{not}} strongly correlated with increased defects.
\end{tcolorbox}

Here we describe the method and results for testing \textbf{H$_{{RQ}_3}$}:
  That changes in ONNX operator sets are correlated with increased defects.
The expected correlated failures in the model converters are incompatibility and type problems (defined in~\cref{tab:taxonomy_causes_concise}).

\subsubsection{Methodology}\label{sec:meth:H1}
We test the hypothesis by checking whether \textit{larger} changes in the ONNX specification correlate with \textit{greater} incidence of failures in DL model converters. 
%We need the size of changes to the ONNX specification. 
To obtain the size of changes to the ONNX specification, we count the number of operator changes
(additions or updates)
per release in the ONNX release
notes (\cref{fig:OpsetOverTime}). % and measure the number of operator changes. 
We approximate the number of failures per release using the GitHub issue creation times of the failures sampled for RQ2.
%To obtain this, we examine the issue creation times of the failures sampled for RQ2. 
We then attribute each issue to the nearest previous ONNX release.

We note four ways in which this measurement approximates:
  (1) DL model converters lag behind ONNX releases (this might cause a failure to be mis-attributed to another release, \ie offset in time);
  (2) Failures might be in any ONNX available release, not just the most recent (possibly inflating the failure rate of a given release);
  (3) our failure analysis data were randomly sampled, possibly under-sampling certain time windows (though we note that our sample comprises $\sim$30-50\% of qualifying issues);
  (4) as noted in the user survey (\cref{sec:mot:results:interopFailureModes}), users may simply revert to a previous operator set without opening a bug report.
For items 1 and 2, issues do not reliably include ONNX versions, so no more accurate data is available.
For item 3, we use the data sampled during RQ2 so that we can check incidences by cause (incompatibility and type problems).

We test the hypothesis qualitatively and quantitatively.
Qualitatively, we inspect a visualization of the hypothesized trend.
We also measure the relationship, assessing the correlation in the number of changes in an ONNX release and the number of failures between its release and the next. 
We use the Spearman correlation, which is a commonly-used and robust metric for measuring a monotonic relationship between two variables~\citep{SW_Metrics_Fenton_Bieman_2014}.

% %%%%%%%%%%%%% Op Opset 
% \begin{figure}[b]
%     \centering
%     \includegraphics[width=0.97\columnwidth]{figures/illustrations/operators_over_time.png}
%     \caption{
%     % \small
%         Additions and updates of operators by ONNX operator set version, from version 1 (2017)---version 18 (2022).
%         %The number added are shown in orange, the number updated, in blue.
%         Version size := sum. %We sum to obtain the release size.
%     }
%     \label{fig:OpsetOverTime}
% \end{figure}

\subsubsection{Results} \label{sec:result:H1}

\begin{figure}[t]
    \centering
    \includegraphics[width=0.75\columnwidth]{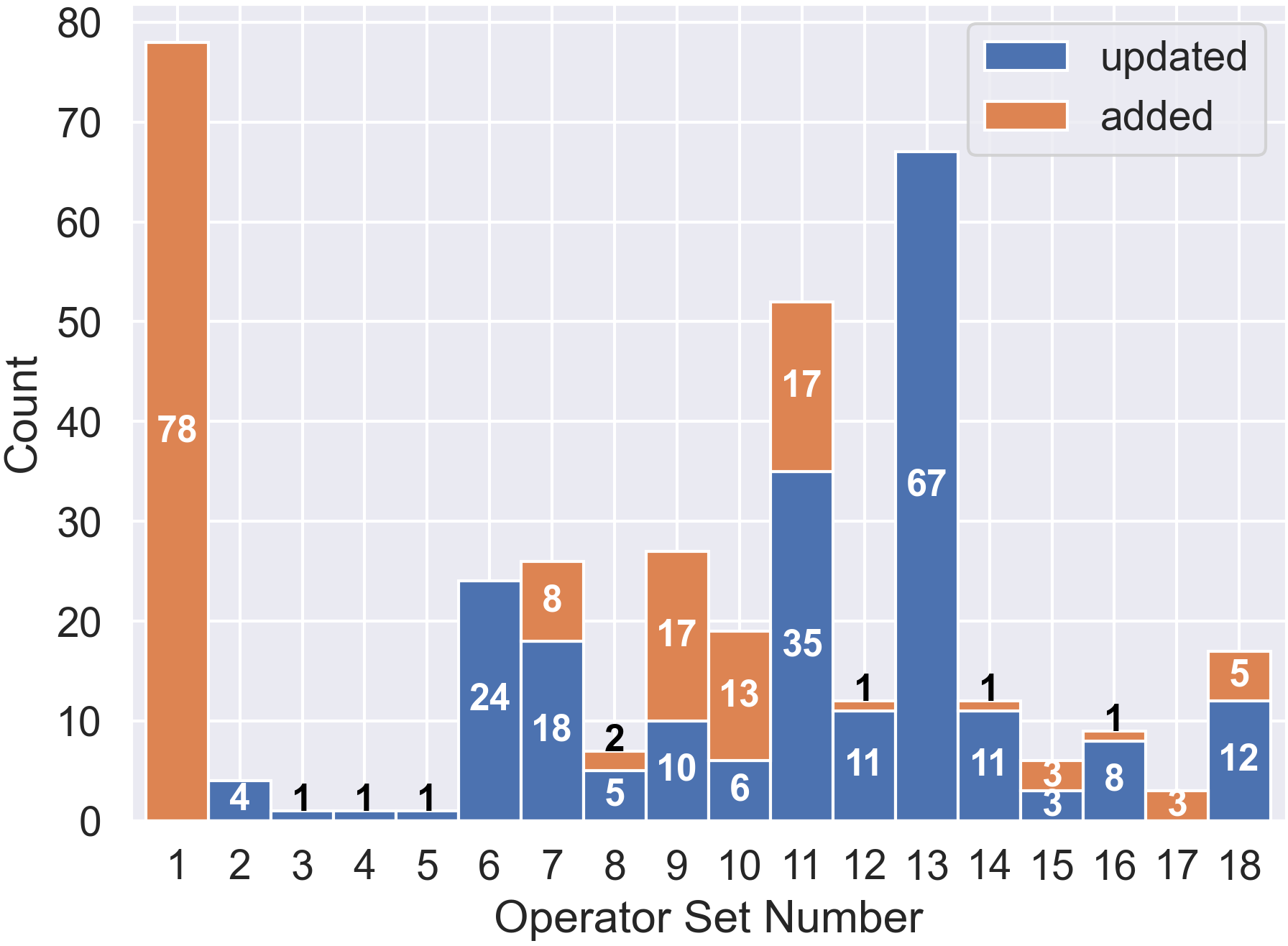}
    \caption{
    % \small
        Additions and updates of operators by ONNX operator set version, from version 1 (2017)---version 18 (2022).
        %The number added are shown in orange, the number updated, in blue.
        Version size := sum of operator additions and updates. %We sum to obtain the release size.
    }
    \label{fig:OpsetOverTime}
\end{figure}

\begin{figure}
    \centering
    \includegraphics[width=0.75\columnwidth]{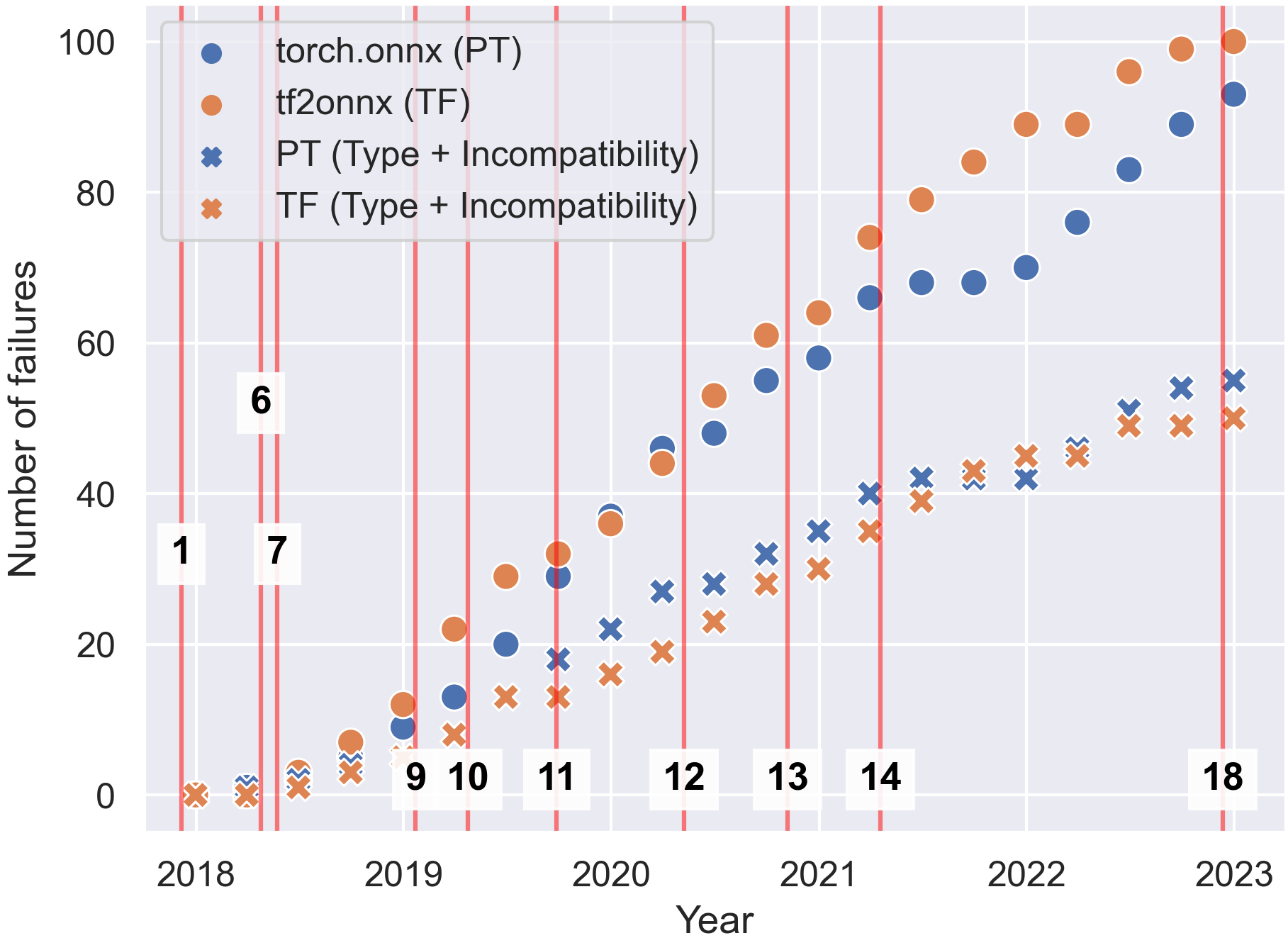}
    \caption{
    % \small
         Cumulative number of failures in the \code{torch.onnx} and \code{tf2onnx} converters, plotted quarterly from 2018-2023.
         %Each point represents cumulative failures up to that quarter, so
         The gap between two points is the number of newly opened issues during that time period.
         The O's track all failures.
         The X's track the subset of Incompatibility and Type Problem failures.
         The annotated vertical lines indicate the release of ONNX operator sets with $\geq 5$ changes. % (omitting operator sets releases with $<5$ changes). 
    }
    \label{fig:FailureVsRelease}
\end{figure}

%As discussed in~\cref{sec:meth:H1}, we measure the size of changes to ONNX using the ONNX release notes. 
%These sizes are shown in Figure 6a for the 18 ONNX releases.
\cref{fig:FailureVsRelease} depicts the hypothesized correlation between ONNX releases and model converter failures from~\cref{sec:theme1}.
After larger ONNX releases (\cref{fig:OpsetOverTime}) we expect more failures (\cref{fig:FailureVsRelease}). %, either in total (O's) or in the Incompatibility and Type subset (X's). 

Qualitatively, in~\cref{fig:FailureVsRelease} we see no discernible increase in the number of failures following larger ONNX updates. %, regardless of the size of that update.
Quantitatively (Spearman), results are similar.
The test yields a weak positive correlation ($\rho = 0.34$). % between the number of changes in an ONNX release and the total number of failures attributed to that release. 
Similarly, Incompatibility and Type Problems 
  %(the most likely types of failures after an update),
  are weakly positively correlated ($\rho = 0.33$).
In the test, we discarded the first release (converters may be unstable) and the most recent release (insufficient data). 
We also merged releases 2–7 because they are too close together (released within a 1-month period). %: releases 2–5 were released on the same date and 6–7 were released less than 1 month apart.

%\vspace{0.1cm}
%\noindent
\paragraph{Conclusion:}
\textit{Given the weak correlations, we do not find evidence to support H$_{{RQ}_3}$.}
There is a fairly steady rate of defects in ONNX model converters, whether the associated ONNX release is large or small.

% \vspace{-0.15cm}
\subsection{RQ4: Do Model Types Affect Failure Rates?}
\label{sec:H2}

\begin{tcolorbox} [width=\linewidth, colback=yellow!30!white, top=1pt, bottom=1pt, left=2pt, right=2pt]
%\textbf{Finding XX}: 
\newfinding Real models convert well: converter crashes and incorrect behavior affected only 5\% of models. Synthetic models show incorrect model behavior more often than Real models: 320/3544 (9\%) of synthetic models vs. 20/1605 real models (1\%).

\newfinding We find support for H$_{{RQ}_4}$. Though incorrect conversions are not directly attributable to the use of unusual operators, they may be attributable to operator sequences. %  We did not observe strong correlation between conversion failures and unusual operator sequences. 

\end{tcolorbox}

For this RQ we analyze models at two scales, \textit{macro} and \textit{micro}. 
In our \textit{macro} scale analysis (\cref{sec:meth:H2}, \cref{sec:result:H2}), we study entire models (all layers).
For our \textit{micro} scale analysis (\cref{sec:meth:H3}, \cref{sec:result:H3}) we study individual layer and sequence patterns from the macro analysis.

\subsubsection{Method for Macro Analysis}\label{sec:meth:H2}
\begin{figure}[b]
    \centering
    \includegraphics[width=0.70\columnwidth]{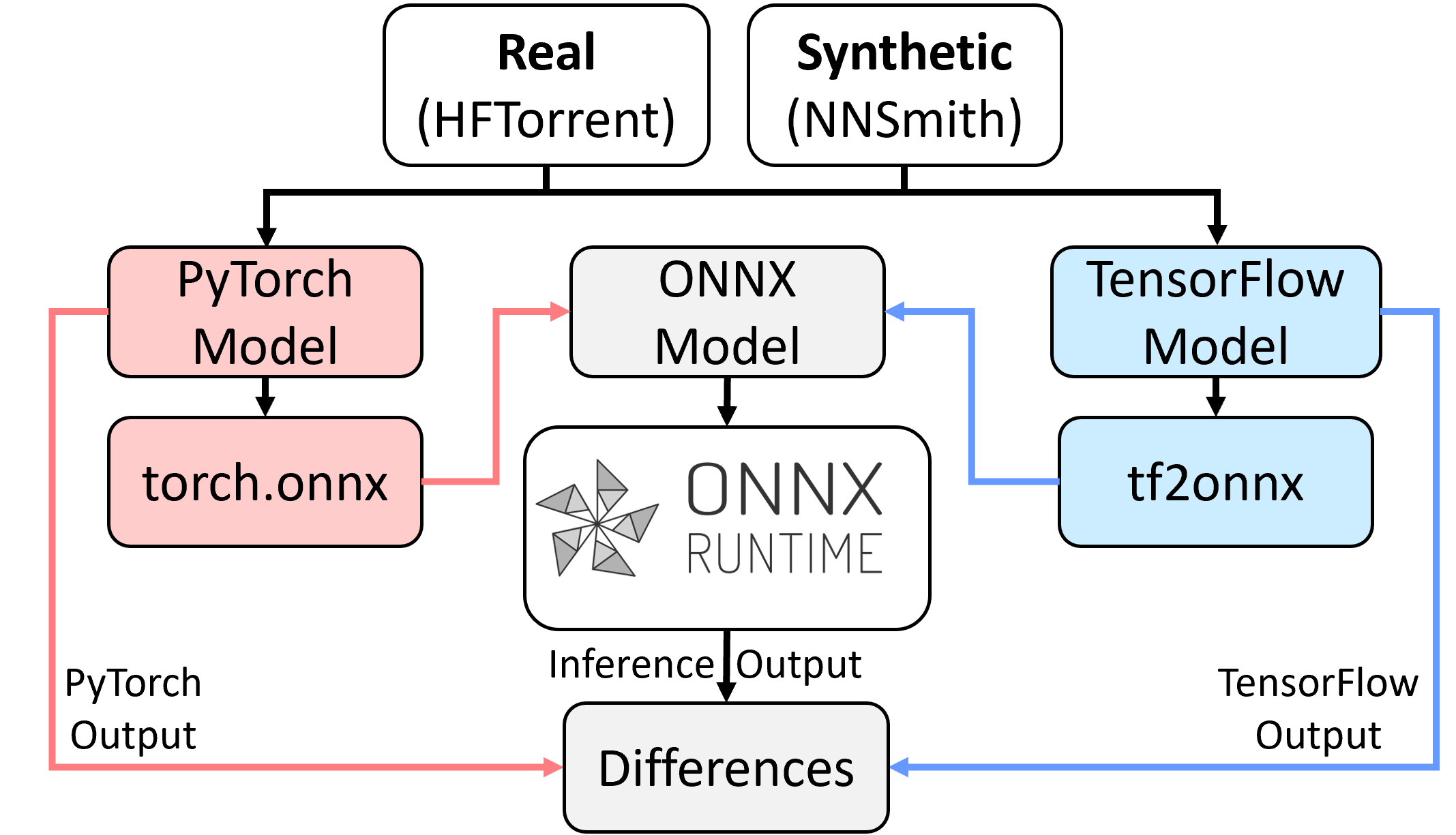}
    \caption{
    % \small
    \textbf{Macro analysis}: \textit{Real} and \textit{Synthetic} models are converted and then differences are measured. % between original and converted models. 
    \ifJOURNALEXTENSION
    Mismatching models are analyzed in our \textit{micro analysis} (\cref{sec:meth:H3}).
    \fi
    For synthetic models, we used \code{tf2onnx} or \code{torch.onnx} directly.
    For real models we used HuggingFace's converter, which does preprocessing and then calls those converters.
%For the real models, we used the HuggingFace converter. This converter performs some preprocessing and then invokes \code{torch.onnx} or \code{tf2onnx}. For synthetic models, we invoked the respective converters directly
    }
    \label{fig:RQ4MacroAnalysis}
\end{figure}

At the macro scale, we test DL converters using two types of models and then analyze the failure-inducing inputs.
% To answer \textbf{RQ4}, we test DL converters with a variety of inputs and analyze the characteristics of failure-inducing inputs. 
We used a differential testing approach (\cref{fig:RQ4MacroAnalysis}).
Differential tests need  
  (1) inputs,
  and
  (2) a difference criterion~\citep{mckeeman_differential_1998}.

%illustrates our testing process.
%We use a differential testing approach.
%
\iffalse
We convert two types of models to capture both typical and unusual input regimes:
  real-world DL models from Jiang \etals{} \code{HFTorrent} dataset~\citep{jiang_empirical_2023_1, jiang_empirical_2023_1, jiang2023challengespracticesdeeplearning}
  and
  synthetic models generated systematically~\citep{NNSmith}. 
%These models represent the spectrum of possible input, both expected and improbable.
We convert each model with its respective converter.
The conversion may fail (\eg crash).
If the converted model loads in the ONNX Runtime, we use differential testing~\citep{mckeeman1998differential} to measure conversion quality. % between original and converted models.
%For failing models, we submit GitHub issues to the respective repositories.
We analyze the characteristics of failing models.
Specifically, we analyze the operator types and operator sequences --- we justify our choices in \cref{todo}. 
And their operator and operator sequences to identify shared features.
Additionally, we compare with models that do not fail and determine if operators or operator sequences are shared.
\fi

%%%%%%%%%%%%% Op Opset 

%%%%%%%%%%%%% Failures as functions of time
% \vspace{-0.05cm}
%%%%%%%%%%%%%

\myparagraph{(1) Inputs} For inputs, we converted both real models and synthetic models.
Real models contain input patterns that the converter expects. 
Synthetic models are more diverse and should exercise edge cases on the converter.
Real and synthetic models came from:
% Real models are inputs that converters are likely to see.
% Synthetic models are inputs that converters would not see (fuzzing).
% Saying the converters would not see synthetic models may make the reviewer ask why they are important.
\begin{itemize}[leftmargin=10pt]
    \item \textit{Real Models:}
        We used the HFTorrent dataset from Jiang \etal~\citep{jiang_empirical_2023_1}.
        At time of experiment, this was the largest available set of real-world DL models, containing \HFTorrentTotalRepos pre-trained models collected from the HuggingFace model registry.\footnote{Future measurements could use the PeaTMOSS dataset~\citep{jiang2024peatmoss} or its successors.}
        We filtered for the \TotalHFModels models with both PyTorch and TensorFlow versions, allowing comparison between the two converters on similar inputs.
        These models represent \HFDistinctArch distinct full architectures (\HFArchitectures backbones).
    \item \textit{Synthetic Models:}
        We systematically generated synthetic DNNs for conversion, using the \code{NNSmith} tool~\citep{NNSmith}.
        \code{NNSmith} generates random valid TensorFlow and PyTorch DNNs that use operators supported by ONNX.
        We added a parameter to \code{NNSmith} to generate DNNs of a given size (\# nodes).
        We then \textit{systematically generated DNNs containing between \GenLowerBound-\GenUpperBound nodes in increments of \GenIncrements}.
        For each node count, we generated DNNs until two conditions were met:
          (1) All \code{NNSmith}-supported operators appeared at least once in the family;
          and
          (2) $\geq$ \GenUpperBound distinct models were created.
\end{itemize}

In initial tests, we found the resulting models were often convertible yet unsupported by ONNX~RunTime.
    \ifJOURNALEXTENSION
    (\cref{tab:RQ4ConversionResults}).
    \fi
To address common errors, we constrained the NNSmith synthesizer.
    % (\cref{tab:RQ4ConversionConstrainedResults}).
    Specifically, we removed \PTOpsRemoved operator out of \PTOps for \code{torch.onnx} and \TFOpsRemoved out of \TFOps for \code{tf2onnx}. 
    Additionally, we confined the tensor data type to \code{float32}.

    There are other DL model generation approaches, \eg 
      MUFFIN~\citep{Muffin_Gu_Luo_Zhou_Wang_2022},
      COMET~\citep{COMET_Li_Cao_Tian_Tsz_On_Li_Wen*_Cheung*_2023},
      and
      LEMON~\citep{LEMON_Wang_Yan_Chen_Liu_Zhang_2020}.
    These tools generate inputs by mutating seed models, typically real models.
    We viewed HFTorrent as a sufficient source of real models, and synthesized with~\code{NNSmith} as a complementary approach. %because it is more random. %\GKT{Ok, I think this is good.}
    %We discuss its limitations in~\cref{sec:Threats}.

\myparagraph{(2) Difference Criterion}
For each model, we attempt to convert it to ONNX using its respective converter.
%\footnote{For the real models, we used the HuggingFace converter. This converter performs some preprocessing and then invokes \code{torch.onnx} or \code{tf2onnx}. For synthetic models, we invoked the respective converters directly.} % --- \code{torch.onnx} for PyTorch and \code{tf2onnx} for TensorFlow. 
If conversion succeeds, we load it into the ONNX~RunTime.
For models that load successfully, we perform inference on both the original model (using PyTorch or TensorFlow) and the converted model (using ONNX Runtime).
For \emph{real} models, we use the inputs provided by the model owners to test the model.
(These are also known as ``dummy inputs'' on the HuggingFace platform.)
For \emph{synthetic} models, we use 100 randomly generated inputs matching the model's input shape.

In both cases, to measure model misbehavior we used a simple community-accepted approach:
  the distance measure and threshold used by the PyTorch exporter. % one distance measure to compare the output tensors from the original and converted models:
This rule is that \textit{between the original and ONNX-converted models, in the inference result tensors, the maximum absolute element-wise difference should be $< 10^{-7}$}
~\citep{pytorch_onnx_verification}.
% ~\citep{huggingface_export_validation}

We considered alternatives to measure behavioral differences.
% We acknowledge that there are many other ways to measure model behavioral differences.
A more general notion of tolerance or ``acceptable error'' could be used, but the choice of tolerance is not well established --- the PyTorch verification tool uses $10^{-7}$, NNSmith uses $10^{-3}$~\citep{NNSmith}.
% , and HuggingFace uses $10^{-7}$.
Openja \etal~\citep{ConvertingDLempirical} proposed using model accuracy and robustness,
\ifJOURNALEXTENSION
\footnote{Though related, there is a distinction between the predictions of the model and the accuracy of a model. The ``accuracy'' of a model is the performance of a model on a dataset, whereas the predictions are the outputs of the model. A model with accuracy differences (relative to an oracle) would necessarily mean that there are differences in the predictions, but the inverse is not necessarily true.}
\fi 
but this method requires training each model on a suitable dataset.
Training substantially increases the cost of the measurement, and the dataset requirement limits the types of models that can be generated.
% \JD{This is a fact, but we also need to give the rationale/reasons why it's hard to decide as well. Methods sections must answer both What and Why. Apply this principle throughout your methods sections, not just here.}
% \TODO{We received many comments about "accuracy loss" / "correct" behaviour in the last review. It would be worthwhile to quantify why this is the case }
  % and the L1 and L2 measures.
%We consider the converted and original models identical if the max absolute difference $< 10^{-3}$.
%This follows the thresholds used by HuggingFace~\citep{huggingface_transformers_serialization}.

\iffalse
% FIGURE 5
\begin{wrapfigure}{r}{0.61\columnwidth}
    \centering
    \includegraphics[width=0.6\columnwidth]{figures/illustrations/conversions.png}
    \caption{
    Conversion test.
    Models are converted using \code{tf2onnx} or \code{torch.onnx} as appropriate.
    If conversion succeeds, differences are measured between original and converted model.
    }
    \label{fig:MethodForRQ4}
    %\vspace{-0.35cm}
\end{wrapfigure}
\fi

\subsubsection{Results for Macro Analysis} \label{sec:result:H2}
\myparagraph{Real Models}
Real models exhibited few converter failures and little incorrect model behavior.
Most issues occurred within the HuggingFace-specific converter.
Converter failures occurred for only \ConverterFailures/\ConverterAttempted models ($\sim$\ConverterFailuresPercentage\%).
Once models reached \code{torch.onnx} and \code{tf2onnx}, over \ConverterSuccessPercentage\% could convert and had equivalent behavior in ONNX Runtime.

One interesting form of failure we observed was models with identical architectures but varying conversion issues.
%We note that the dataset of real models we used included many that shared an architecture and differed only in weights (\ie checkpoints of the same model).
% We were surprised to observe that the different weights alone were enough to produce different behaviors in our evaluation.
For example, in \code{tf2onnx}, various checkpoints of the \code{t5} model
  had 65 HuggingFace conversion errors,
  56 unsuccessful \code{tf2onnx} conversions,
  1 unsuccessful ONNX Runtime load,
  0 incorrect outputs,
  and
  183 successes.
We could not determine the cause(s) of this instability. % It bears further investigation.\GKT{See future plans section?}

% The model authors
% %know of the inconsistent behavior in ONNX and
% attribute this to ``implementation differences''~\citep{sharma2021primera}.

\myparagraph{Synthetic Models} 
Synthetic models exhibit more difficulties and often show incorrect model behaviour.
%
%As described in~\cref{sec:Methods-Theme1}, we systematically generated models using our adaptation of the NNSmith tool.
% , with and without constraining for the ONNX Runtime.
%Synthetic models had more problems. %varied substantially from real models. 
%
% \myparagraph{Difficulties in Converting and Loading:}
%

\textit{Crashes:} 
% In this experiment, we independently identified both known and novel defects.
Synthetic models exhibit
% more
difficulties when converting to ONNX and loading converted models to the ONNX Runtime.
We find \TotalConvCrash/\TotalSynModel ($\sim$23\%) \emph{Unsuccessful Conversions}, almost entirely in synthetic models on \code{tf2onnx}.
These crashes %\TotalConvCrash crashes 
correspond to \UnsucessfulUniqueCrashes unique crash locations.
% , both of which involve \emph{``no complex type support''}~\citep{pytorchissue59246}.
Of the \SuccessConvNN/\TotalSynModel ($\sim$77\%) synthetic models that successfully convert, only \SuccessLoadNN/\SuccessConvNN ($\sim$45\%) successfully load into ONNX Runtime. 
We observed \ORTUnique unique ONNX Runtime errors, of which 6 appear to correspond to open GitHub issues~\citep{pytorchissue78721, onnxruntime9760, onnxruntime17061, onnxruntime11846, onnxruntime12302, onnxruntime12594}.
%returned with \ORTTypeMismatch.
We disclosed the rest to the engineering teams.
% ~\citep{tf2onnx2235, tf2onnx2247, pytorch109806, ONNX Runtime17616}. 
For most disclosures, we have not yet received a response. 
For one disclosure, which described a model that causes the \code{libc++ abi} to terminate unexpectedly during inference with ONNX Runtime, it had been (unintentionally) fixed in the development branch.
The defect still affects older releases. 
\textit{Behavioural Differences:} We observed a large fraction of behavioural differences (incorrect output) with synthetic models. % with constrained synthetic models revealing the most for the \code{torch.onnx} converter.
Compared to real models, which had \FailureHFSemanticTotal instances, synthetic models had \FailureSemanticTotalSyn instances where the inference results exceeded the threshold. % from~\cref{sec:meth:RQ3}.
% Almost all of these instances were observed only for the constrained synthetic models.
The majority of these instances were observed in the \code{tf2onnx} converter.
For both converters, we disclosed such instances to the respective engineering teams (we have not heard a response yet).
A summary of disclosed issues can be found in the artifact.
% \WJ{Do we have the reponses yet?}
% \PJ{We have some. I can add them in.}
% \JD{Do so, and include an anonymized version of the tracking spreadsheet in the artifact. Specifically, please include columns `Date opened', `Issue title', `Summary of engineers' response', `Current issue state', `Action taken', but omit the column `Link to issue'.}
% \PJ{Done.}

%~\footnote{Blinded for submission.} 

{
\renewcommand{\arraystretch}{0.7}
\begin{table}[]
    \caption{
    Results of conversion testing.
    Real models' conversion may fail in the HuggingFace wrapper.
    Both kinds of models may fail in DL model converter, or when the result is fed to ONNX Runtime (ORT).
    %\emph{Crash (converter)} is a crash during model conversion.
    %\emph{Crash (ORT)} is a crash in the loading of a \emph{successfully} converted model into ONNX Runtime.
    \emph{Behavioural Difference}: ORT inference results with difference $> 10^{-7}$. 
    Real models fail rarely, synthetic models often.
    % \JD{Needs percentages}
    }
    % \vspace{-5pt}
   % Reduce font size?
   \small
    \begin{tabular}{l|rr|rr}
    \toprule
    \multirow{2}{*}{\textbf{Outcome}}   &  \multicolumn{2}{c}{\emph{\textbf{\code{tf2onnx}}}} &  \multicolumn{2}{c}{\emph{\textbf{\code{torch.onnx}}}} \\
                    &         \textbf{Real} &  \textbf{Syn.} &      \textbf{Real}  & \textbf{Syn.} \\
    \midrule
    \emph{Start: Total models} &        \TotalHFModels &  \TotalTFConGen &  \TotalHFModels & \TotalPTConGen\\
    % Successful      &        \SuccessHFTF &   \SuccessNNTF &    \SuccessHFPT & \SuccessNNPT\\
    Conv. Fail (HF) &    \CrashesHFTFHugging (\CrashesHFTFHuggingPercent) & N/A & \CrashesHFPTHugging (\CrashesHFPTHuggingPercent) & N/A \\
    \midrule
    Conv. Fail        &          \CrashesHFTF (\CrashesHFTFPercent) & \ConvCrashTFConGen (\ConvCrashTFConGenPercent) &     \CrashesHFPT (\CrashesHFPTPercent) & \ConvCrashPTConGen (\ConvCrashPTConGenPercent) \\
    ORT load Fail  & \CrashesORTHFTF (\CrashesORTHFTFPercent) & \ORTCrashTFConGen (\ORTCrashTFConGenPercent) & \CrashesORTHFPT (\CrashesORTHFPTPercent) & \ORTCrashPTConGen (\ORTCrashPTConGenPercent) \\ 
    Mismatch     &       \FailureHFTFSemantic (\FailureHFTFSemanticPercent)  &  \SemanticTFConGen (\SemanticTFConGenPercent) &     \FailureHFPTSemantic (\FailureHFPTSemanticPercent) & \SemanticPTConGen (\SemanticPTConGenPercent)\\
    % \midrule
    % \midrule
    Successful      &        \TrueSuccessHFTFM (\TrueSuccessHFTFMPercent) & \SuccessfulTFConGen (\SuccessfulTFConGenPercent) &    \TrueSuccessHFPTM (\TrueSuccessHFPTMPercent) & \SuccessfulPTConGen (\SuccessfulPTConGenPercent)\\
    
    % Unique Crashes &   \UnqCrashesHFTF & 0 & \UnqCrashesHFPT & \UnqCrashesNNPT\\
    % \midrule\midrule
    % Diff. Tested &  - & \DiffTestedNNTF & \DiffTestedHFPT & \DiffTestedNNPT\\    
    % \midrule\midrule
    \bottomrule
    \end{tabular}
    \label{tab:RQ4ConversionResults}
%\vspace{-0.15cm}
\end{table}

\ifJOURNALEXTENSION
\begin{table}[h]
  \centering
  \small
  \caption{
  Table describing the issue.
  }
  \begin{tabular}{c|c|c|c|}
    \toprule
    \textbf{Description} & \textbf{Symptom} & \textbf{Location} & \textbf{Conjectured cause} \\
    \hline
    \code{tf2onnx} segfaults during conversion & Crash & ... & protobuf size limit \\
    \code{tf2onnx} segfaults during conversion & Crash & ... & protobuf size limit \\
    
    \bottomrule
  \end{tabular}
  \label{tab:mytable}
\end{table}
}
\fi

\subsubsection{Method for Micro Analysis}\label{sec:meth:H3}
%\paragraph{Model Causal Analysis (\cref{fig:RQ4Microanalysis})} 
%\label{sec:meth:model_analysis}

\begin{figure}[b]
    \centering
    \includegraphics[width=0.8\columnwidth]{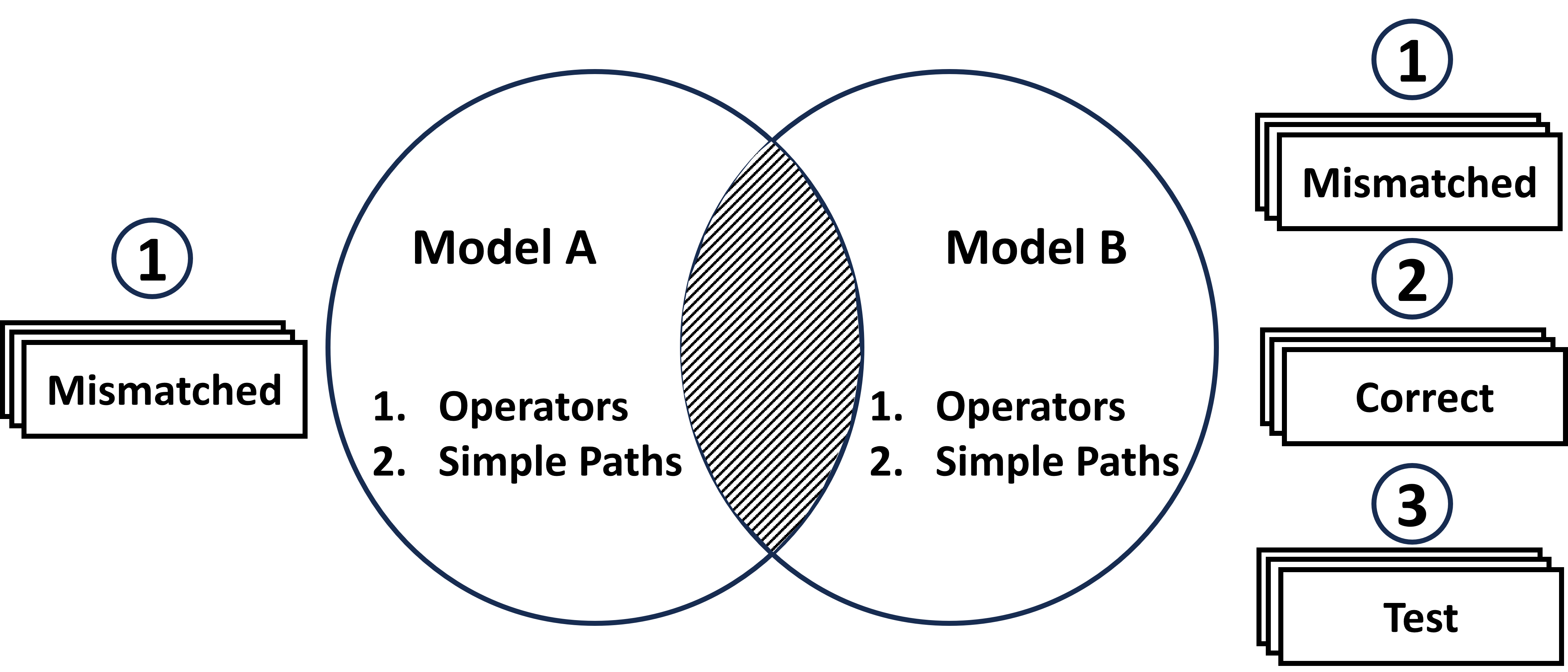}
    \caption{
    % \small
    \textbf{Micro analysis}:
    Given two sets of models, we measure shared operators and simple paths between all model pairs (intersection).
    %The common operators and sequences are depicted in the intersection.
    We compare three sets.
    First, the pairing
    $\circledstep{1} \times \circledstep{1}$ compares mismatched models to themselves.
    Any operators and sequences shared by mismatched models may indicate shared causes of failure.
    Second, $\circledstep{1} \times \circledstep{2}$ compares mismatched and correct models.
    Overlaps suggest a given operator or sequence is sometimes problematic, depending on context.
    %It contains all operators and sequences shared by mismatched and correct models.
    Third, $\circledstep{1} \times \circledstep{3}$ compares mismatched models to those from converter test suites.
    Substantial \textit{non-}overlap implies gaps in test coverage.
    %It contains all operators and sequences shared by mismatched and test suite models.
    % \circledstep{2} is found by comparing models with the set of mismatched models. 
    % It contains all sequences that are shared by mismatching models. \circledstep{3} is found by comparing models from the mismatched and correct model sets.
    % It contains all sequences that are shared between correct and mismatched models.
    % \circledstep{4} is found by comparing models from the mismatched and test suite sets.
    }
    \label{fig:RQ4Microanalysis}
\end{figure}

%The macro analysis shows failure rates but does not explain causes. 
To investigate the causes of the conversion failures we observed, we analyze mismatching models in terms of the operators used.
We examine the individual operators and the sequences of operators.
We compare these to non-failing models (for trends) as well as to the models in the converter test suites (for gaps in testing). 
Test suite models are collected from converters' CI/CD pipelines.
\cref{fig:RQ4Microanalysis} illustrates our approach.
%along two dimensions: \emph{incorrect conversions result from \textbf{operator types} and \textbf{operator sequences}}.
% Given the challenges in interpreting the behaviors of DNNs, particularly across two different representations, 
%Here we explain our method to evaluate them.
% These hypotheses derive from our results in Theme 1, so we defer the rationale to~\cref{sec:res:RQ3}. % (specifically, Findings \TODO{X,Y,Z} in \cref{sec:Results}). 
%\emph{Node Type} and \emph{Algorithmic Errors} are common --- \eg \TODO{Some examples from Theme 1} --- indicating that both the types of operators and their sequences of them play a part in failures.

For operator types, we measure the operators present in each converted model, out of the set of available operators in the ONNX opset.
For operator sequences, we extract the simple paths through each model. 
To illustrate,
  the ONNX model in~\cref{fig:graph_comp} shows 
  three operators (ReduceMax\_0, ArgMax, and ReduceMax) 
  and
  two simple linear paths (ReduceMax\_0$ \rightarrow $ArgMax and ReduceMax\_0 $\rightarrow$ ReduceMax). % GKT: Added spacing to assist in proper wrapping.
Models with similar architectures may have simple paths that are largely identical,
inflating the number of shared sequences. 
%For example, the sequences \code{aaaabd} and \code{aaaabc} are identical except for the last element.
For example, the sequences \code{aaaabd} and \code{aaaabc} are identical except for the last element.
To deal with cases we further reduce the common sequences to the smallest shared subsequences, \ie we recursively find the longest common subsequences for our sequences until we cannot find any smaller subsequences~\citep{algo_cormen2022introduction}.
After reduction we find that sequences are between 3 to 5 operators long.

\myparagraph{Evaluation Criteria for H$_{{RQ}_4}$} \label{sec:meth:eval_criteria}
% Given this data, we test the two hypotheses.
%Given this data, we evaluate the effect of the dimensions of analysis.
To evaluate the impact of \emph{operator types}, we compare the operators of mismatched and correct models.
If failing and correct models use different operators, we conclude that specific operators may cause failures. 
To evaluate the impact of \emph{operator sequences}, we identify shared operator sequences belonging \emph{only} to mismatched models, and their frequency. %determine how common they are. 
If mismatching models often share common simple paths of operators, that will support H$_{{RQ}_4}$.
H$_{{RQ}_4}$ would be further strengthened if operator sequences are rarely shared with correct models.

\subsubsection{Results for Micro Analysis}\label{sec:result:H3}
%As described in~\cref{sec:meth:H3} we analyze both the operator type and operator sequences for various sets of models,
We focused on synthetic models that converted successfully but had mismatching behavior.
There were 330 such models across TensorFlow and PyTorch.
We compared them to correct models and models from converter test suites.
% (discussed in~\cref{sec:res:RQ4}).

% \JD{State which hypothesis is being considered and what your conclusion is}
% We find that incorrect conversions do not result simply from operator types.
Incorrect conversions cannot be explained solely by operator types.
%For operator types, we found that incorrect conversions are not related simply to the types of nodes present.
%\cref{tab:RQ4AnalysisResultsNodes} contains the results of our node analysis.
%We collect and compare the operator types in the mismatched, correct, and test suite models.
Of the 154 ONNX operators, 
  mismatched models contain 58 (\code{torch.onnx}) and 54 (\code{tf2onnx}) unique operators 
  while
 correct models contain 59 (\code{torch.onnx}) and 52 (\code{tf2onnx}) unique operators. 
% (3) test suite models contain 16 (\code{torch.onnx}) and 54 (\code{tf2onnx}) unique operator types.
%There is a significant overlap between operators in correct and mismatched models.
All but 1-2 operators are shared, indicating that for synthetic models, operator types do not predict mismatch. % but 1 (\code{torch.onnx}) and 2 (\code{tf2onnx}). %, indicating that for synthetic models operator types may not be determinant of mismatch.
% Test suite models on the other hand shared 12 operators (\code{torch.onnx}) and 33 (\code{tf2onnx}) with mismatched models.
% For \code{tf2onnx}: 
% (1) correct models contain 52 unique operator types; 
% (2) mismatched models contain 54 unique operator types;
% (3) test suite models contain 57 unique operator types.
% There is a significant overlap between operators in correct and mismatched models.
% Almost all operators are shared save for 1, indicating that at least for synthetic models node types may not be determinant of mismatch.
% Test suite models on the other hand only shared 35 operators with mismatched models.

In contrast, our analysis of shared operator sequences supports hypothesis H$_{{RQ}_4}$.
As shown in~\cref{tab:RQ4AnalysisResultsSeq} an overwhelming majority of mismatched models share unusual operator sequences.
$312/320$ of the mismatching models tested share at least one operator sequence that never occurred in the correct models.
Further, we assess how often the sequences are shared by many models.
We find that for \code{torch.onnx}, only 1 of the 131 sequences is shared by more than 3 models, and for \code{tf2onnx} no sequence is shared by more than 11 models.
This indicates the failing models will often contain common operator patterns, suggesting families of sequences that cause errors.
Finally, our comparison of test suite and mismatching models ($\circledstep{3}$ in \cref{tab:RQ4AnalysisResultsSeq}) shows that the failing models share few sequences with the models used in converter test suites (which are real models rather than synthetic ones), suggesting a gap in test coverage.
Further analysis is left to future work.

{
\renewcommand{\arraystretch}{0.5}
\begin{table}
    \centering
    \caption{
    Model sequence analysis results for the synthetic models that converted but mismatched (``Mismatch'' in~\cref{tab:RQ4ConversionResults}).
    $\circledstep{1}$ : Sequences shared by mismatched models. 
    $\circledstep{2}$ : Sequences shared by mismatched and correct models;
    $\circledstep{3}$ : Sequences shared by mismatched and test suite models.
    $\circledstep{1} \setminus \circledstep{2}$ : sequences present only in mismatched models.
    % \circledstep{4} - \circledstep{2} are all unique sequences that are \emph{only} present in test suite models.
    \textit{Reduced}: smallest shared subsequences (\cref{sec:meth:H3}).
    }
    % \vspace{-3pt}
   % Reduce font size?
   \small
    \begin{tabular}{c|rr|rr}
    \toprule
    \multirow{2}{*}{\textbf{Set}}   &  \multicolumn{2}{c}{\emph{\textbf{\code{tf2onnx}}}} &  \multicolumn{2}{c}{\emph{\textbf{\code{torch.onnx}}}} \\
                    &   \textbf{Unique} & \textbf{Reduced} &      \textbf{Unique} & \textbf{Reduced} \\
    \midrule
    Total Models & \multicolumn{2}{r|}{220} & \multicolumn{2}{r}{100}\\
    \midrule
    $\circledstep{1}$ Mism. $\cap$ Mismatch  &    2,125 & 1,126 & 980 & 635 \\
    $\circledstep{2}$ Mism. $\cap$ Correct  &    1,050 & 508 & 4,243 & 2,988 \\
    $\circledstep{3}$ Mism. $\cap$ Tests &    35 &  35 & 2 & 2\\
    \midrule
    $\circledstep{1} \setminus \circledstep{2}$ &      1,527 & 862 & 176 & 131\\
    % \circledstep{2} - \circledstep{4} &       2,135 & 1,097 & 979 & 637\\
    \midrule
    \# models with $\circledstep{1} \setminus \circledstep{2}$  
    &  & 216 &  & 96 \\
    %& & & & \\
    \bottomrule
    \end{tabular}
    \label{tab:RQ4AnalysisResultsSeq}
%\vspace{-0.15cm}
\end{table} 
}

\ifJOURNALEXTENSION
Constraining synthetic models is effective at revealing incorrect model behavior post-conversion, this is for the \code{torch.onnx} converter.
The effect of constraining models results in more models being able to be loaded into ORT, as seen in \cref{tab:RQ4ConversionResults}.
We find that 728 of the 7,192 synthetically generated models exhibit inference result differences, with the majority being by synthetic models that were constrained.
Constrained \code{torch.onnx} models more often revealed incorrect model behavior, with \SemanticPTConGen/\SuccLoadedPTConGen ($\sim$37\%) as compared to unconstrained which had \FailureNNPTSemantic/\SuccLoadedPTSyn ($\sim$24\%).
We do not measure this for \code{tf2onnx} where though more models successfully were loaded into ONNX Runtime this did not result in measuring more models which exhibited incorrect behavior.
\fi

\ifJOURNALEXTENSION
% Present Data
\subsubsection{Issues with Model Loading in Compilers}\TODO{Merge with 5.4.3 and talk about constrained generation}
In attempting to do this we find that only \TotalAttempted of the \TotalSuccessForDiff that successfully converted were able to load into ORT --- mainly belonging to \emph{real} models.  
We discuss the implications of this with respect to converter testing in \cref{sec:dis:conv_test}.
\fi

\section{Discussion and Future Work}

\subsection{Validating DL Model Converters}\label{sec:dis:conv_test}

\iffalse
\begin{tcolorbox} [width=1.0\linewidth, colback=blue!07!white, top=1pt, bottom=1pt, left=2pt, right=2pt]
% \textbf{OQ1}:
% % How can we identify failures in DL model converters?
% How can we better test DL model converters?

\textbf{OQ1}: How can we better localize faults in converters?

\textbf{OQ2}:
What tools would reduce failures in current converters? % mitigate coupling between frameworks and runtimes?

\textbf{OQ3}: How much error is tolerable as a result of conversion?
\end{tcolorbox}
% \vspace{-0.10cm}
\fi
We find DL model converters are robust to new operator releases (\cref{sec:H1}).
The weak correlation found between ONNX releases and issues implies that test suites are sufficient for catching issues that may come along with new ONNX releases --- modulo the shortcomings of the approximations listed in~\cref{sec:meth:H1}.

Converter test suites share little in common with failing models, often missing critical model structures (\cref{sec:result:H3}).
Specifically, \textbf{model converters need better testing techniques for behavioral changes}.
Though this critical failure mode occurred in $\sim$1/3 of historical failures (\cref{sec:res:RQ1}) and affected \IncorrectTotalPercentage\% of the models we tested (\cref{sec:result:H2}), the existing test suite models share little in common with models affected by this failure mode.
For example, only 37 operator sequences are shared between mismatching models and test suite models (\cref{tab:RQ4AnalysisResultsSeq}). 
Introducing \textbf{operator sequence coverage metrics may help improve the quality of test suites}, to promote the diversity of the models tested.
%key to improving the quality of test suites, since they can ensure that test suite models are sufficiently diverse.
These metrics could also be incorporated into fuzzers.
Since some nodes or sequences are more likely to cause failures (\cref{tab:RQ4AnalysisResultsSeq}), a coverage-guided fuzzing approach could help identify them.

\ifJOURNALEXTENSION
In light of our results and converter test suite examination, we believe that testing tools for model converters are an important direction for future research. 
As such, we propose a tripartite testing framework that can be used in conjunction with existing methods: (1) Converters need to be tested from the ``framework-side" \ie if test models can be converted at all, (2) from the ``runtime-side", \ie testing whether converted models can be used by runtimes, and (3) incorrect behavior of successfully converted and used models needs to be tested.
Our results in~\cref{sec:res:RQ4} conceptually demonstrate this type of testing.
For example, using real and unconstrained synthetic models will prove effective at identifying (1) and (2) --- "framework-side"  and "runtime-side" issues. 
Whereas constrained synthetic models can use to identify the incorrect behavior of converted models.
\fi

\ifJOURNALEXTENSION
\PJ{The results of this testing would evolve over time too. For example, initially, new operators would cause issues in (1), then later in (2) and finally in (3). As frequencies in each part shift towards 3 that would mean that you becoming more interoperable}
\fi

\ifJOURNALEXTENSION
In light of our results and converter test suite examination, we believe that \WJ{future research on DL converter testing can be focused on different testing methods. (remove the rest of this sentence.)}testing tools for model converters are an important direction for future research. 
We believe that converter testing can be improved by using diverse \emph{real-world} (pre-trained models) --- utilizing the pre-trained model infrastructure within the DL ecosystem\footnote{The broader PyTorch team did so in their development of PyTorch 2.0~\citep{pytorch_get_started}.} --- and the \emph{synthetic} models using DL model generation techniques.
Synthetic model generation technologies also have room for improvement, with existing model generation techniques such as NNSmith \citep{NNSmith} being insufficient due to a majority of synthetic models being unable to be tested (\cref{tab:RQ4ConversionResults}).
We found that synthetic model generation must be constrained from both the framework side and the runtime side. 
\WJ{This is about end-to-end test, we need one more paragraph on unit testing and differential testing.}

% \myparagraph{Testing subjects.}
Putting together our recommendations we propose a tripartite testing framework for model converters: (1) Converters need to be tested from the ``framework-side" \ie crashes need to be induced when converting a model, (2) from the runtime side, \ie crashes need to be induced when loading the model to the runtime, and (3) incorrect behavior need to be induced. Each part of this framework ensures the correctness of a specific converter property --- (1) ensures input model compatibility, (2) ensures converted model validity, and (3) ensures converted model behavioral correctness.

\WJ{Here our prose inform that NNSmith is insufficient. We can add some discussion about what is the ``sufficient'' converter testing. I put a draft here: ``Using synthetic models generated by NNSmith, soem part of the test suites might be repetitive (Is this true?). We highlight that it is more sufficient to generate specific test suites for DL converters, \eg layer-wise or operator-wise. In such case, we can not only avoid the repetitive testing but could also effectively generate test suites based on the latest changes to converters.''.}
\PJ{I think what would be considered proper converter testing is tripartite: (1) You need to test converters from the framework side, i.e., testing stuff that would make the converter crash, (2) Test from the downstream side, is the converter producing stuff that is correct for downstream us, i.e., ORT crashes, and finally, test converted behavior, i.e., incorrect behaviors. This is what }
\fi

% Fuzzing model converters could also improve the relatively simple end-to-end tests in model converter test suites (\cref{sec:dis:conv_test}), but converter-specific fuzzing tools are lacking. 
% Tools such as NNSmith are designed for the testing of compilers --- we found they need to be unnecessarily constrained for the testing of converters (\cref{sec:meth:RQ3}. % to reduce marred with false positives.
% As guidance, we note that to test DL interoperability software, a fuzzer must satisfy two constraints: the generated input must be a legal converter input (to exercise the system) that produces a legal runtime input (so that differential testing is possible).
% Fuzzers that consider both constraints would have fewer false positives. % and provide a direct way to test converters.
% % Fuzzers specifically for converters would minimize the number of false positives (\cref{tab:RQ4ConversionResults})

Identifying behavioral changes after model conversion is critical, this is typically done with an end-to-end test. 
End-to-end testing carries with it the need to assess the outcome: is the converted model acceptable, and
%Some error is expected during conversion.
how much error is tolerable~\citep{jiang2023challengespracticesdeeplearning,huggingface_transformers_serialization}?
Engineers currently use a variety of heuristic tolerances derived from experience (\cref{sec:meth:H2}).
Theoretical and empirical examination of the expected and acceptable tolerance would benefit DL model converter testing.
%many test suites utilize end-to-end tests to verify the numerical agreement between original and converted models, but we were unable to find a source for thresholds used by HuggingFace~\citep{huggingface_transformers_serialization}. 

%\vspace{-0.08cm}
%\vspace{-7pt}

% \vspace{-0.1cm}
\subsection{Comparison to Prior Studies}

\subsubsection{DL Model Converters}
Openja \etal evaluated DL model converters by converting 5 models~\citep{ConvertingDLempirical}.
% They evaluated model conversion to the ONNX and CoreML formats. 
% For ONNX they use the \code{torch.onnx} and \code{tf2onnx} converters.
We extend their findings in two directions.
First, we conducted a failure analysis (\cref{sec:theme1}) to study \emph{how} converters fail.
Second, we convert models systematically (\cref{sec:theme2}) and analyze them to understand \emph{why} they fail.
We considered both realistic and synthetic models.
For realistic models, rather than picking 5 real-world models, we used \TotalRealModels models from a large model corpus.
For synthetic models, we adapted a DL model generation tool to conduct a bounded systematic exploration of converter behaviors.
We omitted measurements of model size, adversarial robustness, and prediction accuracy, to focus instead on measuring the common failure modes (crashing and behavioral differences) identified in our failure analysis.
Our analysis reveals that converters can successfully convert many real model but synthetic models are more prone to failure, moreover analyzing the synthetic models reveals that particular operator sequences co-occur with failures. 
%We can therefore report on the risks of DL model converters: both the criticality and the likelihood of failures.

\subsubsection{DL Compilers}

Shen \etal studied failures in DL compilers~\citep{DLcompilerbugs}, including the components that load models from DL frameworks and from common intermediary formats such as ONNX.
Many of their results generalized to DL model converters:
  (1) we were able to adapt their taxonomies of failure symptoms and causes to DL model converters (\cref{sec:Methods-DataSelection},~\cref{tab:taxonomy_causes_concise});
  and
  (2) as shown in~\cref{tab:symptoms}, we found a similar distribution of failure symptoms for DL model converters.
However, the causes of failures between the two contexts differed:
  as noted in~\cref{sec:RQ1:causes},
  \emph{Incompatibilities} were more common in DL model converters,
  while
  \emph{Algorithmic Errors} were more common in DL compilers.
  
This causal asymmetry may be attributable to differences in the requirements of DL model converters and DL compilers.
The purpose of DL model converters is \emph{interoperability} (\cref{B:Interoperability}), making compatibility failures a focus and reducing the need for optimizations.
Hardware-specific optimizations are left to the consumer of the model.
Conversely, DL compilers must provide both compatibility and hardware-specific optimizations.
%Conversely, although DL compilers need to ingest models compatibly, they must also provide hardware-specific optimizations.
%The convertibility and semantic correctness of models is the key requirement as such \emph{Incompatibility} and \emph{Node Type Problem} causes are prominent.
%These causes represent failures at  

\subsubsection{Methodological Innovations on the ``Bug Study''}
Our study began with a survey that offered valuable insights into the problem domain, guiding our design of the subsequent failure study, including the selection of the interoperability tool and the creation of a failure taxonomy (\cref{sec:theme1}).
This methodological progression, from survey to failure study, and then to hypothesis testing about the failures, diverges from the typical failure study in software engineering.
Those studies typically focus on the ``middle step''~\citep{amusuo_reflections_2022}, understanding the root causes of defects and the conditions under which they manifest (\eg~\citep{islam2019comprehensive, Zhang2019CommonChallengesinDevelopingDLApplications}). %, rather than examining the broader implications and systematically testing interventions to mitigate these failures. 
% which often stops at the failure analysis. 
Our hypothesis testing phase, in particular, adds understanding by providing an estimate of failure rates, an aspect seldom addressed in standard failure studies. 
%For example, the likelihood of failure as measured by our \textit{macro analysis} (\cref{sec:meth:H2}) paints a different picture than the failure analysis.
We suggest that this holistic approach, integrating a survey, failure study, and hypothesis testing, may be a useful methodology for future failure studies.

\ifJOURNALEXTENSION
\subsection{Does It Matter Who Owns A Converter?}\label{sec:disc:conv_own}

Our study considered the DL model converters for the two most popular DL frameworks: \code{torch.onnx} (PyTorch) and \code{tf2onnx} (TensorFlow).
Many of our results were comparable between these two converters, suggesting that they will generalize to, and can thus inform the engineering of, other DL model converters.
The most notable disparity between the two converters was in the causes of failures.
As~\cref{tab:causes} shows,
  \emph{Algorithmic Errors} occurred 3x as often in the TensorFlow converter (18\%) than the PyTorch converter (6\%).
  % and these often occurred
  % in the graph optimization component (\cref{tab:location}).
Failures within the \emph{Graph Optimization} stage are more prevalent in the \cref{tab:location} stage for \code{tf2onnx}.

We conjecture an explanation for this disparity:
 the owners of the PyTorch ONNX model converter may have greater expertise in the PyTorch model representation, and this manifests with a lower failure rate in converter components that perform complex logic over that representation.
We examined the two DL model converter repositories and find some support for this conjecture.
The \code{torch.onnx} converter is maintained by the PyTorch engineering team, and the converter's implementation is integrated with internal PyTorch interfaces.
In contrast, the \code{tf2onnx} converter is maintained by the ONNX organization, and the implementation relies on the public TensorFlow interfaces.
We therefore suggest further study of the potential effect of ownership of interoperability software, particularly in the common intermediary pattern.

\JW{This section ought to be emphasized, this in my opinion is very significant for people who want to use these converters, as doing research in the maintenance team}
\fi

% \subsection{On the relations of changes and failures}

% \vspace{-0.2cm}
\section{Threats to Validity} \label{sec:Threats}

We discuss three types of threats to validity~\citep{wohlin2012experimentation}.
Taking into consideration the criticisms of Verdecchia \etal~\citep{verdecchia2023threats}, we focus on substantive threats that might influence our findings.

\JD{This needs to be updated for the new framing.}
\JD{Wrong RQ numbers etc}

\textbf{Construct Threats} are potential limitations of how we operationalized concepts.
We mitigated definitional concerns for RQ2 by adapting existing taxonomies in our failure analysis.
We defined failures consistently with other works (GitHub issues closed with a repair), though we note there are other means of failure discovery~\citep{Aranda2009SecretLifeofBugs}.
To answer RQ3, we conservatively defined failure in terms of clear misbehaviors --- when ONNX converter output is incompatible with ONNX Runtime, or the model's behavior changes.
This may mask some ``Crash'' failures in the ONNX converter.
In addition, different measures of behavioral difference are possible, such as the L1- or L2- distance.
We used the measure recommended by PyTorch.
\GKT{I could use a few words here. Is HuggingFace using the L1- or L2- distance? If not, we may need a one-liner to describe the distance measure being used.}

%\vspace{-5pt}
\textbf{Internal threats} are those that affect cause-effect relationships.
We specifically allowed respondents to leave answers blank to improve the response rate.
As a result, the results for interoperability tool use in RQ1 may be biased as only 48/92 respondents use an interoperability tool (\cref{sec:meth:sampling}).
However, of the 48 respondents who reported using interoperability tools, 98\% (47/48) gave analyzable data.

Our failure characterization in RQ2 was manual. %, with the threat of subjectivity.
We mitigated subjectivity via interrater agreement, on a pilot sample and a subsequent tranche during the full analysis (\cref{sec:meth:class}).
In~\cref{sec:meth:H1} we noted several approximations in our test of hypothesis H$_{{RQ}_3}$.
We tested hypothesis H$_{{RQ}_4}$ indirectly,
using a frequency analysis
of operators and sequences in mismatched models
rather than directly testing the discovered subsequences. 
% Our test for \textbf{H2} relies on the frequency of occur
% To test our hypothesis --- if incorrect conversions result from the presence of certain operator types and operator sequences ---  we use models that are mismatching and infer from them whether operator types or sequences cause incorrect conversion.
% and then try to determine if they contain unique operators and sequences 
%To mitigate the risk of subjectivity,
%  we measured interrater agreement to assess the consistency of taxonomies, and found strong agreement in a pilot study.
%We therefore used a single rating in answering RQ1-2, with two raters working on distinct failures.
%by utilizing drift-testing as mentioned in \cref{sec:meth:class} and only utilizing a single rater because a high inter-rater agreement was measured.
%Our measurement of the relationship between changes and failures did not take into account the fact that failures may lag changes. 

%\vspace{-5pt}
\textbf{External threats} may impact generalizability.
%The survey instrument is part of a broader investigation of DL interoperability.
For RQ1, the surveyed population (HuggingFace users) may not fully reflect all ONNX users.
However, it is an appropriate population, and we sampled at a confidence level of 90\%.
%The results for RQ1 may not fully reflect the use of specific interoperability tools.
%To mitigate this, 
For RQ2-4, we examined two DL model converters for one interoperability framework (ONNX).
Our results may not extend to other model converters nor other interoperability tasks.
As mitigation, our data suggest that ONNX is the most popular DL interoperability tool.
For ONNX, our results were similar across the two converters, suggested generalizability in two dimensions:
  DNN modeling approaches,
  and
  converter owners (\cref{sec:Methods-DataSelection}).
With regard to our methodology for RQ3-4, we generated synthetic models with the NNSmith tool.
Other methods~\citep{Muffin_Gu_Luo_Zhou_Wang_2022, COMET_Li_Cao_Tian_Tsz_On_Li_Wen*_Cheung*_2023, LEMON_Wang_Yan_Chen_Liu_Zhang_2020} might have different results.
%, mutating seed sets of models (maybe more realistic) rather than generating them entirely at random as NNSmith does.

\ifJOURNALEXTENSION
\WJ{Do we know the coverage of DL operators using the realistic and synthetic models? The results may also not extend to other operators (if they are not existed in HF models and NNSmith could not generate them).}
\fi

%This is in part reflected by the fact that our precision measures were high (\TorchPrecision, and \TFPrecision) while our recall measures were low (\TorchRecall and \TFRecall), indicating that we had a higher rate of false negatives as a result of our filter.
%Our study focuses on the converter for the ONNX intermediate format. 
%Our findings may not extend to other model converters or other intermediaries.

% \vspace{-0.2cm}
\section{Conclusion} \label{sec:Conclusions}

%The deep learning (DL) software ecosystem is expanding.
DL model converters play a crucial role in providing interoperability between DL frameworks, registries, and runtimes.
Understanding the nature of failures in DL model converters enables engineers to make informed decisions when relying on them.
We conducted the first failure analysis of DL model converters, considering the PyTorch and TensorFlow converters for the popular ONNX intermediate representation. % for DL interoperability. 
The most common symptoms of failure are crashes and, perhaps more concerningly, models that misbehave on certain inputs.
Of the five stages of a typical model converter, one stage (node conversion) accounts for $\sim$75\% of the defects.
%In the latest version of the two studied converters, we found that
%Real models convert mostly without error, but synthetic models expose erroneous converter behavior.
A deductive description of the causes of erroneous converter behavior remains elusive --- individual operators are not predictive of failure; sequences of operators may be correlated. %Controlled measurement is challenging.
%Test suites of model converters are weaker than either (1) a large corpus of real models, or (2) synthetic generation of models.
% George comment: For now, I am not having a new paragraph here in an effort to recover some space.
Our findings suggest that in ONNX, engineers can rely on model converters but should validate the result for behavioral consistency.
Through a mix of positive and negative results, we exposed several directions for further improvement of DL model converters.
The main opportunities are new measurements:
  a new architectural coverage measure for a DL model converter,
  and
  a refined measure of tolerance after conversion.

\iffalse
\JD{Below is the old stuff that is confusing}
We studied ONNX model converter failures
  retrospectively, via failure analysis,
  and
  prospectively, via testing converters using pre-trained and randomly generated models.
This methodological combination measures both criticality and likelihood of failures, allowing us to discuss the risks of converters.
In previously-reported failures, the criticality of failures is high: users reported incorrect model behaviors in one-third of cases.
Prospectively, in current implementations the likelihood of encountering a failure is moderate: models crash or behave incorrectly \IncorrectCrashTotalPercent\% of the time, usually on synthetic models.
  %$\sim$90\% of reported failures in DL model converters manifest as crashes ($\sim$55\%) and incorrect models (33\%).
%and incompatibilities of converters with ONNX have often been a source of failures. 
We conclude that DL model converters are stable enough to use, but that users should test the converted models carefully, particularly after new releases of or significant updates to the ONNX specification.
Our results motivate research in the development and testing of DL interoperability software.
\fi

\ifARXIV
\section{Acknowledgements
} \label{sec:Ackowledgements}
This work is supported by Cisco, Google, and by NSF \#2107020, \#2107230, \#1813935, and \#2104319. 
%We thank Purdue's PurPL center for financial support. % (\JD{IS THIS APPROPRIATE HERE? I FORGET WHO WAS FUNDED ON WHAT.})
%We thank S. Sreekanth and Mahir for feedback on the manuscript.
\else

\fi

\ifARXIV
\else

% \vspace{-0.2cm}
\section{Data-Availability Statement} \label{sec:DataAvailability}

This paper is accompanied by an artifact~\citep{jajal_2024_12667479} to support replication and reproduction.
The artifact includes
  the survey data for Theme 1,
  the failure analysis data for Theme 2,
  and
  the processed results of data associated with the hypothesis evaluation for Theme 3.
The artifact also includes the modified NNSmith model generator, and the specific synthetic models used in the study.
The source code for the artifact is available at  {\url{https://github.com/PurdueDualityLab/issta24-onnx-artifact}}.

% It omits raw synthetic models due to storage limits.
\fi

\section*{Acknowledgements
} \label{sec:Ackowledgements}
This work is supported by Cisco and Google, as well by NSF awards \#2107020, \#2107230, \#1813935, and \#2104319. 

%\clearpage

%A supplement is on HotCRP, including code and tables/figures omitted for space.
%An anonymized supplement is available at: \url{https://github.com/Anon-FSE/FSE-24}, containing
%\JD{describe, X, Y, Z.}.
%We will submit to the artifact track, for ``available + reusable'' badges.

\iffalse
We perceive no ethical concerns related to our study.
This study involved no human subjects.
\fi

\JD{Here is a list of problematic references. They are missing URLs or have weird URLs or should have the venue expanded. They are numbers: 11, 22, 24, 51, 58, 70. ALSO, numbers 58 and 59 are the same reference (Petroski), so please only cite it with one cref name.}

\ifARXIV
\else
  \balance
\fi
% \pagebreak
% \newpage

% SPACE HACK: ieeetr cites in order of appearance, not alphabetical, so we end up with more compressed [1-5]-type references.
%\bibliographystyle{acm}
% \bibliographystyle{ieeetr}
\bibliographystyle{ACM-Reference-Format}  
\bibliography{bib/references, bib/DualityLab, bib/WenxinZotero}

% \ifJOURNALEXTENSION
\ifAppendix

\pagebreak
\appendix
\section{Summary of Appendices}\label{appendix:a}
\begin{itemize}
    \item \cref{appendix:a}: Appendix Summary.
    \item \cref{appendix:theme2}: Taxonomies and Extended Tables from \cref{sec:theme1}.
    \item \cref{appendix:theme3}: Additional Tables from \cref{sec:theme2}.
\end{itemize}

\section{Theme 2 -- Failure Analysis (Of ONNX)}\label{appendix:theme2}
\subsection{Taxonomies}
\begin{itemize}
    \item \cref{tab:taxonomy_symptoms_appendix}: Adapted taxonomy of failure symptoms.
    \item \cref{tab:taxonomy_causes_appendix}: Adapted  taxonomy of failure causes.
\end{itemize}

\begin{table}[b]
%\begin{wraptable}{r}{0.70\columnwidth}
\caption{
\small
    Adapted taxonomy of failure symptoms~\citep{DLcompilerbugs}.
    \emph{Emphasis}: wording changed.
    \emph{Hang} symptom omitted (unobserved).
    This is a tabular version of symptoms presented in~\cref{sec:Methods-Theme1}.
}
  % Reduce font size?
  \small
  \footnotesize
\begin{tabular}{cp{6cm}}
\toprule
\textbf{Symptom} & \textbf{Definition} \\
\toprule
Crash           & The \emph{model converter} terminates unexpectedly during \emph{conversion}, usually producing an error message. \\
Wrong \emph{Model}     & The \emph{model converter} behaves unexpectedly without crashing, producing a wrong intermediate or final result. \\
Bad Performance & The time cost or memory consumption of the \emph{model converter} is much larger than developer/user expectations (\eg the performance achieved on the previous version during regression testing or performance achieved by certain hardware). \\
Build Failure   & Installation of \emph{model converter} or dependencies fails. \\
Unreported      & The symptom cannot be identified by analyzing the corresponding issues including code changes, discussions, and related issues. \\
%JD: Below this is the Hang symptom that we did not observe
\midrule
\midrule
Hang            & This symptom means that the \emph{converter} keeps running for a long period of time without producing the expected result. \\
\bottomrule
\end{tabular}
\label{tab:taxonomy_symptoms_appendix}
%\vspace{-0.35cm}
\end{table}

\subsection{Symptom and Cause Data}
\begin{itemize}
    \item \cref{tab:causes_appendix}: Distribution of causes in \code{tf2onnx} and \code{torch.onnx}.
    \item \cref{tab:cau_sym_appendix}: Joint distribution of causes and symptoms (unabridged). 
\end{itemize}

\section{Theme 3 -- Deeper Causes}\label{appendix:theme3}
\cref{tab:RQ4AnalysisResultsNodes_appendix} contains results of model operator analysis as described in~\cref{sec:meth:H3}.

\begin{table}[b]
    \footnotesize
    \caption{\small
        Distribution of causes in \code{tf2onnx} and \code{torch.onnx}.
        The top-5 causes in terms of frequency are shown. 
        The \emph{Others} are: \emph{Concurrency}, \emph{Documentation}, \emph{Incorrect Numerical Computation}, \emph{Misconfiguration}, \emph{Testing}, \emph{Incorrect Exception Handling}, \emph{Typos}, and \emph{Incorrect Assignment} (each has $<3$\% representation).
        The final column shows the cause distribution from Shen \etal~\citep{DLcompilerbugs}.
        Note the larger proportion of \emph{Incompatibility} failures in DL model converters (first row).
        }
    % Space
    % \small
    \begin{tabular}{l|c|r|r||r||p{1.6cm}}
    \toprule
    \multicolumn{2}{c|}{\textbf{Causes}}                          &  \textbf{TF} &  \textbf{PT} &  \textbf{Total} & \textbf{Comp. to DL compilers}~\citep{DLcompilerbugs}\\
    \toprule
    \multirow{3}{*}{Incompatibility}    
                                         &      External &       \textbf{23} &       \textbf{32} &     \textbf{55 (\IncompExt)} &  \ShenIncompExt (\ShenIncompExtPercent)\\  
    &      Internal &        2 &        0 &      2 (\IncompInt) & \ShenIncompInt (\ShenIncompIntPercent) \\
                                         &      Resource &        0 &        0 &      0 (\IncompRes) & \ShenIncompRes (\ShenIncompResPercent)\\  
    \midrule
    \multirow{3}{*}{Type Problem}        &          Node &       \textbf{21} &       \textbf{25} &     \textbf{46 (\TypeNode)} & \ShenTypeNode (\ShenTypeNodePercent)\\
                                         &  Conventional &        3 &        2 &      5 (\TypeConv) & \ShenTypeConv (\ShenTypeConvPercent)\\
                                         &        Tensor &        1 &        2 &      3 (\TypeTensor) & \ShenTypeTens(\ShenTypeTensPercent)\\  
    \midrule
    \multicolumn{2}{c|}{Algorithmic Error}               &       \textbf{18} &        \textbf{6} &     \textbf{24 (\Algo)} & \ShenAlgo (\ShenAlgoPercent) \\
    \midrule
    \multicolumn{2}{c|}{Shape Problem}                   &       9 &        12 &     21 (\Shape) & \ShenShape (\ShenShapePercent) \\
    \midrule
    \multicolumn{2}{c|}{API Misuse}                        &        6 &        6 &     12 (\API) & \ShenAPI (\ShenAPIPercent) \\
    % \midrule
    % \multicolumn{2}{c|}{Misconf.}                        &        4 &        2 &      6 (\Misconf)\\
    % \midrule
    % \multicolumn{2}{c|}{Testing}                         &        1 &        3 &      4 (\Testing)\\
    % \midrule
    % \multicolumn{2}{c|}{Inc. Except. Handling}           &        1 &        2 &      3 (\IncExceptHandling)\\
    % \midrule
    % \multicolumn{2}{c|}{Typo}                            &        1 &        2 &      3 (\Typo)\\
    % \midrule
    % \multicolumn{2}{c|}{Inc. Assign.}                    &        0 &        2 &      2 (\IncAssign) \\
    % \midrule
    % \multicolumn{2}{c|}{Inc. Num. Comp.}                 &        0 &        0 &      0 (0\%)\\
    % \midrule
    % \multicolumn{2}{c|}{Concurrency}                     &        0 &        0 &      0 (0\%)\\
    % \midrule
    % \multicolumn{2}{c|}{Documentation}                   &        0 &        0 &      0 (0\%)\\
    \midrule
    \multicolumn{2}{c|}{Others}                          &       \BinnedOtherTF & \BinnedOtherPT &     \BinnedOther (\BinnedOtherPercentage) & \ShenOthers(\ShenOthersPercent) \\
    % \midrule 
    \midrule
    \midrule
    \multicolumn{2}{c|}{\emph{Total}}                           &      \emph{100} &      \emph{100} &    \emph{200 (100\%)} &  \emph{\ShenTotalCauses (\ShenTotalCausesPercent)}\\  
    \bottomrule
    \end{tabular}
    \label{tab:causes_appendix}
\end{table}

\begin{table}[b]
    \centering
% \begin{wraptable}{r}{0.45\columnwidth}
    \caption{\small
    Model Operator Analysis Results.
    As described in~\cref{sec:meth:H3}: 
    Mismatched $\circledstep{1}$, all operators that mismatched models contain; 
    Correct $\circledstep{2}$ all operators that correct models contain;
    Test Suite $\circledstep{3}$ all operators that test suites contain.
    $\circledstep{1} \setminus \circledstep{2}$ are all operators in $\circledstep{1}$ but not in $\circledstep{2}$.
    $\circledstep{2} \setminus \circledstep{1}$ are all operators in $\circledstep{2}$ but not in $\circledstep{1}$.
    $\circledstep{1} \setminus \circledstep{3}$ are all operators in $\circledstep{1}$ but not in $\circledstep{3}$.
    $\circledstep{3} \setminus \circledstep{1}$ are all operators in $\circledstep{3}$ but not in $\circledstep{1}$.
    % \circledstep{4} - \circledstep{2} are all unique sequences that are \emph{only} present in test suite models.
    % The reduced sequences are found by applying the reduction as described in~\cref{sec:meth:model_analysis}
    }
   % Reduce font size?
   \small
    \begin{tabular}{c|r|r}
    \toprule
    \textbf{Model Type}   &  \emph{\textbf{\code{tf2onnx}}} &  \emph{\textbf{\code{torch.onnx}}} \\
    \midrule
    Mismatched $\circledstep{1}$ &   54  & 58 \\
    Correct $\circledstep{2}$ &    52  & 59 \\
    Test Suite $\circledstep{3}$ &   54 & 16\\
    \midrule
    $\circledstep{1} \setminus \circledstep{2}$ &   0  & 1 \\
    $\circledstep{2} \setminus \circledstep{1}$ &   2  & 0 \\
    $\circledstep{1} \setminus \circledstep{3}$ &   21  & 46 \\
    $\circledstep{3} \setminus \circledstep{1}$ &   21  & 4 \\
    \bottomrule
    \end{tabular}
    \label{tab:RQ4AnalysisResultsNodes_appendix}
%\vspace{-0.15cm}
\end{table}

\begin{table*}[h!]
    %\small
    \caption{
    Unabridged joint distribution of causes and symptoms.
    %We bold notable occurrences and drop the \emph{Concurrency} and \emph{Documentation} rows because they contain all zeros. 
    The majority of \emph{Crashes} result from \emph{Incompatibilities} and \emph{Type Problems}.
    \emph{Algorithmic Errors} that result in \emph{Wrong Models} occur more often in \code{tf2onnx}.
    \cref{tab:cau_sym_abridged} is derived from this by binning the top-5 causes in terms of frequency, with the rest binned as \emph{Other}.
    The \emph{Others} are \emph{Concurrency}, \emph{Docs}, \emph{Incorrect Numerical Computation}, \emph{Misconfig}, \emph{Testing}, \emph{Incorrect Exception Handling}, \emph{Typos}, and \emph{Incorrect Assignment}.
    Rare symptoms are likewise binned as \emph{Other}.
    }
    % Space
    %\small
\begin{tabular}{l|cc|cc|cc|cc|cc||cc}
\toprule
\multirow{2}{*}{\backslashbox[15em]{Causes}{Symptoms}} & \multicolumn{2}{c|}{Crash} & \multicolumn{2}{c|}{Wrong Model} & \multicolumn{2}{c|}{Bad Performance} & \multicolumn{2}{c|}{Build Failure} & \multicolumn{2}{c||}{Unreported} & \multicolumn{2}{c}{Total} \\
{} &    PT &  TF &         PT &  TF &              PT & TF &            PT & TF &         PT &  TF &    PT &   TF \\
\midrule
API Misuse                      &     5 &   6 &          1 &   0 &               0 &  0 &             0 &  0 &          0 &   0 &     \emph{6} &    \emph{6} \\
Incompatibility                 &    \textbf{28} &  \textbf{19} &          3 &   4 &               0 &  0 &             0 &  0 &          1 &   2 &    \emph{32} &   \emph{25} \\
Type Problem                    &    \textbf{14} &   \textbf{8} &         \textbf{13} &  \textbf{17} &               1 &  0 &             0 &  0 &          1 &   0 &    \emph{29} &   \emph{25} \\
Shape Problem                   &     4 &   5 &          7 &   4 &               0 &  0 &             0 &  0 &          1 &   0 &    \emph{12} &    \emph{9} \\
Incorrect Numerical Computation &     0 &   0 &          0 &   0 &               0 &  0 &             0 &  0 &          0 &   0 &     \emph{0} &    \emph{0} \\
Incorrect Assignment            &     2 &   0 &          0 &   0 &               0 &  0 &             0 &  0 &          0 &   0 &     \emph{2} &    \emph{0} \\
Incorrect Exception Handling    &     1 &   1 &          1 &   0 &               0 &  0 &             0 &  0 &          0 &   0 &     \emph{2} &    \emph{1} \\
Misconfiguration                &     0 &   1 &          0 &   0 &               0 &  0 &             2 &  2 &          0 &   1 &     \emph{2} &    \emph{4} \\
% Concurrency                     &     0 &   0 &          0 &   0 &               0 &  0 &             0 &  0 &          0 &   0 &     0 &    0 \\
Algorithmic Error               &     3 &   4 &          \textbf{3} &  \textbf{10} &               0 &  2 &             0 &  0 &          0 &   2 &     \emph{6} &   \emph{18} \\
Typo                            &     2 &   0 &          0 &   0 &               0 &  0 &             0 &  1 &          0 &   0 &     \emph{2} &    \emph{1} \\
Testing                         &     1 &   0 &          1 &   0 &               0 &  0 &             0 &  0 &          1 &   1 &     \emph{3} &    \emph{1} \\
% Documentation                   &     0 &   0 &          0 &   0 &               0 &  0 &             0 &  0 &          0 &   0 &     0 &    0 \\
Others                          &     2 &   6 &          1 &   0 &               0 &  0 &             0 &  0 &          1 &   4 &     \emph{4} &   \emph{10} \\
\midrule\midrule
\emph{Total}                    &    \emph{62} &  \emph{50} &         \emph{30} &  \emph{35} &        \emph{1} &  \emph{2} &    \emph{2} &  \emph{3} &          \emph{5} &  \emph{10} &   \emph{100} &  \emph{100} \\
\bottomrule
\end{tabular}
\label{tab:cau_sym_appendix}
\end{table*}

\begin{table*}[b]
    \caption{
    Adapted taxonomy of failure causes~\citep{DLcompilerbugs}.
    \textit{API Misuse}, \textit{Type Problem}, and \textit{Algorithmic Error} are high-level categories with sub-causes underneath them.
    This table is an unabridged version of~\cref{tab:taxonomy_causes_concise}.
%    Taxonomy of failure causes, adapted from~\citep{DLcompilerbugs}.
    \JD{NB: Some of these categories could be SUBTRACTED rather than ADDED in our case, \eg API Misuse?. To do this, I need an updated Table 4. Put the subtracted ones BELOW a double midrule.}
    }
    \small
    % \begin{tabular}{c|>{\arraybackslash}p{12.5cm}}
    \begin{tabular}{c|p{12cm}}
    \toprule
    {\textbf{Causes}} & \textbf{Definition}\\
    \midrule\midrule
    \textbf{API Misuse}       &       High-level category that encapsulates \textit{Wrong API}, \textit{Condition missing/redundancy}, and \textit{API missing/redundancy}.              \\
    \cmidrule{2-2}
    \textbf{Wrong API}  & The developer or user uses the wrong API or wrong arguments in an API. \\
    \cmidrule{2-2}
    \textbf{Condition missing/redundancy} & Developer fails to use (or redundantly uses) a condition check for an API. \\
    \cmidrule{2-2}
    \textbf{API missing/redundancy} &  Developers fail to use (or redundantly use) an API.\\
    \midrule
    \textbf{Incompatibility} &      High-level category that encapsulates \textit{Internal}, \textit{External}, and \textit{Resource}.  \\
    \cmidrule{2-2}
    \textbf{Internal} &  There are API compatibility issues within a model converter caused by API evolution. \\
    \cmidrule{2-2}
    \textbf{External} &  There are API compatibility issues between a model converter and third-party libraries (e.g., ONNX, TensorFlow, ONNX Runtime). \\  
    \cmidrule{2-2}
    \textbf{Resource} &  There are compatibility issues between model converters and external resources. The characteristics of the target device are incompatible with the model converter. \\  
    \midrule
    \textbf{Type Problem}        &          \\
    \cmidrule{2-2}
    \textbf{Node} &  A model converter works on a computational graph, where nodes represent the atomic DL operators (such as convolution and pooling) and edges represent the tensors. Each node takes zero or more tensors as input and produces a tensor as output. \\
    \cmidrule{2-2}
    \textbf{Tensor} &  This relates to the types of tensors. Specifically, a tensor is a multi-dimensional matrix containing elements of a single data type. \\ 
    \cmidrule{2-2}
    \textbf{Conventional} &  There are also conventional variables widely used in the development of traditional software systems. This subcategory refers to the problem involving the types of conventional variables. \\  
    \midrule
    % \multicolumn{2}{c|}{Shape Problem}                   &       9 \\
    \textbf{Shape Problem} & This cause covers, broadly, the shape of inputs and outputs a node can have. These occur during the operation of shape matching, shape transformation, shape inference, layout transformation, etc. \\ 
    \cmidrule{2-2}
    \textbf{Tensor Shape Problem} &  This is related to tensor shape or layout. The tensor shape is the number of elements in each dimension. Layout describes how the shape is represented in memory.  \\
    \midrule
    \textbf{Incorrect Numerical Computation} &  This involves incorrect numerical computations, values, or usage, such as incorrect operators or operands, dividing by 0, and missing or redundant operands. \\
    \midrule
    \textbf{Incorrect Assignment} &  This involves a variable being incorrectly initialized or assigned, or a variable lack initialization. \\
    \midrule
    \textbf{Incorrect Exception Handling} &  This category occurs due to incorrect exception handling. For example, an exception is not thrown when it should be, an example is thrown when it should be, or an incorrect/imprecise exception message is thrown. \\
    \midrule
    \textbf{Misconfiguration} &  This category occurs due to misconfiguration in the model converter. This includes dependency versioning and misconfigured environments. \\
    \midrule
    \textbf{Concurrency}  &  This is a result of incorrect operations on concurrency-oriented structures (e.g., locks, threads, and critical regions). \\
    \midrule
    \textbf{Algorithmic Error}   &  High-level category that encapsulates \textit{Incorrect Optimization} and \textit{Incorrect Tracing}. \\ %This refers to the implementation logic of an algorithm.   \\
    \cmidrule{2-2}
    \textbf{Incorrect Optimization} &  There is an issue with optimizations that occur in the model converter. Examples: incorrect fusing, dead-code elimination, et cetera. \\
    \cmidrule{2-2}
    \textbf{\emph{Incorrect Tracing}} &        There is an issue with the tracing of DL models. This occurs in frameworks that use dynamic models. Dynamic models need to be “traced” to find the nodes in the computational graph prior to the conversion. \\  
    \midrule
    \textbf{Typo}                            &  This is due to trivial mistakes (e.g. slips \citep{Norman_2013}) by developers, e.g., “default” is mistakenly written as “defualt”. \\
    \midrule
    \textbf{\emph{Testing}}                         &        This class contains issues that are a result of incorrect, outdated, or missing documentation. \\
    \midrule
    \textbf{\emph{Documentation}}                   &        This class contains issues related to software tests such as unit tests. This covers “flaky” tests, missing tests, broken tests, or the addition of new tests. \\
    \midrule
    \textbf{Others}                          &       This cause is relegated to causes that occur infrequently and do not belong to the other causes. \\
    \bottomrule
    \end{tabular}
    \label{tab:taxonomy_causes_appendix}
\end{table*}

\fi
\end{document}